\documentclass{elsarticle}

\usepackage[utf8]{inputenc}
\usepackage{graphicx}
\usepackage[T1]{fontenc}    
\usepackage{hyperref}       
\usepackage{url}            
\usepackage{booktabs}       
\usepackage{amsfonts}       
\usepackage{nicefrac}       
\usepackage{microtype}      
\usepackage{xcolor}         
\usepackage{multirow}

\usepackage{amsmath,amssymb,amsfonts}
\usepackage{amsthm}
\usepackage{natbib}

\DeclareMathOperator*{\argmin}{arg\,min}

\hypersetup{
    colorlinks=true,
    linkcolor=blue,
    citecolor=blue,
    urlcolor=magenta,
    pdftitle={GA-Agent: Large Language Models as Hyperparameter Optimizers for Evolutionary Controller Synthesis},
    pdfauthor={Anonymous Authors},
}

\usepackage{mdframed}
\usepackage{xcolor, soul}
\usepackage{listings}          
\newmdenv[
backgroundcolor=yellow!20,
linewidth=0pt,
leftmargin=0pt, rightmargin=0pt,
innertopmargin=0pt, innerbottommargin=0pt,
innerrightmargin=0pt, innerleftmargin=0pt,
skipabove=5pt, skipbelow=5pt,
nobreak=true
]{shadedbox}
\usepackage{colortbl}
\usepackage{soul}
\definecolor{lightgreen}{rgb}{0.56, 0.93, 0.56}
\sethlcolor{yellow}
\newmdenv[
backgroundcolor=red!20,
linewidth=0pt,
leftmargin=0pt, rightmargin=0pt,
innertopmargin=0pt, innerbottommargin=0pt,
innerrightmargin=0pt, innerleftmargin=0pt,
skipabove=5pt, skipbelow=5pt,
nobreak=true
]{shadedbox2}
\lstdefinestyle{promptstyle}{
    basicstyle=\ttfamily\footnotesize,
    backgroundcolor=\color{gray!8},
    frame=single,
    framerule=0.4pt,
    rulecolor=\color{gray!50},
    breaklines=true,
    breakatwhitespace=false,
    columns=fullflexible,
    keepspaces=true,
    showstringspaces=false,
    tabsize=2,
    captionpos=b,
}
\usepackage{tocloft}   

\journal{Robotics and Autonomous Systems}

\begin{document}

\begin{frontmatter}

\title{GA-Agent: Large Language Models as Hyperparameter Optimizers for Evolutionary Controller Synthesis}

\author[a1]{Mohammad Narimani}
\author[a1]{Seyyed Ali Emami}
\affiliation[a1]{organization={Department
of Aerospace Engineering, Sharif University of Technology},
            city={Tehran},
            postcode={14588-89694},
            state={Tehran},
            country={Iran}}
\affiliation[a1]{organization={Artificial Intelligence in Design and Complex Systems (AIDACS) Group, Sharif University of Technology},
            city={Tehran},
            state={Tehran},
            country={Iran}}

\begin{abstract}
Tuning proportional-integral-derivative (PID) controllers to satisfy competing performance objectives, namely, low tracking error, fast settling, limited overshoot, and moderate control effort, remains a labor-intensive task that demands significant domain expertise. Genetic algorithms (GAs) offer a principled, gradient-free approach to optimizing controller gains against a weighted fitness function, yet their practical success depends critically on meta-level design choices: population size, generation budget, gain bounds, and fitness weights. These hyperparameters are conventionally set via manual trial-and-error or computationally expensive cascade (bilevel) optimization, both of which highlight a fundamental tension: the GA excels at dense numerical search, yet configuring it requires high-level, context-dependent reasoning that is inherently semantic. We propose \emph{GA-Agent}, a framework that decouples these two modes by assigning them to complementary computational substrates. A standard GA handles low-level optimization of PID gains, while a large language model (LLM) agent operates at the meta-level, observing outcomes of completed GA runs, diagnosing performance gaps relative to user-specified control objectives, and proposing updated GA configurations. The architecture follows established agentic design patterns: structured memory of past attempts, quantitative goal translation, resource-aware termination, and outcome-driven routing. We evaluate GA-Agent on eight control case studies spanning diverse plant dynamics (DC motor, inverted pendulum, aircraft pitch, autonomous underwater vehicle, and others). GA-Agent achieves $100\%$ success on all benchmarks, consistently outperforming a Regular GA with fixed hyperparameters in both solution quality and sample efficiency, and matches or surpasses a Cascade-GA baseline while reducing function evaluations by one to two orders of magnitude, typically converging within one to three optimization attempts. A comprehensive sensitivity analysis demonstrates robustness across LLM backbones and memory configurations, with a compact memory buffer (size~2--3) and cost-effective models (DeepSeek-V4-Flash at ${\approx}$\$$0.002$ per run) achieving superior performance.
\end{abstract}


\end{frontmatter}

\tableofcontents

\section{Introduction}
\label{sec:introduction}

Feedback controller design remains a cornerstone of modern engineering, with Proportional-Integral-Derivative (PID) controllers deployed in the vast majority of industrial control loops. Despite their conceptual simplicity, tuning PID gains to satisfy multiple, often competing, performance objectives---low tracking error, fast settling, limited overshoot, and moderate control effort---is a nontrivial task that typically demands significant domain expertise. Classical analytical methods (e.g., Ziegler--Nichols, root-locus shaping) provide useful starting points but seldom yield satisfactory performance across all metrics simultaneously, particularly for nonlinear or poorly modeled plants.

Genetic algorithms (GAs) offer a compelling alternative: they impose minimal assumptions on the cost landscape, handle nonconvex and multimodal fitness functions, and naturally accommodate the mixed-integer structure of PID-like parameterizations~\cite{Duriez.2017}. A single GA run optimizes a weighted fitness function
$J = \sum_i w_i \cdot (\text{achieved}_i)^2$
over the space of controller gains, where the weights $w_i$ encode the designer's relative preference among performance metrics such as mean squared error (MSE), settling time, overshoot, and control effort. However, the practical success of any GA run depends critically on design choices that sit \emph{outside} the search loop: population size, generation budget, gain-range bounds, and the very weights $w_i$ that shape the fitness landscape. These meta-level hyperparameters are conventionally set through one of two strategies:
\begin{enumerate}
    \item \emph{Manual trial-and-error}, in which a human expert iteratively adjusts the GA configuration based on inspection of closed-loop responses---a process that is time-consuming, difficult to reproduce, and heavily reliant on individual experience; or
    \item \emph{Cascade (hierarchical) optimization}, in which a higher-level optimizer (itself often a GA) searches over the hyperparameter space of a lower-level GA~\cite{Duriez.2017}. While principled, this approach multiplies the computational cost by the population size and generation count of the outer loop, easily requiring one to two orders of magnitude more function evaluations (NFEs) than a single well-configured GA run.
\end{enumerate}
Both strategies highlight a fundamental tension: the low-level GA excels at dense, gradient-free numerical search, yet configuring it demands the kind of high-level, context-dependent reasoning---interpreting transient-response shapes, diagnosing failure modes, balancing competing objectives---that is inherently \emph{semantic} rather than numerical.

Recent advances in large language models (LLMs) and agentic AI~\cite{gulli.2025} suggest a natural resolution to this tension. LLMs have demonstrated strong capabilities in in-context reasoning, strategy formulation, and iterative refinement from textual and numerical feedback~\cite{Liu.2402,Zhou.2025.ProPS,Ma.2310,Liu.2025}. A growing body of work has begun to explore LLMs as components within optimization and control pipelines---either as direct numerical optimizers of controller parameters~\cite{Narimani.2025,Zhou.2025.ProPS}, reward-function designers~\cite{Xie.2309,Ma.2310}, or advisory modules within adaptive control architectures~\cite{Aghaee.2026,Aydin.2026,Nosrati.2026}. 
Related actor-critic deep reinforcement learning approaches have also been applied to PID tuning for similar underactuated systems~\cite{Udekwe.2024}. 
However, approaches that task the LLM with directly proposing controller gains conflate semantic reasoning (at which LLMs excel) with dense numerical search (at which population-based methods excel), while purely advisory roles underutilize the LLM's capacity for closed-loop, goal-directed decision-making.

In this paper, we propose \textbf{GA-Agent}, a framework that \emph{decouples} these two modes of reasoning by assigning them to complementary computational substrates. A standard GA handles the low-level, gradient-free optimization of PID gains against a weighted fitness function $J$, while an LLM agent operates at the \emph{meta-level}---observing the outcomes of completed GA runs, diagnosing performance gaps with respect to user-specified control objectives, and proposing updated GA configurations (population size, generation count, gain bounds, and fitness weights $w_i$) for the next optimization attempt. The LLM thus plays a role analogous to the high-level optimizer in a cascade-GA scheme, but replaces population-scale exploration with informed, single-shot proposals grounded in semantic understanding of the control problem. This division of labor is the central thesis of our work.

The GA-Agent architecture is structured around established agentic design patterns~\cite{gulli.2025}: \emph{memory} (a buffer of past attempts and their outcomes provides context for the LLM's decisions), \emph{goal setting} (the user's textual control objective and quantitative targets are translated into fitness-function weights), \emph{resource awareness} (computational budget and API cost constraints govern termination), and \emph{routing} (a rule-based mechanism determines whether to invoke the LLM for another attempt or terminate the optimization). A preprocessing agent (WarmUp-Agent) translates the user's problem specification---plant dynamics, control objectives, and simulation parameters---into a structured configuration that initializes the optimization loop. Each cycle of LLM reasoning followed by a complete GA run constitutes one \emph{attempt}; the framework typically converges within one to three attempts.

We evaluate GA-Agent on ten control case studies spanning diverse plant dynamics (e.g., DC motor, inverted pendulum, aircraft pitch, autonomous underwater vehicle). Performance is assessed using a weight-independent baseline cost $\mathcal{L}$ (Eq.~\ref{eq:baseline}) that enables fair cross-run comparison regardless of the fitness weights used in each attempt, alongside success scores and cumulative NFEs. We benchmark against both a Regular GA with fixed, manually tuned hyperparameters and a Cascade-GA with a high-level evolutionary outer loop.

\paragraph{Contributions.} The main contributions of this work are:
\begin{enumerate}
    \item We introduce \textbf{GA-Agent}, an agentic framework that decouples low-level numerical optimization from high-level meta-configuration by pairing a genetic algorithm (GA) with a large language model (LLM) for automated PID controller synthesis. The LLM operates exclusively at the meta-level—iteratively observing GA outcomes and proposing updated hyperparameters and fitness weights—while the GA performs dense numerical search over controller gains. To our knowledge, this is the first work to formalize and systematically evaluate this decoupled LLM--GA architecture for control engineering.

    \item We provide a \textbf{systematic empirical evaluation} across eight control case studies with diverse plant dynamics (DC motor, inverted pendulum, aircraft pitch, AUV, CSTR, ball-and-beam, 2-DoF helicopter, two-link manipulator), demonstrating that GA-Agent (i) achieves $100\%$ success on all benchmarks by jointly satisfying multiple performance targets, (ii) consistently outperforms a Regular GA with fixed hyperparameters in both solution quality and sample efficiency (33\% average NFE reduction), and (iii) matches or surpasses a Cascade-GA hierarchical baseline while reducing function evaluations by one to two orders of magnitude on individual benchmarks ($33\times$ on AUV, $12\times$ on 2-DoF helicopter).

    \item We conduct a \textbf{comprehensive sensitivity analysis} across LLM backbone choice, memory buffer size, and prompt formulation. GA-Agent is robust across these axes. Lightweight models perform best at compact buffers ($B=2$), while heavier models benefit from slightly larger buffers ($B=3$). DeepSeek-V4-Flash emerges as the Pareto-optimal backbone, achieving $100\%$ success at the lowest per-run cost (\$$0.0019$--$0.0026$ depending on exact configuration).
\end{enumerate}

\paragraph{Paper organization.} Section~\ref{sec:related_work} reviews related work on evolutionary methods in control, agentic AI, and LLM-based optimization. Section~\ref{sec:methodology} formalizes the control problem, the GA-based optimization pipeline, and the GA-Agent architecture, drawing explicit analogies between the LLM agent and the high-level optimizer in a cascade scheme. Section~\ref{sec:results} presents comparative results against Regular GA and Cascade-GA baselines, along with the sensitivity analysis. Section~\ref{sec:discussion} discusses implications, limitations, and future directions. Appendices~A--D provide detailed per-case-study results, full comparative tables, and extended sensitivity analyses.

\section{Related Work}
\label{sec:related_work}

This section reviews three interrelated threads of prior work that motivate and contextualize our approach: evolutionary and machine-learning methods for controller design, the emergence of agentic AI paradigms and their design patterns, and recent efforts that leverage large language models for optimization and control engineering.

\subsection{Machine Learning and Evolutionary Methods in Control}
\label{sec:rw_mlc}

The application of machine learning to feedback control design has gained considerable traction over the past decade. Duriez et al.~\cite{Duriez.2017} provide a comprehensive treatment of \emph{Machine Learning Control} (MLC), highlighting the promise of evolutionary search---particularly genetic algorithms (GAs) and genetic programming---for discovering control laws that are difficult to obtain through classical linear techniques. GAs are attractive because they impose minimal assumptions on the cost landscape, can handle nonconvex and multimodal fitness functions, and naturally accommodate the mixed-integer structure of PID-like parameterizations. However, the practical effectiveness of a GA run depends critically on design choices that sit \emph{outside} the search loop: population size, generation budget, gain-range bounds, and the weighting of competing performance objectives. These meta-level hyperparameters are typically set through laborious trial-and-error by a human expert or, more systematically, through nested (cascade) optimization schemes in which a higher-level optimizer tunes the lower-level GA~configuration. Both approaches incur significant computational or cognitive cost, motivating the search for automated, sample-efficient alternatives---a gap that the present work aims to fill.

\subsection{From Large Language Models to Agentic AI}
\label{sec:rw_agentic}

The AI landscape has shifted rapidly from static prompt--response pipelines to autonomous, tool-augmented agents. Gulli~\cite{gulli.2025} traces this trajectory from basic LLM workflows through Retrieval-Augmented Generation (RAG) to full-fledged \emph{Agentic AI}, in which teams of specialized agents collaborate to achieve complex goals. The accompanying design-pattern catalog---including \emph{memory}, \emph{routing}, \emph{tool use}, \emph{goal setting}, and \emph{resource-aware optimization}---provides reusable abstractions that inform the architecture of our GA-Agent (see Section~\ref{sec:methodology}).

A substantial body of work has studied the synergy between reinforcement learning (RL) and LLMs. Pternea et al.~\cite{Pternea.2402} propose a taxonomy with three interaction classes: RL-for-LLM, LLM-for-RL, and joint RL+LLM frameworks. On the RL-for-LLM side, Wang et al.~\cite{Wang.2412} survey how RL fine-tuning---including RLHF and direct preference optimization---has become the dominant paradigm for aligning LLM outputs with human intent. More relevant to our setting is the converse direction, \emph{LLM-enhanced RL}, surveyed by Cao et al.~\cite{Cao.2404} and demonstrated in domain-specific applications by Cai et al.~\cite{Cai.2505}. These works identify four roles an LLM can play within an RL loop: \emph{information processor}, \emph{reward designer}, \emph{decision-maker}, and \emph{generator}. Our agent most closely resembles the \emph{reward designer} and \emph{decision-maker} roles: it adjusts the fitness-function weights $w_i$ (analogous to reward shaping) and proposes GA configurations (analogous to policy-level decisions) based on textual and numerical feedback from previous optimization attempts.

The reward-design role has been explored explicitly by Xie et al.~\cite{Xie.2309} in \emph{Text2Reward}, which uses an LLM to generate executable dense reward code from natural-language goal descriptions, and by Ma et al.~\cite{Ma.2310} in \emph{Eureka}, which performs evolutionary optimization over reward programs using GPT-4. Both works demonstrate that LLMs can craft reward signals that match or exceed hand-crafted alternatives. Our approach shares the spirit of iterative reward refinement but operates at the level of scalar fitness weights within a fixed cost structure rather than synthesizing reward code from scratch, trading generality for reliability in safety-relevant control applications.

At the intersection of LLMs and evolutionary computation, Wang et al.~\cite{Wang.2508} draw micro-level parallels between transformer mechanisms and EA operators---token representation versus individual encoding, attention versus selection, and so forth---and survey emerging work on evolutionary fine-tuning and LLM-enhanced EAs. These conceptual correspondences further motivate our coupling of an LLM agent with a GA optimizer, viewed as complementary reasoning modalities rather than competing ones.

\subsection{LLMs in Optimization and Control Design}
\label{sec:rw_llm_control}

\paragraph{LLM-based hyperparameter optimization.}
Liu et al.~\cite{Liu.2402} introduce \emph{AgentHPO}, an LLM agent that autonomously processes task descriptions, runs experiments with specific hyperparameter configurations, and iteratively refines them based on historical trial data. Evaluated on twelve machine-learning tasks, AgentHPO matches or surpasses the best human trials while substantially reducing the number of required evaluations. Our GA-Agent follows a similar closed-loop philosophy---observe results, reason about failures, propose improved configurations---but specializes it to the control-engineering domain by incorporating domain-specific metrics (settling time, overshoot, control effort) and operating over the joint space of GA hyperparameters \emph{and} fitness-function weights.

\paragraph{LLM as a numerical optimizer.}
Zhou et al.~\cite{Zhou.2025.ProPS} demonstrate, in \emph{Prompted Policy Search} (ProPS), that an LLM placed at the center of a policy-optimization loop can perform competitive numerical optimization in-context, directly proposing policy-parameter updates from reward feedback and natural-language strategy hints. ProPS outperforms classical RL baselines on a majority of Gymnasium tasks, establishing that in-context numerical reasoning is a viable optimization mechanism. Narimani and Emami~\cite{Narimani.2025} extend this idea to control engineering with \emph{AgenticControl}, a multi-agent LLM framework that directly tunes PID gains through an actor--critic dialogue between coordinated agents. While AgenticControl demonstrates the feasibility of LLM-driven controller synthesis, directly optimizing gains with an LLM conflates semantic reasoning (at which LLMs excel) with dense numerical search (at which population-based methods such as GAs excel). GA-Agent decouples these concerns: the GA handles the combinatorial, gradient-free search over PID gains, while the LLM agent operates at the meta-level---adjusting the GA's hyperparameters and reshaping its fitness landscape between attempts. This division of labor is the central thesis of our work and is evaluated empirically against both a fixed-hyperparameter GA baseline and a cascade-GA alternative (Section~\ref{sec:results}).

\paragraph{LLMs in the control-systems literature.}
A growing number of studies explore LLM integration into control workflows from complementary angles. Eslami and Yu~\cite{Eslami.2026} provide a control-theoretic foundation for agentic systems, formalizing a five-level hierarchy of agency---from fixed control laws to runtime synthesis of objectives---within an augmented closed-loop representation; their analysis highlights stability and safety implications of increasing agent authority, a concern our routing-based termination mechanism partially addresses. Aghaee and Shaker~\cite{Aghaee.2026} propose \emph{RB-LLM Control}, in which structured logical rules guide an LLM to adapt control inputs for UAV systems under disturbances and model uncertainty, achieving significant tracking-error reductions over classical baselines. Ayd{\i}n et al.~\cite{Aydin.2026} develop the \emph{MRAC-LLM Toolbox}, integrating LLM advisory guidance into model-reference adaptive control for reference-model selection and adaptation tuning, while preserving classical MRAC stability guarantees. Raval et al.~\cite{Raval.2025} present \emph{Circuit-AI}, a self-hosted LLM agent for power-electronics control-loop prototyping on edge hardware, emphasizing air-gapped deployment for defense and industrial settings. Taheri et al.~\cite{Taheri.2025} introduce \emph{BarrierBench}, a benchmark of 100 dynamical systems for evaluating LLM-guided barrier-certificate synthesis, demonstrating that agentic coordination and retrieval-augmented generation can achieve over 90\% success rates in formal safety verification.

\paragraph{Positioning of this work.}
Table~\ref{tab:related_positioning} summarizes how GA-Agent relates to the most relevant prior works along key design dimensions. Unlike approaches that employ LLMs as direct numerical optimizers of controller parameters~\cite{Zhou.2025.ProPS,Narimani.2025}, GA-Agent delegates low-level search to a GA and reserves the LLM for higher-level, semantically rich decisions---hyperparameter selection, fitness-weight adaptation, and strategy reasoning---where language models offer the greatest advantage. Compared to AgentHPO~\cite{Liu.2402}, our framework is specialized to the control domain, incorporates domain-specific metrics (settling time, overshoot, control effort) and a structured baseline cost for cross-run evaluation, and benchmarks against both fixed-hyperparameter and cascade-GA baselines. Relative to other LLM-in-the-loop control works~\cite{Aghaee.2026,Aydin.2026,Eslami.2026,Raval.2025}, GA-Agent targets offline design-time optimization of controller parameters rather than online adaptation.

\begin{table}[t]
\centering
\caption{Positioning of GA-Agent relative to closely related works. \emph{HPO}: hyperparameter optimization; \emph{Direct}: LLM directly proposes controller/policy parameters; \emph{Meta}: LLM operates on the optimizer's configuration rather than the controller parameters.}
\label{tab:related_positioning}
\resizebox{\textwidth}{!}{%
\begin{tabular}{lcccc}
\toprule
\textbf{Method} & \textbf{LLM Role} & \textbf{Optimizer} & \textbf{Domain} \\
\midrule
ProPS~\cite{Zhou.2025.ProPS} & Direct & LLM (in-context) & General RL \\
AgenticControl~\cite{Narimani.2025} & Direct & LLM (multi-agent) & PID Control \\
AgentHPO~\cite{Liu.2402} & Meta (HPO) & Wrapped ML trainer & General ML \\
Eureka~\cite{Ma.2310} & Reward code & RL (PPO) & Robotics \\
Text2Reward~\cite{Xie.2309} & Reward code & RL & Manipulation \\
RB-LLM~\cite{Aghaee.2026} & Decision-maker & Rule-based & UAV Control \\
MRAC-LLM~\cite{Aydin.2026} & Advisory & MRAC & Adaptive Control \\
\midrule
\textbf{GA-Agent (Ours)} & Meta (HPO + weights) & GA & PID Control \\
\bottomrule
\end{tabular}%
}
\end{table}

\section{Methodology}
\label{sec:methodology}

\begin{figure*}[t]
    \centering
    \includegraphics[width=\textwidth]{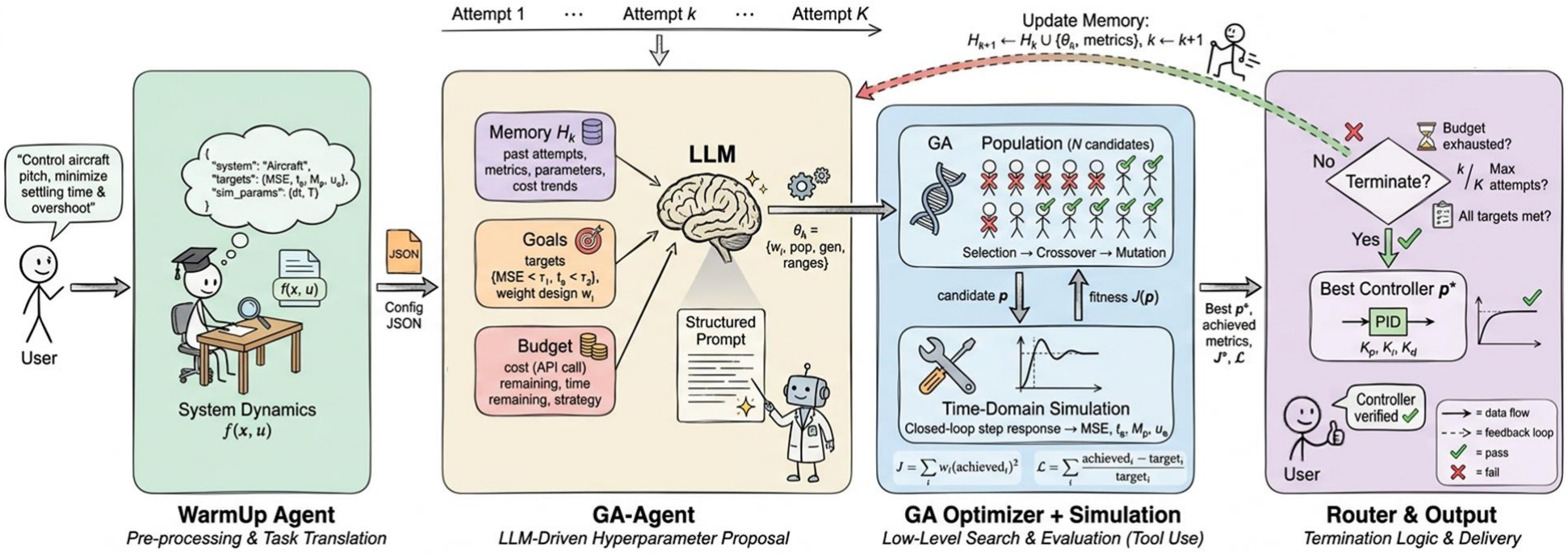}
    \caption{Architecture of the GA-Agent framework. The User provides a control task to the WarmUp Agent, which produces a structured configuration. The GA-Agent (LLM) iteratively proposes GA hyperparameters $\theta_k$ informed by Memory, Goals, and Budget. The GA Optimizer executes the search and the Simulation Tool evaluates candidates. The Router inspects termination criteria and either loops back (updating Memory) or emits the final controller.}
    \label{fig:overview}
\end{figure*}

Fig.~\ref{fig:overview} provides a high-level overview of the proposed \emph{GA-Agent} framework. The system operates as an iterative LLM-in-the-loop optimization pipeline in which a language model proposes genetic algorithm (GA) hyperparameters based on accumulated optimization history, user-specified control objectives, and available computational resources. Each iteration executes a complete GA optimization of PID parameters, evaluates controller performance through time-domain simulation, and feeds the resulting metrics back to the agent for subsequent decision-making.

The remainder of this section formalizes the control problem, describes the bilevel optimization structure underlying the GA-Agent, and details the architecture of the agentic workflow.

\subsection{Problem Formulation}
\label{sec:problem}

We consider the regulation problem for a single-input, single-output (SISO) dynamical system:
\begin{equation}
    \dot{x}(t) = f\!\bigl(x(t),\, u(t)\bigr), \qquad y(t) = h\bigl(x(t)\bigr),
    \label{eq:dynamics}
\end{equation}
where $x \in \mathbb{R}^{n}$ is the state vector, $u \in \mathbb{R}$ is the scalar control input ($m=1$ for the designated actuation channel; see footnote\textsuperscript{\ref{fn:multichannel}}), $y \in \mathbb{R}$ is the controlled output, and $f(\cdot)$ describes the---possibly nonlinear---system dynamics. The output map $h(\cdot)$ is likewise allowed to be nonlinear. The control objective is to regulate $y(t)$ to a reference setpoint~$r$ while satisfying transient and steady-state performance specifications.

We employ a proportional--integral--derivative (PID) control law:
\begin{equation}
    u(t) = K_p\, e(t) \;+\; K_i \!\int_{0}^{t}\! e(\tau)\, d\tau
          \;+\; K_d\, \dot{e}(t),
    \label{eq:pid}
\end{equation}
where $e(t) = r - y(t)$ is the tracking error and $p = [K_p,\, K_i,\, K_d]^{\!\top} \!\in \mathbb{R}^{3}$ are the controller gains to be determined.%
\footnote{\label{fn:multichannel}For multi-input plants, a single input channel is designated for closed-loop control; the remaining inputs are held at their trim values.}

\noindent\textbf{Remark (Derivative term implementability).}\;
The ideal derivative term $K_d\dot{e}(t)$ in Eq.~\eqref{eq:pid} is approximated numerically in simulation via a backward finite difference: $D_k = (e_k - e_{k-1})/\Delta t$, where $e_k$ denotes the tracking error at time step~$k$ and $\Delta t$ is the fixed simulation step size.  No derivative filter or anti-windup is applied; the control signal is saturated by hard clipping to the actuator limits $[u_{\min},\, u_{\max}]$ before being passed to the plant dynamics.  This choice keeps the number of free parameters fixed at three ($K_p$, $K_i$, $K_d$), simplifying the GA search space while remaining consistent with standard embedded PID implementations. Details are given in \ref{app:ode}.

\paragraph{Performance metrics.}
Controller quality is assessed via four metrics computed from the closed-loop unit step response simulated over a finite horizon~$[0, T]$ with time step $\Delta t$:

\begin{enumerate}
    \item \textbf{Mean Squared Error (MSE):}\;
          $\displaystyle\mathrm{MSE} = \frac{1}{N}\sum_{n=1}^{N} e(t_n)^{2}$,

    \item \textbf{Settling Time} ($t_s$): the earliest time after which
          $|y(t) - y_{\mathrm{ss}}|$ remains within $2\%$ of the
          steady-state value~$y_{\mathrm{ss}}$; if $y(t)$ never enters this
          band before the end of the simulation horizon, the fallback
          $t_s = T$ is applied as a worst-case penalty,

    \item \textbf{Percent Overshoot} ($M_p$):
          $\displaystyle M_p = \frac{\max_{t}\, y(t) - y_{\mathrm{ss}}}%
                                    {|r - y(0)|}
          \times 100\%$,

    \item \textbf{Normalized Control Effort:}\;
          $\displaystyle u_{\mathrm{eff}} =
          \frac{1}{T}\int_{0}^{T} \left|\frac{u(t)}{u_{\max}}\right| dt
          \;\in [0,\,1]$,
\end{enumerate}
where $N = T / \Delta t$ is the number of simulation steps and $u_{\max}$ is the actuator saturation limit. The summation index~$n$ in the MSE relation is distinct from the attempt index~$k$ used in the bilevel formulation that will be defined later. 

We collect these into the metric set $\mathcal{M} = \{\mathrm{MSE},\; t_s,\; M_p,\; u_{\mathrm{eff}}\}$, with fixed cardinality $|\mathcal{M}| = 4$ throughout this work.

\subsection{Optimization Formulation}
\label{sec:optimization}

\subsubsection{Low-Level GA}
\label{sec:low_level_ga}

Given a hyperparameter vector $\theta = \bigl(\{w_i\}_{i \in \mathcal{M}},\; N_{\mathrm{pop}},\; N_{\mathrm{gen}},\; [\underline{p},\, \overline{p}]\bigr)$, comprising objective weights, population size, generation count, and PID gain search bounds, a standard genetic algorithm (GA) solves
\begin{equation}
    p^{\!*}(\theta)
      = \argmin_{p\, \in\, [\underline{p},\, \overline{p}]}
        J(p;\, \theta)
      = \argmin_{p}\;
        \sum_{i \in \mathcal{M}} w_i \cdot
        \bigl(\mathrm{achieved}_i(p)\bigr)^{2},
    \label{eq:fitness}
\end{equation}
where $w_i \geq 0$ are the objective weights and $\mathrm{achieved}_i(p)$ denotes the $i$-th metric obtained by simulating the closed-loop system with PID gains~$p$. 

\noindent\textbf{Remark (Fitness scaling).}\;
The fitness $J = \sum_i w_i(\mathrm{achieved}_i)^2$ squares raw metric values that have heterogeneous units and magnitudes, e.g., $\mathrm{MSE} \approx 10^{-3}$ vs.\ $t_s \approx 10^{0}$\,s.  Without explicit normalization, one metric can numerically dominate regardless of~$w_i$.  In GA-Agent, this is handled implicitly: by observing achieved values alongside their deviations from targets in the memory context, the LLM infers the relative numerical magnitudes of each metric and sets weights accordingly. 
The agent is informed in the system prompt that only weight \emph{ratios} matter (doubling all weights has no effect), and the reasoning trace allows it to document and correct cross-metric imbalances across attempts.

\subsubsection{Baseline Cost}
\label{sec:baseline_cost}

Because the fitness~$J$ depends on the weight vector~$w$, it cannot be used for fair comparison across different hyperparameter configurations. We therefore define a weight-independent \emph{baseline cost}:
\begin{equation}
    \mathcal{L}(p)
      = \sum_{i \in \mathcal{M}}
        \frac{\mathrm{achieved}_i(p) - \mathrm{target}_i}
             {\mathrm{target}_i},
    \label{eq:baseline}
\end{equation}
where $\mathrm{target}_i > 0$ is a strictly positive, fixed reference value
specified per case study (see \ref{app:case_studies}).

\noindent\textbf{Remark (Interpretation of $\mathcal{L}$).}\;
$\mathcal{L}$ is a \emph{sum} of relative deviations from targets, not a maximum or $\ell_\infty$ norm. A controller that badly fails one metric while exceeding all others may therefore still achieve $\mathcal{L} < 0$. This design reflects the intent to reward \emph{average} progress across all objectives simultaneously. By contrast, the success score (defined below) uses a strict per-metric binary criterion and is immune to this trade-off. Together, $\mathcal{L}$ and the success score provide complementary views: the former measures continuous improvement, the latter measures discrete target satisfaction. A negative~$\mathcal{L}$ indicates that the achieved metrics are, on average, below their respective targets, i.e., the controller exceeds the design specifications on average.

Each metric contributes equally to a \emph{success score}: metric~$i$ is considered \emph{passed} if $\mathrm{achieved}_i \leq \mathrm{target}_i$, contributing $100\% / |\mathcal{M}| = 25\%$ toward a maximum of $100\%$.

The cumulative number of function evaluations (NFE) is tracked across attempts as
\begin{equation*}
    \mathrm{NFE}_{\mathrm{total}} = \sum_{k=1}^{K} \mathrm{NFE}^{(k)},
\end{equation*}
where each \emph{attempt}~$k$ corresponds to one complete GA optimization cycle triggered by a hyperparameter configuration, $K$ is the total number of attempts, and $\mathrm{NFE}^{(k)} = N_\mathrm{pop}^{(k)} \cdot N_\mathrm{gen}^{(k)}$ is the number of fitness evaluations (i.e., closed-loop simulations) in attempt~$k$.

\subsubsection{The Hyperparameter Tuning Problem}
\label{sec:bilevel}

The quality of the low-level GA solution~$p^{\!*}(\theta)$ depends critically on the hyperparameters~$\theta$.

Since the GA is stochastic, $p^{\!*}(\theta)$ is a random variable; the
outer problem is formally an expectation minimization:
\begin{equation}
    \theta^{*}
      = \argmin_{\theta \,\in\, \Theta}\;
        \mathbb{E}\!\bigl[\mathcal{L}\!\bigl(p^{\!*}(\theta)\bigr)\bigr],
    \label{eq:bilevel}
\end{equation}
where the expectation is over the GA's stochastic initialization and operators.  In practice, each outer-loop evaluation observes a single noisy realization of~$\mathcal{L}$; the LLM agent's in-context reasoning is particularly suited to this sparse, noisy-observation setting, as it can extrapolate informed configurations from a small number of evaluations.

\noindent\textbf{Hyperparameter space $\Theta$.}\;
The hyperparameter vector~$\theta$ lives in a mixed discrete--continuous space: population size $N_{\mathrm{pop}} \in \mathbb{Z}_{>0}$, generation budget $N_{\mathrm{gen}} \in \mathbb{Z}_{>0}$, fitness weights $w_i \geq 0$ (free positive reals; no simplex constraint is imposed), and component-wise bounds $\underline{p} < \overline{p}$ for the inner PID gain search. 

The inherently mixed discrete--continuous nature of~$\Theta$, together with the non-differentiable and stochastic outer-loop objective, motivates the use of an LLM for meta-level reasoning instead of a gradient-based optimizer (which would require continuous relaxations and struggle with the semantic/discrete aspects of hyperparameter selection).

\begin{figure*}[t]
    \centering
    \includegraphics[width=0.5\textwidth]{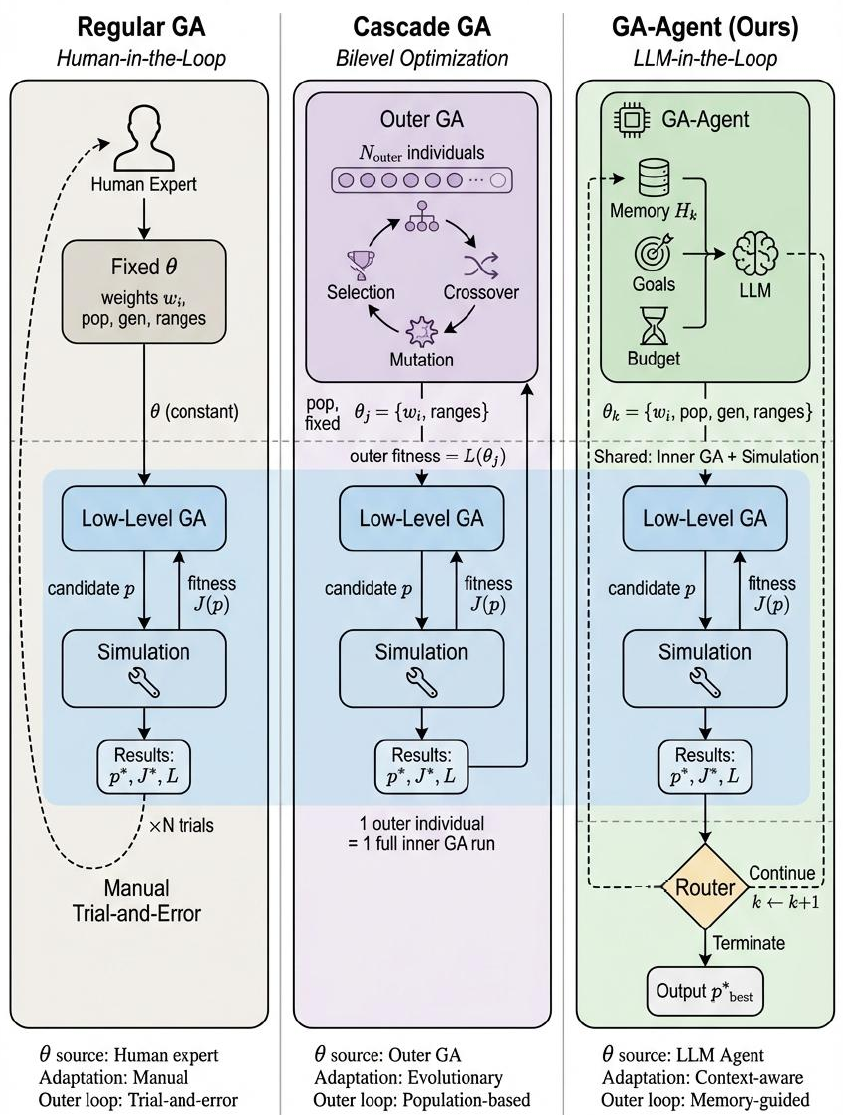}
    \caption{%
        Comparison of the three PID tuning paradigms. All approaches share
        the same inner-loop GA and simulation infrastructure (blue region).
        They differ in how the hyperparameters~$\theta$ are selected and
        adapted (top region of each panel): \emph{Regular~GA} relies on
        manual, fixed configurations set by a human expert;
        \emph{Cascade~GA} employs a population-based outer evolutionary
        loop; and \emph{GA-Agent}~(ours) uses an LLM that conditions on
        accumulated optimization history~$\mathcal{H}_k$, goals, and
        budget status to propose a single informed configuration per
        attempt.}
    \label{fig:paradigm_comparison}
\end{figure*}

We consider three approaches to solving---or approximating---this outer problem, compared in Fig.~\ref{fig:paradigm_comparison}.

\paragraph{(A) Manual tuning.}
In common practice, a control engineer manually selects the weights~$w_i$, GA configuration ($N_{\mathrm{pop}}$, $N_{\mathrm{gen}}$), and search ranges $[\underline{p},\, \overline{p}]$ based on domain expertise and iterative trial-and-error. This process is labor-intensive, non-reproducible, and scales poorly across diverse systems.
In our experiments, the \emph{Regular~GA} baseline uses a single, fixed configuration~$\theta$ set to represent a general-purpose practitioner default: gain bounds are conservatively wide (sufficient to cover all ten case studies), fitness weights are uniform ($w_i = 1.0$ for all $i \in \mathcal{M}$), and GA size parameters are chosen without system-specific tuning. Actuator saturation via hard clipping prevents the GA from being misled by unrealistically large gain values even when broad bounds are used. The full baseline configuration is detailed in \ref{app:repro:ga:regular}.

\paragraph{(B) Cascade GA.}
To fully automate the outer loop, a second (outer) GA can search over the hyperparameter space~$\Theta$. Each candidate~$\theta$ in the outer population triggers a complete inner-GA run, and the outer fitness is evaluated as~$\mathcal{L}(p^{\!*}(\theta))$.
The outer GA optimizes the fitness weights $\{w_i\}_{i\in\mathcal{M}}$ and the PID gain range upper bounds $(\bar{K}_p,\, \bar{K}_i,\, \bar{K}_d)$; the inner-loop population size and generation count are held fixed as cascade-level constants rather than being included in the outer search, as these parameters tend to saturate (maximize) when evolved---yielding diminishing returns at high computational cost. Full configuration details are given in \ref{app:repro:ga:cascade}.
While fully automated, this cascade incurs substantial computational cost: with an outer population of $N_{\mathrm{pop}}^{\mathrm{out}}$ individuals evolved over $N_{\mathrm{gen}}^{\mathrm{out}}$ generations, the total NFE scales as $\mathcal{O}\!\bigl(N_{\mathrm{pop}}^{\mathrm{out}} \cdot N_{\mathrm{gen}}^{\mathrm{out}} \cdot N_{\mathrm{pop}} \cdot N_{\mathrm{gen}}\bigr)$, rendering it computationally expensive even for moderately sized configurations.

\paragraph{(C) GA-Agent (proposed).}
We propose replacing the outer GA with a large language model (LLM) agent that iteratively configures the low-level GA. Rather than evaluating an entire population of outer candidates per generation, the LLM leverages in-context reasoning to propose a \emph{single}, informed hyperparameter configuration per attempt, conditioned on the full optimization history.
Formally, the GA-Agent implements a policy $\pi_{\mathrm{LLM}}$:
\begin{equation}
    \theta_{k+1}
      = \pi_{\mathrm{LLM}}\!\bigl(\mathcal{H}_k,\;
        \mathcal{G},\; \mathcal{B}_k\bigr),
    \label{eq:llm_policy}
\end{equation}
where $k$ is the attempt index,
\begin{equation*}
    \mathcal{H}_k
      = \bigl\{(\theta_j,\, p^{\!*}_j,\, \mathcal{L}_j,\,
        \mathrm{metrics}_j,\, \mathrm{stats}_j)
        \bigr\}_{j=\max(1,\,k-B+1)}^{k}
\end{equation*}
is the optimization history (\emph{memory}), truncated to the most recent $B$ attempts to avoid exceeding the LLM's context window (the effect of $B$ is studied in \ref{app:buffer_size}), $\mathcal{G}$ encodes the system description, control objectives, and fixed targets (\emph{goal}), and $\mathcal{B}_k$ represents the remaining computational time and monetary budgets (\emph{budget awareness}).

\noindent\textbf{Initialization ($\theta_1$).}\;
Eq.~\eqref{eq:llm_policy} defines $\theta_{k+1}$ for $k \geq 1$ given prior history. The initial configuration~$\theta_1$ is obtained from a dedicated \emph{initial LLM call} in which the memory context is empty and the agent receives only the goal specification~$\mathcal{G}$ and the initial budget status~$\mathcal{B}_0$.  In the comparative experiments of Section~\ref{sec:results}, both a manually specified~$\theta_1$ and an LLM-generated~$\theta_1$ are evaluated; within each trial the initial~$\theta_1$ is held identical across all three methods to ensure a fair comparison.  In the full architecture, $\theta_1$ is intended to be provided by the WarmUp-Agent (Section~\ref{sec:warmup}).

Drawing a direct analogy to~\eqref{eq:bilevel}, both the Cascade~GA and the GA-Agent aim to minimize~$\mathcal{L}$ over the hyperparameter space~$\Theta$: the Cascade~GA does so via population-based evolutionary search, while the GA-Agent does so via sequential LLM reasoning conditioned on accumulated evidence. The key advantage is \emph{sample efficiency}: where the Cascade~GA requires $\mathcal{O}(N_{\mathrm{pop}}^{\mathrm{out}} \cdot N_{\mathrm{gen}}^{\mathrm{out}})$ inner-loop evaluations to update the outer configuration once, the GA-Agent proposes a single informed configuration per attempt. 

\subsection{GA-Agent Architecture}
\label{sec:architecture}

\begin{figure}[t]
    \centering
    \includegraphics[width=0.7\columnwidth]{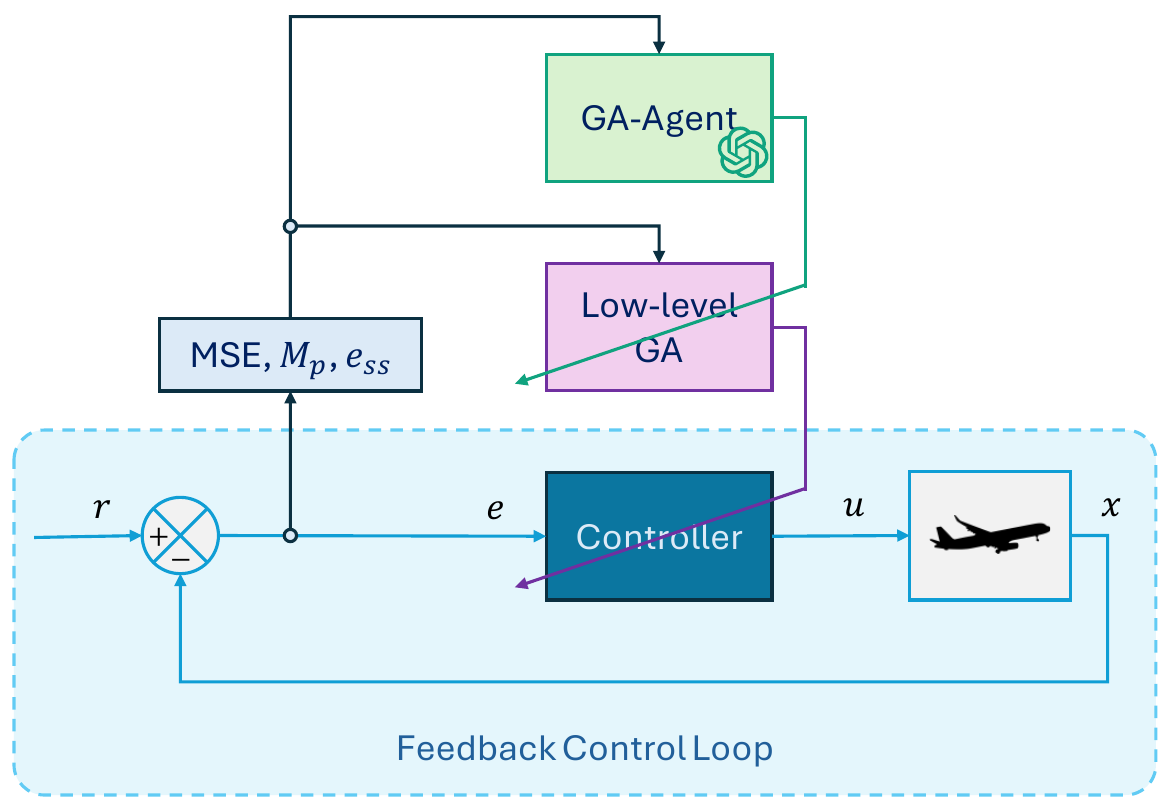}
    \caption{System architecture. The inner feedback loop implements PID control of the plant. The \emph{low-level GA} optimizes PID gains by minimizing a weighted fitness cost~$J$. The \emph{GA-Agent} (LLM) tunes the GA hyperparameters---weights, population size, generation count, and search ranges---between successive optimization attempts.}
    \label{fig:block_diagram}
\end{figure}

The GA-Agent is realized as a modular agentic system following established agentic design patterns~\cite{gulli.2025, yao2022react, wang2024survey}. The complete architecture and information flow are illustrated in Fig.~\ref{fig:overview} (system-level) and Fig.~\ref{fig:block_diagram} (closed-loop structure). The system comprises four principal modules---Memory, Goal Setting, Budget Awareness, and Routing---plus an inner-loop GA Optimizer and a Simulation Tool, described below.

\paragraph{Memory ($\mathcal{H}_k$).}
A structured rolling buffer maintains a fixed-length history of the $B$ most recent attempts. For each attempt $j \leq K$, the buffer stores:
\begin{enumerate}
    \item the hyperparameters~$\theta_j$ (population size, generation budget, gain bounds, fitness weights);
    \item the best PID gains found~$p^{\!*}_j$ and their exploration statistics (population mean $\mu_j$, standard deviation $\sigma_j$, min/max achieved values, and flags indicating boundary saturation);
    \item all four achieved control metrics (MSE, settling time, percent overshoot, normalized control effort) together with the corresponding GA fitness $J_j$ and baseline cost $\mathcal{L}_j$;
    \item the LLM's natural-language reasoning and diagnostic summary from attempt~$j$, which is re-injected into the next prompt's feedback section to maintain semantic continuity;
    \item resource consumption metadata (wall-clock duration, LLM input/output token counts).
\end{enumerate}
At each new attempt, the serialized history is injected into the LLM prompt, enabling the agent to identify performance trends, detect parameter boundary saturation, and avoid repeating unsuccessful configurations. The buffer size~$B$ is a tunable hyperparameter whose effect is studied in \ref{app:buffer_size}.

\noindent\emph{Boundary saturation.}\;
Boundary saturation occurs when the GA's best solution lies within $5\%$ of a boundary of the search range $[\underline{p},\, \overline{p}]$, indicating that the true optimum may lie outside the current search box. Saturation flags are included in the memory buffer, and the agent is explicitly instructed to diagnose the cause before acting: if the saturating metric is improving, the range is instructed to be expanded; if stagnant or worsening, the upper bound is advised to be restricted. A ceiling-chasing exception rule handles the pathological case in which a parameter persistently hits the newly restricted bound across multiple consecutive attempts: at that point, expansion is mandated and the restriction strategy is abandoned. The exact prompt instructions encoding these rules are reported in \ref{app:prompts}.

\noindent\emph{Population statistics.}\;
Storing population-level mean, standard deviation, minimum, and maximum allows the agent to distinguish between a GA that has converged prematurely (low variance, solutions clustered away from boundaries) and one that is still exploring (high variance), thereby informing the choice of population size or mutation intensity in the next attempt.

\paragraph{Goal setting ($\mathcal{G}$).}
The system description---including plant dynamics, control objective, and fixed performance targets~$\{\mathrm{target}_i\}_{i \in \mathcal{M}}$---is encoded as structured text within the prompt context. These fixed targets define the baseline cost~\eqref{eq:baseline} and are \emph{not} modifiable by the agent. Instead, the agent adjusts the GA fitness weights~$\{w_i\}$ to steer the low-level optimizer toward meeting these targets. This separation between fixed evaluation criteria and tunable optimization weights is central to the approach: it allows the agent to reshape the fitness landscape across attempts while maintaining a consistent comparison metric.

For instance, if percent overshoot $M_p$ consistently exceeds its target while the other metrics are satisfied, the agent increases~$w_{M_p}$ in the next attempt to penalize high-overshoot controllers more heavily in the GA fitness. 

\paragraph{Budget awareness ($\mathcal{B}_k$).}
The agent receives real-time information about two resource budgets:
\begin{itemize}
    \item \textbf{Computational time budget:} wall-clock time consumed by GA runs
    \item \textbf{Monetary cost budget:} API costs incurred by LLM inference calls, proportional to prompt and completion token counts but independent of the GA configuration.
\end{itemize}
The prompt includes the current percentage of each budget consumed, the cost/time of the last attempt, and strategy guidelines (e.g., use small GA configurations for early exploration; reserve budget for a final convergence run). This enables the agent to plan a resource-aware exploration--exploitation schedule across the available attempts.

The exploration-first heuristic embedded in the prompt guidelines is a design choice. The ablation study in \ref{app:ablation} evaluates four prompt-language variants (all encoding the same strategic information but with different phrasing); systematically varying the underlying strategy guidelines is left as future work.

\paragraph{Simulation tool.}
The simulation is invoked \emph{deterministically} by the GA optimizer---not by the LLM---at every fitness evaluation. This contrasts with standard LLM tool-use frameworks~\cite{schick2023toolformer} where the model \emph{decides} when to call a tool. The tool integrates the closed-loop dynamics~\eqref{eq:dynamics}--\eqref{eq:pid} over the specified horizon $[0, T]$ using a fixed-step forward Euler scheme (\ref{app:ode}) and returns the metric vector~$\{\mathrm{achieved}_i\}_{i \in \mathcal{M}}$.

\paragraph{Routing.}
A rule-based routing module governs the agent's control flow after each attempt. The optimization loop terminates when any of the following conditions is satisfied:
\begin{enumerate}
    \item the maximum number of attempts~$K_{\max}$ is reached;
    \item the computational time budget is exhausted;
    \item the monetary cost budget is exhausted; or
    \item all performance targets are satisfied, i.e., $\mathrm{achieved}_i \leq \mathrm{target}_i$ for all $i \in \mathcal{M}$.
\end{enumerate}
Any satisfied condition is sufficient to halt the loop.
Upon termination under any condition, the controller with the lowest baseline cost~$\mathcal{L}$ across \emph{all} $K$ completed attempts is returned as the final solution.
If none of these conditions hold, the results of the current attempt are appended to the memory buffer~$\mathcal{H}_k$ and a new LLM inference call is initiated for attempt~$k+1$.

\subsection{Prompt Design}
\label{sec:prompt}

\begin{figure}[t]
    \centering
    \includegraphics[width=0.5\columnwidth]{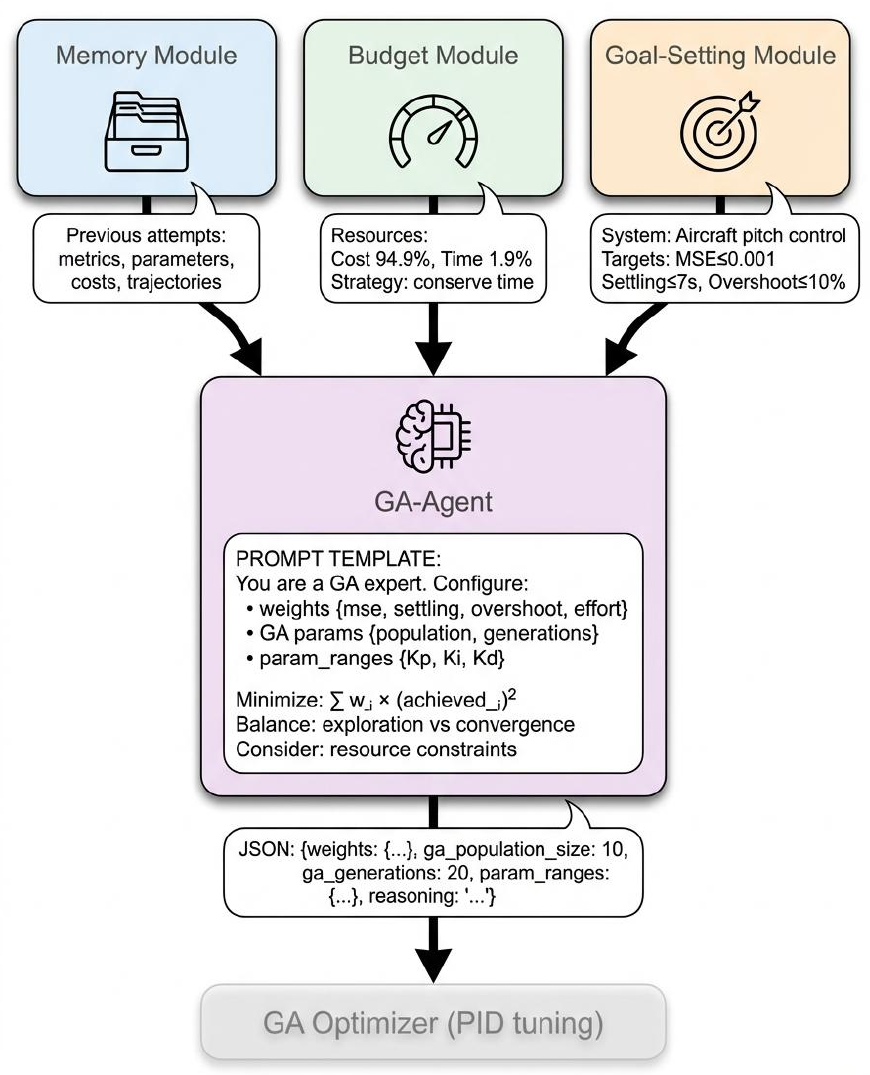}
    \caption{Prompt design workflow. The Memory, Budget, and Goal modules contribute structured information to a prompt template. The LLM produces a JSON configuration specifying the weights, GA parameters, and search ranges for the next optimization attempt.}
    \label{fig:prompt_template}
\end{figure}

The prompt provided to the LLM at each attempt is assembled from the
modules described above. As illustrated in Fig.~\ref{fig:prompt_template},
the prompt comprises:
\begin{enumerate}
    \item A \textbf{system prompt} that defines the agent's role as a GA expert, specifies the mathematical structure of the fitness cost~$J$ and baseline cost~$\mathcal{L}$, describes each tunable hyperparameter and its effect, and prescribes the JSON output format;
    \item The \textbf{goal specification}~$\mathcal{G}$, including the system name, physical description, control objective, and fixed performance targets;
    \item The \textbf{memory}~$\mathcal{H}_k$, serialized as a structured per-attempt summary containing the hyperparameters used, achieved metrics (with population-level statistics), the LLM's reasoning trace from that attempt (reproduced verbatim to maintain reasoning continuity), obtained PID gains (with boundary-saturation warnings), and cost trajectories;
    \item The \textbf{budget status}~$\mathcal{B}_k$, with percentage-based resource indicators, per-attempt consumption, trend analysis, and resource-aware strategy recommendations.
\end{enumerate}
The LLM returns a structured JSON object containing the updated weights $\{w_i\}$, GA configuration ($N_{\mathrm{pop}}$, $N_{\mathrm{gen}}$), parameter ranges~$[\underline{p},\, \overline{p}]$, and a natural-language reasoning trace justifying the choices. The reasoning trace is reproduced verbatim in the subsequent prompt's memory block, providing the agent with a continuous record of its own past reasoning.

\noindent\textbf{JSON output validation.}\;
LLM outputs are validated against a predefined JSON schema; fields are checked for correct types, non-negative weights, valid positive integer GA parameters, and properly ordered gain bounds ($\underline{p} < \overline{p}$). Malformed or out-of-range responses trigger a re-query with an error message appended to the prompt, up to a maximum of three retries.

\subsection{Pre-Processing}
\label{sec:warmup}

Before the main optimization loop, the user's problem specification must be translated into a structured configuration defining: the plant dynamics (as a callable Python function), state and input dimensions, input/output channel indices, reference setpoint~$r$, actuator limits~$[u_{\min},\, u_{\max}]$, simulation parameters ($\Delta t$, $T$), and initial performance targets. In a fully autonomous pipeline, a dedicated \emph{WarmUp-Agent} performs this translation, including preliminary open-loop simulations to determine an appropriate simulation horizon.

\noindent\textbf{Remark (Scope of Automation).}\;
\emph{The WarmUp-Agent is described for completeness of the proposed architecture. All experiments in this paper use manually specified configurations (detailed in \ref{app:case_studies}) to ensure a fair, reproducible comparison across methods. 
In the manual-configuration experiments, the initial hyperparameter configuration~$\theta_1$ is either~(a) set to a general-purpose default held identical across all methods for baseline comparisons, or~(b) generated by an initial LLM call with no memory context (the GA-Agent's natural initialization scheme). Both variants are evaluated and reported in Section~\ref{sec:results}.}

\subsection{Implementation Details}
\label{sec:impl_details}

\paragraph{Software stack and dependencies}
All experiments are implemented in Python~3.11.10. The agentic workflow is orchestrated with LangGraph~v0.3.27, which manages the iterative LLM-in-the-loop execution graph, including routing logic, memory updates, and termination conditions. LLM inference is accessed via the OpenAI Python SDK~v1.84.0 using the standard \texttt{/v1/chat/completions} endpoint. Numerical simulation relies on NumPy~v2.2.4; the fixed-step forward Euler integrator is implemented directly in NumPy without external ODE solver libraries. The GA optimizer is implemented using PyGAD~v3.5.0; detailed operator parameters and configuration are provided in \ref{app:repro:ga:regular} (Regular GA) and \ref{app:repro:ga:cascade} (Cascade GA).

\paragraph{LLM backbone and decoding parameters.}
All main experiments use \textbf{DeepSeek-V4-Flash} as the backbone LLM at a fixed sampling temperature of $T_{\mathrm{samp}} = 0$ (deterministic greedy decoding). All other API decoding parameters (top-$p$, frequency penalty, presence penalty, etc.) are left at OpenAI SDK defaults. Structured JSON output is enforced via prompt instruction only; no constrained-decoding API feature is used. The sensitivity analysis in \ref{app:sensitivity:models} evaluates performance across the following alternative LLM backbones: Gemini-2.5-Flash, Grok-4.1-Fast, Grok-3-Mini-Beta, Claude-Haiku-4.5, O3-Mini, GPT-4.1-Mini.

\paragraph{Token consumption and API costs.}
Token usage is summarized in \ref{app:impl:cost}. Average per-run totals range from approximately 4.7k input / 4.3k output tokens (GA-Agent) to 8.0k input / 6.2k output tokens (GA-Agent+ LLM~Init), with 1.9--3.3 LLM calls per run depending on the variant. Total API costs across all experiments are reported in \ref{app:impl} (Table~\ref{tab:api_costs}).

\paragraph{Hardware and runtime environment.}
Experiments were executed on an Intel Core i7-14700HX processor with 20 cores and 28 threads (hyperthreading enabled), equipped with 32.0 GB of RAM. No GPU acceleration was used; all computations (GA fitness evaluations, LLM API calls, and state integration) are CPU-bound.د
GA runs are \emph{not} parallelized across the population; all candidate evaluations within a generation execute sequentially on a single core.

\paragraph{Simulation parameters.}
All closed-loop dynamics simulations use the custom fixed-step forward Euler integrator described in \ref{app:ode}. Step sizes ($\Delta t$), simulation horizons ($T$), and control constraints are fixed per case study and are summarized in Table~\ref{tab:sim_params}. 

\paragraph{Reproducibility.}
Full implementation details including hyperparameter grids, seed management, fitness function specifications, and control objective weights are provided in \ref{app:repro}. 

\subsection{Case Studies and Evaluation Protocol}
\label{sec:case_studies}

\begin{figure}[t]
    \centering
    \includegraphics[width=0.7\columnwidth]{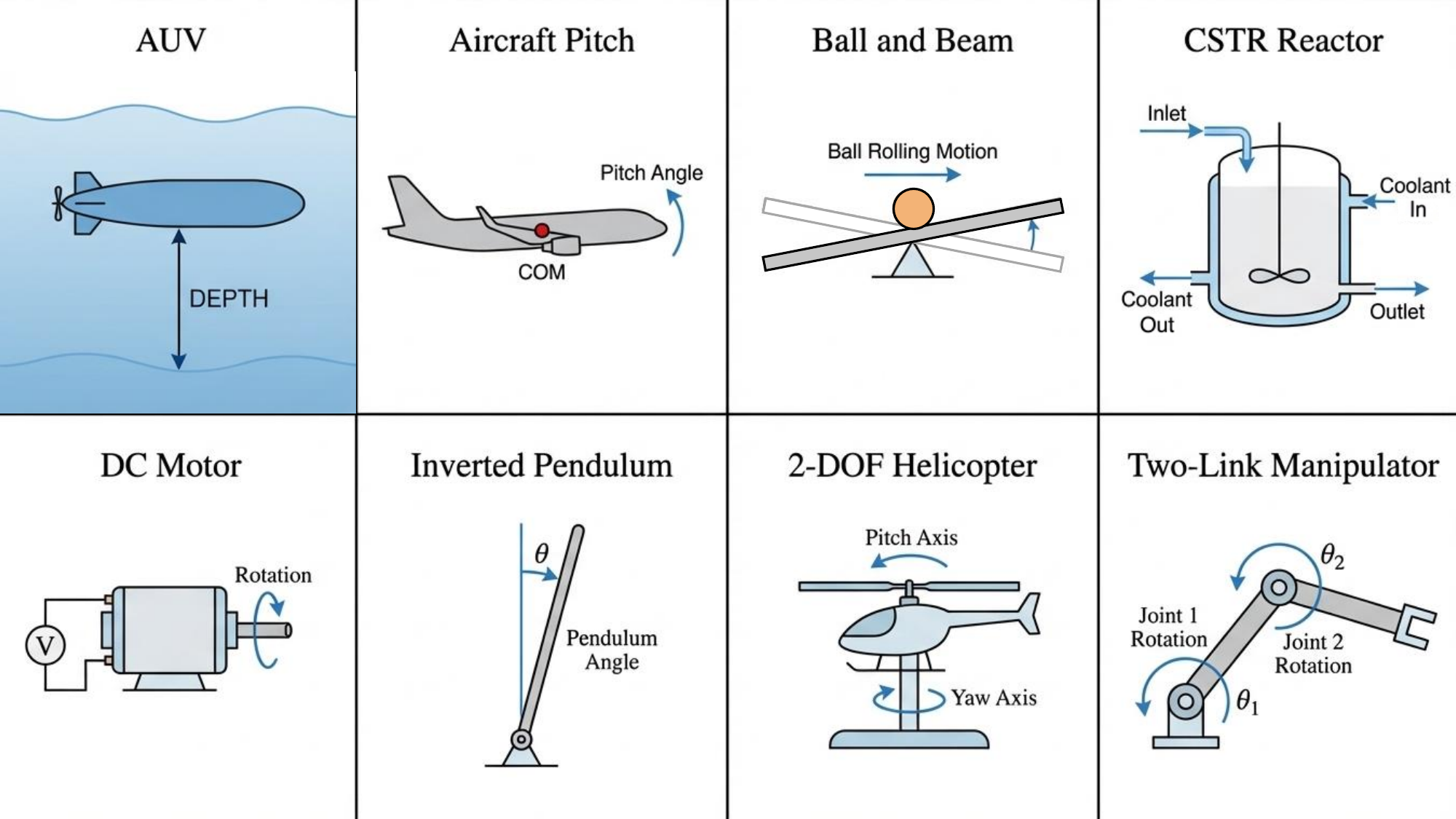}
    \caption{The eight control case studies used for evaluation, covering aerial, underwater, robotic, and process control domains.}
    \label{fig:case_studies}
\end{figure}

We evaluate the proposed GA-Agent on eight control case studies spanning diverse engineering domains, illustrated in Fig.~\ref{fig:case_studies}. 
\begin{enumerate}
    \item Autonomous Underwater Vehicle (AUV) depth control~\citep{vahid2016modeling},
    \item Transport aircraft pitch attitude regulation~\citep{stevens2015aircraft},
    \item Ball-and-beam position control~\citep{ahmad2023modeling},
    \item Continuous stirred-tank reactor (CSTR) temperature regulation~\citep{esfandyari2013adaptive},
    \item DC motor speed control~\citep{Ogata2010},
    \item Cart--pole (inverted pendulum) stabilization~\citep{Tedrake2023},
    \item Two-Degree-of-Freedom (2-DoF) helicopter~\citep{luo2017optimal},
    \item Two-link planar manipulator~\citep{Spong2006}.
\end{enumerate}
All plant models are used in their \emph{nonlinear} form without linearization: the PID control law~\eqref{eq:pid} is applied directly to the full nonlinear dynamics~\eqref{eq:dynamics}. This is noteworthy for the inherently nonlinear systems (inverted pendulum, 2-DoF helicopter, two-link manipulator), where PID performance on the nonlinear model may differ significantly from a linearized approximation; no such simplification is made in any of the eight case studies.

Each case study specifies a plant model, control objective, actuator constraints, simulation parameters, and fixed performance targets. Detailed configurations are provided in \ref{app:case_studies}.

For each case study, we compare three methods under identical resource budgets:
\begin{itemize}
    \item \textbf{Regular GA}: A standard low-level GA with fixed, manually-set hyperparameters~$\theta$ held constant across all attempts.
    \item \textbf{Cascade GA}: A bilevel GA in which an outer evolutionary loop optimizes~$\theta$ while the inner GA optimizes~$p$.
    \item \textbf{GA-Agent} (ours): The low-level GA with LLM-driven hyperparameter adaptation as described in Sections~\ref{sec:architecture}--\ref{sec:prompt}.
\end{itemize}

\noindent\textbf{Resource budget specification.}\;
All three methods operate under the following shared budget constraints: maximum attempts $K_{\max} = 12$, wall-clock time limit of $1{,}000$\,seconds per experiment, and an LLM API cost cap of \$0.05 per experiment. The Regular GA uses a \emph{single} attempt with fixed~$\theta$; its NFE is $\mathcal{O}(N_\text{pop} \cdot N_\text{gen})$ and represents a single budget allocation (not $K_{\max}$ repeated runs). 

\noindent\textbf{Baseline hyperparameter setting.}\;
The Regular GA's fixed~$\theta$ and the Cascade GA's outer-loop configuration are set to general-purpose defaults without case-study-specific fine-tuning, ensuring that the comparison reflects practitioner-level performance rather than an oracle-tuned baseline.

\noindent\textbf{Statistical evaluation.}\;
Each (case study, method) pair is evaluated over $5$ independent runs with distinct random seeds. The same seed is used for all three methods within each trial so that performance differences cannot be attributed to population initialization (\ref{app:repro:ga:regular}). Results are reported as mean~$\pm$~standard deviation.

All methods are evaluated using the baseline cost~$\mathcal{L}$ \eqref{eq:baseline}, the success score (percentage of targets met), and the cumulative NFE as a measure of computational efficiency. 
We now move on to the comparative results of our proposed work against the Regular~GA and the Cascade~GA baselines.

\section{Results}
\label{sec:results}

This section presents the empirical evaluation of GA-Agent across eight control case studies. Two variants are evaluated: \textbf{GA-Agent} (system-blind), in which the LLM meta-optimizer receives only numerical performance feedback from completed GA runs, and \textbf{GA-Agent+} (system-aware), which additionally supplies the LLM with a structured description of the plant name, control objective, and system dynamics. 
Both variants are benchmarked against (i) a Regular~GA with fixed hyperparameters (Section~\ref{subsec:results-b3}) and (ii) a Cascade-GA hierarchical optimizer (Section~\ref{subsec:results-c3}). Results are reported using two complementary metrics: the weight-independent baseline cost~$\mathcal{L}$ (lower is better) and the \emph{success score} (the fraction of the four performance targets jointly satisfied by the synthesized controller, expressed as a percentage; higher is better), alongside cumulative function evaluations~(NFE). All main experiments use DeepSeek-V4-Flash as the LLM backbone at a total API cost of \$0.0745 across all runs (${\approx}$\$0.0019 per run). A sensitivity analysis over prompt variants, memory buffer sizes, and LLM backbones is summarized in Section~\ref{subsec:sensitivity}. 

\subsection{Performance of GA-Agent vs.\ Regular GA}
\label{subsec:results-b3}

\begin{figure}[h]
    \centering
    \includegraphics[width=0.85\linewidth]{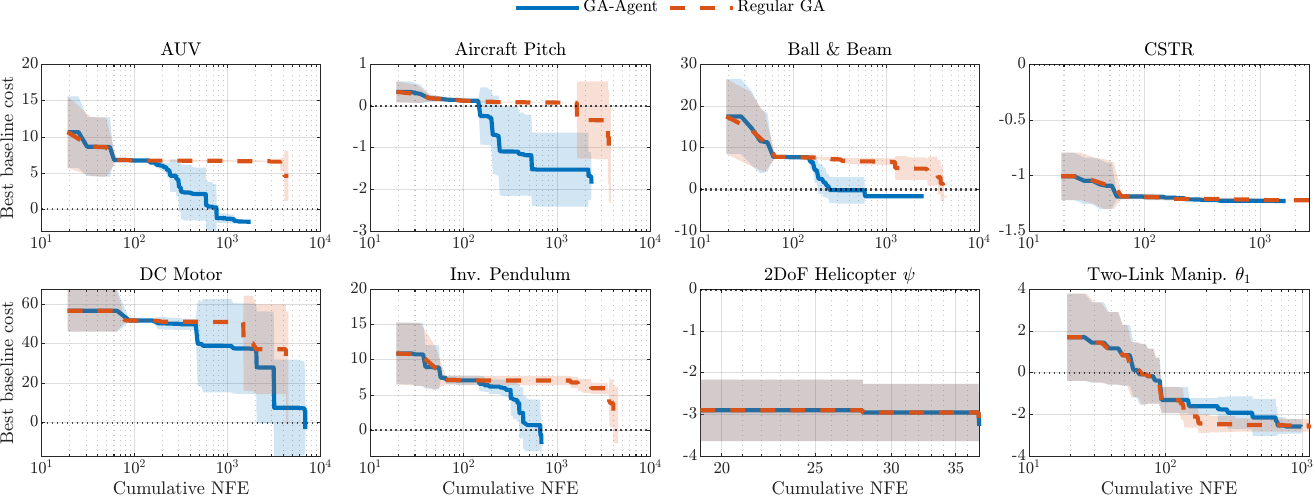}
    \caption{Distribution of baseline cost $\mathcal{L}$ across all GA attempts for GA-Agent and Regular~GA. Each point represents the best individual from a single low-level GA run; lower values indicate better controller performance. GA-Agent attains consistently negative, low-variance costs across all eight case studies. The Regular~GA exhibits high variance and records large positive costs on DC~Motor ($27.79 \pm 28.54$) and AUV ($4.63 \pm 3.45$), confirming the benefit of iterative LLM-driven hyperparameter adaptation over fixed configurations.}
    \label{fig:batch-baseline}
\end{figure}

GA-Agent is evaluated in its system-blind configuration under \emph{manual initialization}: a lightweight warm-start ($N_{\mathrm{pop}}=10$, $N_{\mathrm{gen}}=10$, equal fitness weights $w_i=1$, gain bounds $K_p\!\in\![0,200]$, $K_i\!\in\![0,50]$, $K_d\!\in\![0,50]$) seeds the first GA run; the LLM agent then proposes updated hyperparameters from the second attempt onward. The \textbf{Regular~GA} baseline uses fixed hyperparameters throughout: $N_{\mathrm{pop}}=10$, $N_{\mathrm{gen}}=500$, equal fitness weights, and identical gain bounds (see \ref{app:repro:ga:regular} for the complete specification). Both methods operate over the same search space; GA-Agent adapts population size, generation budget, search bounds, and fitness weights between attempts, while the Regular~GA cannot. Figure~\ref{fig:batch-baseline} shows the distribution of baseline cost~$\mathcal{L}$ across all attempts; Table~\ref{tab:llm_ga_vs_regular_ga_score} reports the final best success score averaged over five independent runs per case study.

\paragraph{Success Score.}
GA-Agent achieves a perfect success score of $100\%$ with zero variance on all eight case studies, confirming that LLM-driven hyperparameter adaptation reliably steers the inner GA toward configurations that jointly satisfy all performance targets. The Regular~GA, constrained to fixed hyperparameters, reaches $100\%$ only on the two structurally simpler benchmarks (2DoF~Helicopter and Two-Link~Manipulator), where a single preset configuration already suffices. The largest deficits appear on DC~Motor ($55.0 \pm 41.1\%$), where the high standard deviation indicates that most runs stall at suboptimal configurations under fixed settings, and on Aircraft~Pitch ($65.0 \pm 13.7\%$). Averaged across all eight benchmarks, the mean success score improves from $81.3\%$ (Regular~GA) to $100\%$ (GA-Agent).

\paragraph{Sample Efficiency.}
GA-Agent achieves better or equal success on all eight case studies and uses fewer NFE on six of eight (with one tie). The largest reductions occur on Inverted~Pendulum ($698$ vs.\ $4{,}518$~NFE, ${\approx}6.5\times$) and AUV ($1{,}803$ vs.\ $4{,}518$~NFE, ${\approx}2.5\times$), where the LLM rapidly identifies effective search boundaries. On average, GA-Agent converges at $2{,}135$~NFE versus $3{,}210$~NFE for the Regular~GA, a $33\%$ reduction accompanied by a $18.7$~percentage-point improvement in success score.
\textit{DC~Motor} requires qualification: GA-Agent expends $6{,}911$~NFE---$1.5{\times}$ the Regular~GA's $4{,}518$~NFE---yet achieves $100\%$ success compared to only $55.0 \pm 41.1\%$ for the Regular~GA. The additional cost is accompanied by a decisive improvement in reliability, indicating that DC~Motor's challenging search landscape requires iterative LLM-driven refinement that a fixed hyperparameter configuration cannot resolve regardless of its evaluation budget.

\paragraph{Baseline Cost Stability.}
Figure~\ref{fig:batch-baseline} visualizes the distribution of baseline cost~$\mathcal{L}$ across all GA-Agent attempts. GA-Agent attains consistently negative, low-variance costs across all case studies (e.g., $-1.716 \pm 0.12$ for AUV; $-3.328 \pm 0.53$ for DC~Motor), reflecting stable convergence behavior across independent runs. The Regular~GA, by contrast, exhibits substantially higher variance and records large positive cost values on the most challenging plants (DC~Motor: $27.79 \pm 28.54$; AUV: $4.63 \pm 3.45$), consistent with its failure to satisfy performance targets reliably under fixed hyperparameters. These results are achieved without any system-level context: the LLM operates solely on numerical performance feedback, demonstrating that iterative outcome-driven adaptation is sufficient for reliable meta-optimization.

\begin{table}[htbp]
\centering
\caption{Final best success score (mean $\pm$ std) and total NFE for GA-Agent and Regular~GA, averaged over 5 independent runs. \colorbox{lightgreen}{\textbf{Bold green}} indicates the better result per row; ties are left unhighlighted.}
\label{tab:llm_ga_vs_regular_ga_score}
\small
\begin{tabular}{lcc}
\toprule
\textbf{Case Study} &
\textbf{GA-Agent} &
\textbf{Regular GA} \\
& \textbf{Score $\pm$ Std (NFE)} & \textbf{Score $\pm$ Std (NFE)} \\
\midrule
Aircraft Pitch              & \cellcolor{lightgreen}\textbf{100.00 $\pm$ 0.00~(2366)}  & 65.00  $\pm$ 13.69~(3772) \\
AUV                         & \cellcolor{lightgreen}\textbf{100.00 $\pm$ 0.00~(1803)}  & 75.00  $\pm$ 0.00~(4518)  \\
Ball \& Beam                & \cellcolor{lightgreen}\textbf{100.00 $\pm$ 0.00~(2568)}  & 95.00  $\pm$ 11.18~(4518) \\
CSTR                        & \cellcolor{lightgreen}\textbf{100.00 $\pm$ 0.00~(1670)}  & 75.00  $\pm$ 0.00~(2656)  \\
DC Motor                    & \cellcolor{lightgreen}\textbf{100.00 $\pm$ 0.00~(6911)}  & 55.00  $\pm$ 41.08~(4518) \\
Inv.\ Pendulum              & \cellcolor{lightgreen}\textbf{100.00 $\pm$ 0.00~(698)}   & 85.00  $\pm$ 13.69~(4518) \\
2DoF Helicopter $\psi$      & 100.00 $\pm$ 0.00~(46) & 100.00 $\pm$ 0.00~(46) \\
Two-Link Manip.\ $\theta_1$ & \cellcolor{lightgreen}\textbf{100.00 $\pm$ 0.00~(1018)}  & 100.00 $\pm$ 0.00~(1135)  \\
\midrule
\textbf{Avg} & 100.00 $\pm$ 0.00~(2135) & 81.25 $\pm$ 16.64~(3210) \\
\bottomrule
\end{tabular}
\end{table}

\subsection{Comparison with Cascade-GA}
\label{subsec:results-c3}

\begin{figure}[h]
    \centering
    \includegraphics[width=0.85\linewidth]{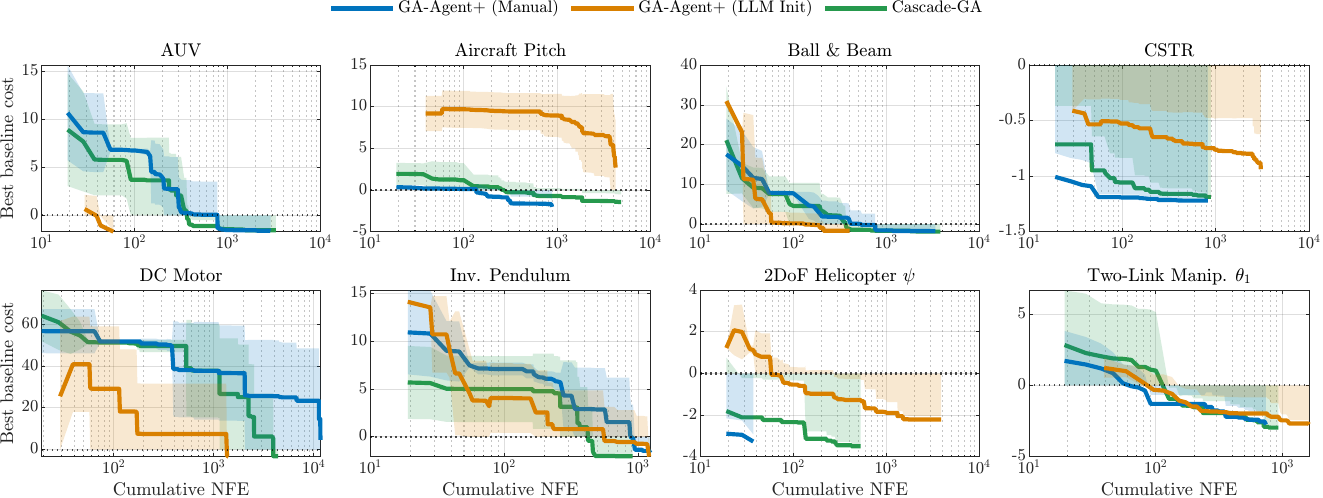}
    \caption{Per-attempt baseline cost $\mathcal{L}$ for GA-Agent+ (Manual~Init), GA-Agent+ (LLM~Init), and Cascade-GA across all eight case studies. GA-Agent+ variants converge within one to three attempts in most cases, while Cascade-GA explores a full population of hyperparameter configurations at the outer evolutionary level. Cases where Cascade-GA achieves a lower baseline cost do not necessarily correspond to higher success scores (see Table~\ref{tab:3way_score_comparison} and the discussion in the text).}
    \label{fig:cascade-multicase}
\end{figure}

We benchmark \textbf{GA-Agent+} (system-aware) against Cascade-GA, a hierarchical optimizer in which a high-level GA ($N_{\mathrm{pop}}^{\mathrm{out}}=5$, $N_{\mathrm{gen}}^{\mathrm{out}}=10$) iteratively tunes the hyperparameters of a low-level GA ($N_{\mathrm{pop}}=10$, $N_{\mathrm{gen}}=10$) across a population of candidate configurations. GA-Agent+ is evaluated under two initialization strategies:
\begin{itemize}
    \item \textbf{Manual Init}: the first GA run uses a fixed preset configuration; the LLM proposes updated hyperparameters from the second attempt onward.
    \item \textbf{LLM~Init}: the LLM proposes the configuration for every attempt, including the first, using a dedicated initialization prompt that receives no prior performance history.
\end{itemize}
Both variants use the system-aware prompt (plant name, control objective, and system description provided to the LLM).
Table~\ref{tab:3way_score_comparison} reports final success scores and NFE;
Figure~\ref{fig:cascade-multicase} shows per-attempt baseline cost trajectories across all case studies.

\paragraph{Success Score.}
All three methods achieve high mean success rates: $96.9\%$ for GA-Agent+ (Manual~Init), $96.3\%$ for Cascade-GA, and $95.0\%$ for GA-Agent+ (LLM~Init). GA-Agent+ (Manual~Init) is the only method to reach $100\%$ on Aircraft~Pitch, doing so at $419$~NFE, while Cascade-GA achieves only $75\%$ at $4{,}655$~NFE. GA-Agent+ (LLM~Init) achieves $100\%$ on DC~Motor at $387$~NFE, while Cascade-GA reaches $95\%$ at $3{,}007$~NFE, confirming that context-driven LLM initialization can circumvent the challenging landscape of this plant. Ball~\&~Beam is the only benchmark where Cascade-GA achieves a strictly higher success score than GA-Agent+ (Manual~Init) ($100\%$ vs.\ $95\%$), though GA-Agent+ (LLM~Init) also achieves $100\%$ there at $275$~NFE. The three methods are therefore competitive on success rates; their primary point of differentiation is sample efficiency.

\paragraph{Sample Efficiency.}
Both GA-Agent+ variants require substantially fewer NFE than Cascade-GA on most benchmarks.
The most pronounced reductions are:
\begin{itemize}
    \item \textit{AUV}: GA-Agent+ (LLM~Init) converges at $70$~NFE vs.\ $2{,}306$~NFE for Cascade-GA at identical $100\%$ success (${\approx}33\times$ fewer evaluations).
    \item \textit{2DoF~Helicopter}: GA-Agent+ (Manual~Init) reaches $100\%$ at $35$~NFE vs.\ $411$~NFE for Cascade-GA (${\approx}12\times$ reduction).
    \item \textit{Ball~\&~Beam}: GA-Agent+ (LLM~Init) achieves $100\%$ at $275$~NFE vs.\ $1{,}559$~NFE for Cascade-GA (${\approx}6\times$ reduction).
\end{itemize}
Averaged across all eight benchmarks, GA-Agent+ (LLM~Init) requires $1{,}147$~NFE per run, compared to $1{,}550$ for GA-Agent+ (Manual~Init) and $1{,}813$ for Cascade-GA, a $37\%$ reduction over the hierarchical baseline.
GA-Agent+ therefore matches or improves upon Cascade-GA's success rates while reducing the number of function evaluations by up to one to two orders of magnitude on individual benchmarks.

\paragraph{Baseline Cost vs.\ Success Score.}
Figure~\ref{fig:cascade-multicase} shows that Cascade-GA achieves the lowest baseline cost~$\mathcal{L}$ on three benchmarks (Ball~\&~Beam, 2DoF~Helicopter, and Two-Link~Manipulator), while GA-Agent+ (Manual~Init) yields the best cost on two (Aircraft~Pitch and CSTR) and GA-Agent+ (LLM~Init) on three (AUV, DC~Motor, and Inverted~Pendulum). Comparing cost and success score jointly reveals an important nuance: a lower aggregate baseline cost does not always translate into a higher success score, because $\mathcal{L}$ is a continuous relative-error aggregate that can be dominated by easily improved metrics, whereas the success score measures whether all performance targets are \emph{jointly} satisfied. Success score is therefore the more informative practical metric for controller synthesis evaluation, as it directly reflects the binary outcome of each performance specification.

\paragraph{Effect of System Context (GA-Agent vs.\ GA-Agent+).}
GA-Agent and GA-Agent+ share identical GA configurations; the sole distinction is whether the LLM prompt includes a structured plant description. Table~\ref{tab:gaagent_vs_gaagentplus_both} (\ref{appx:d}, Section~\ref{subsubsec:context-ablation}) reports a direct per-case cost comparison of the two variants under Manual~Init. On most benchmarks the two variants perform comparably, confirming that numerical feedback alone is generally sufficient for reliable meta-optimization. Two exceptions stand out: on Aircraft~Pitch, system context yields lower cost with reduced variance ($-2.02 \pm 0.15$ vs.\ $-1.86 \pm 0.28$), suggesting that domain knowledge helps the LLM generate more consistent proposals for this plant. On DC~Motor, the effect is sharply reversed: GA-Agent+ (Manual~Init) records a large positive, high-variance cost ($4.64 \pm 17.72$) while GA-Agent consistently achieves $-3.33 \pm 0.53$, revealing seed-dependent behavior in which system-level descriptions can introduce reasoning pathways that interact adversely with specific plant dynamics.

\begin{table*}[htbp]
\centering
\caption{Final best success score (mean $\pm$ std) and total NFE for Cascade-GA, GA-Agent+ (Manual~Init), and GA-Agent+ (LLM~Init), averaged over 5 independent runs. \colorbox{lightgreen}{\textbf{Bold green}} = best result per row; \colorbox{yellow!25}{yellow} = second best. When all methods achieve equal success scores, the cell with the lowest NFE is highlighted green.}
\label{tab:3way_score_comparison}
\small
\resizebox{\textwidth}{!}{%
\begin{tabular}{lccc}
\toprule
\textbf{Case Study} &
\textbf{Cascade-GA} &
\textbf{GA-Agent+ (Manual Init)} &
\textbf{GA-Agent+ (LLM~Init)} \\
 & \textbf{Score $\pm$ Std (NFE)} & \textbf{Score $\pm$ Std (NFE)} & \textbf{Score $\pm$ Std (NFE)} \\
\midrule
AUV
  & 100.00 $\pm$ 0.00~(2306)
  & \cellcolor{yellow!25}100.00 $\pm$ 0.00~(1434)
  & \cellcolor{lightgreen}\textbf{100.00 $\pm$ 0.00~(70)} \\
Aircraft Pitch
  & 75.00 $\pm$ 0.00~(4655)
  & \cellcolor{lightgreen}\textbf{100.00 $\pm$ 0.00~(419)}
  & \cellcolor{yellow!25}75.00 $\pm$ 30.62~(3072) \\
Ball \& Beam
  & \cellcolor{yellow!25}100.00 $\pm$ 0.00~(1559)
  & 95.00 $\pm$ 11.18~(1459)
  & \cellcolor{lightgreen}\textbf{100.00 $\pm$ 0.00~(275)} \\
CSTR
  & \cellcolor{yellow!25}100.00 $\pm$ 0.00~(867)
  & \cellcolor{lightgreen}\textbf{100.00 $\pm$ 0.00~(415)}
  & 85.00 $\pm$ 22.36~(2217) \\
DC Motor
  & \cellcolor{yellow!25}95.00 $\pm$ 11.18~(3007)
  & 80.00 $\pm$ 32.60~(7441)
  & \cellcolor{lightgreen}\textbf{100.00 $\pm$ 0.00~(387)} \\
Inv.\ Pendulum
  & 100.00 $\pm$ 0.00~(861)
  & \cellcolor{yellow!25}100.00 $\pm$ 0.00~(685)
  & \cellcolor{lightgreen}\textbf{100.00 $\pm$ 0.00~(466)} \\
2DoF Helicopter $\psi$
  & \cellcolor{yellow!25}100.00 $\pm$ 0.00~(411)
  & \cellcolor{lightgreen}\textbf{100.00 $\pm$ 0.00~(35)}
  & 100.00 $\pm$ 0.00~(1848) \\
Two-Link Manip.\ $\theta_1$
  & \cellcolor{yellow!25}100.00 $\pm$ 0.00~(840)
  & \cellcolor{lightgreen}\textbf{100.00 $\pm$ 0.00~(513)}
  & 100.00 $\pm$ 0.00~(843) \\
\midrule
\textbf{Avg}
  & 96.25 $\pm$ 8.76~(1813)
  & 96.88 $\pm$ 7.04~(1550)
  & 95.00 $\pm$ 9.64~(1147) \\
\bottomrule
\end{tabular}
}
\end{table*}

\subsection{Sensitivity Analysis Summary}
\label{subsec:sensitivity}

\begin{figure}[h]
    \centering
    \includegraphics[width=0.95\linewidth]{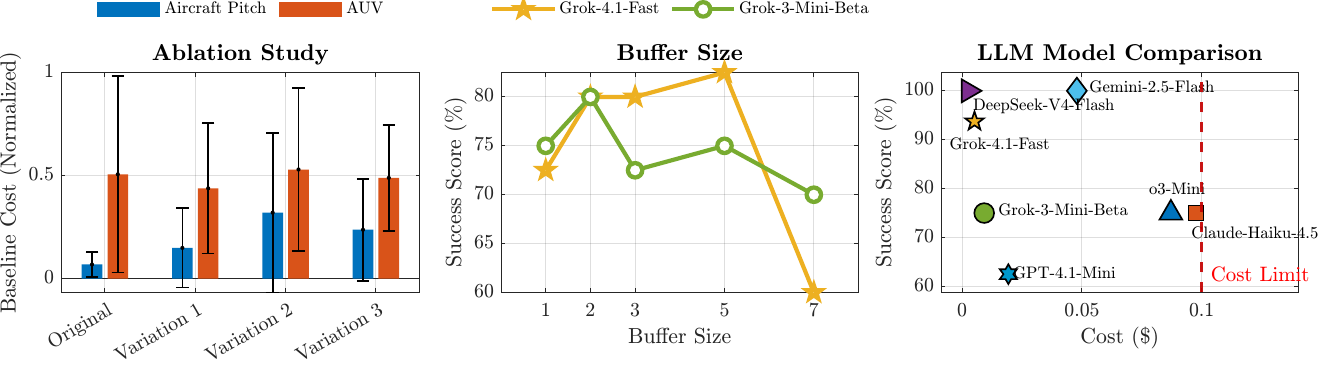}
    \caption{Summary of sensitivity analyses from \ref{appx:d}: prompt ablation (left, \ref{app:ablation}), memory buffer size (center, \ref{app:buffer_size}), and LLM model comparison (right, \ref{app:sensitivity:models}). The left panel reports normalized baseline cost across four prompt variants on AUV (orange) and Aircraft~Pitch (blue). The center panel shows success score vs.\ buffer size for Grok-4.1-Fast and Grok-3-Mini-Beta. The right panel plots success score vs.\ average cost per run for seven frontier LLMs; the red dashed line marks the \$0.10 cost budget.}
    \label{fig:sensitivity-summary}
\end{figure}

\ref{appx:d} presents three complementary studies evaluating the robustness of GA-Agent to:
(i)~prompt formulation (Ablation Study, \ref{app:ablation}),
(ii)~memory buffer size (\ref{app:buffer_size}), and
(iii)~choice of LLM backbone (\ref{app:sensitivity:models}).
Figure~\ref{fig:sensitivity-summary} summarizes the main findings.

\paragraph{Prompt Ablation.}
Four prompt formulations (Original, Variation~1--3) are evaluated on two representative case studies (AUV and Aircraft~Pitch) using DeepSeek-V4-Flash. The Original and Variation~1 (Prose) both attain $100\%$ success; Variation~2 (XML) records the lowest success rate and highest variance. Normalized baseline cost varies across variants: the Original achieves the lowest mean ($0.288 \pm 0.309$) and Variation~2 the highest ($0.425 \pm 0.148$), a $47\%$ relative spread. These results confirm that GA-Agent is largely robust to prompt surface format, with performance driven primarily by the semantic content of the feedback structure rather than by markup conventions.


\paragraph{Effect of Buffer Size.}
Buffer sizes $B \in \{1, 2, 3, 5, 7\}$ are evaluated for Grok-4.1-Fast and Grok-3-Mini-Beta. Neither model benefits monotonically from a larger buffer: Grok-4.1-Fast peaks at $82.5\%$ success at $B=5$, while Grok-3-Mini-Beta peaks at $80.0\%$ at $B=2$. Both models exhibit a declining trend in success rate at larger buffer sizes, and the overall success rate falls from $73.8\%$ at $B=1$ to $65.0\%$ at $B=7$.

This pattern reflects a trade-off between richer historical context and increased prompt length: lightweight models saturate quickly and are harmed by the additional context before heavier models are, as their smaller capacity makes it harder to extract the relevant signal from a longer history. A compact buffer of $B=2$ is recommended as the default for lightweight models (e.g., Grok-3-Mini-Beta, GPT-4.1-Mini), as it captures sufficient recent history while minimizing token cost per LLM call. For heavier models (e.g., Grok-4.1-Fast, Gemini-2.5-Flash), $B=3$ is the recommended default.

\paragraph{LLM Model Comparison.}
Seven frontier LLMs are evaluated as drop-in meta-optimizer backbones. Gemini-2.5-Flash and DeepSeek-V4-Flash both achieve $100\%$ success; the remaining five models range from $62.5\%$ (GPT-4.1-Mini) to $93.8\%$ (Grok-4.1-Fast). Monetary cost spans nearly two orders of magnitude: DeepSeek-V4-Flash is the cheapest at \$$0.0026 \pm 0.0014$ per run, while Claude-Haiku-4.5 is the most expensive at \$$0.0976 \pm 0.0242$. DeepSeek-V4-Flash is the clear Pareto-optimal choice, delivering $100\%$ success at the lowest cost; Gemini-2.5-Flash is the preferred alternative when latency and success rate are prioritized over cost. The cost--success correlation across models is moderately negative, indicating that higher model cost does not imply better optimization performance.

\section{Discussion and Remarks}
\label{sec:discussion}

\paragraph{System-Blind vs.\ System-Aware Adaptation.}
The results reveal a nuanced trade-off between GA-Agent and GA-Agent+. System-blind operation (GA-Agent, Manual~Init) is surprisingly robust: it achieves $100\%$ success on all eight benchmarks by relying on broad initial gain bounds and iterative LLM-driven refinement from numerical feedback alone. Its main limitation appears on plants that prefer small gain values (e.g., DC~Motor), where the broad preset forces the LLM to learn appropriate gain magnitudes over multiple attempts, consuming more NFE than necessary. Adding system context under LLM~Init resolves this: domain knowledge allows the LLM to propose practically appropriate gain ranges from the very first attempt ($365$~NFE on DC~Motor). However, LLM~Init introduces a complementary failure mode on plants that require atypically high gains, such as Aircraft~Pitch, whose low control authority demands large $K_p$ values that a cold-start LLM consistently underestimates, yielding $75\%$ success and high variance. Manual~Init sidesteps this by deferring gain-range specialization entirely to the adaptation loop, at the cost of slower initial convergence on gain-sensitive systems. Overall, numerical performance feedback alone is a powerful driver of adaptation, while system context selectively accelerates or occasionally degrades performance depending on whether the ideal gain requirements of the specific plant align with the LLM's prior.

\paragraph{A Parallelization Opportunity.}
The complementary failure modes of Manual~Init and LLM~Init motivate a straightforward improvement: run both initialization strategies in parallel for the first one to two attempts and continue from whichever yields the lower $\mathcal{L}$ or higher success score. This hedge would likely eliminate the failure of LLM~Init on Aircraft~Pitch while preserving its efficiency gains on DC~Motor ($365$~NFE) and AUV ($88$~NFE for success). Since LLM calls are inexpensive (\$$0.003$ per query for DeepSeek-V4-Flash), the marginal overhead of a parallel cold-start is negligible relative to the GA evaluation budget. More broadly, this \emph{best-of-$k$-starts} strategy is straightforward to implement within the existing agentic loop and could be extended to a portfolio of initialization heuristics, further improving robustness across diverse plant dynamics.

\paragraph{LLM Backbone Characteristics.}
Lightweight backbones (e.g., Grok-3-Mini-Beta, GPT-4.1-Mini) exhibit greater sensitivity to memory buffer size and perform best at $B=2$, where sufficient recent history is available without exceeding the model's effective in-context reasoning capacity. Heavier models (e.g., Gemini-2.5-Flash, Grok-4.1-Fast) are more robust across buffer sizes and can leverage slightly larger buffers (recommended $B=3$, with Grok-4.1-Fast peaking empirically at $B=5$), consistent with their stronger capacity to extract relevant signal from longer histories.

For cost-sensitive deployment, DeepSeek-V4-Flash is the Pareto-optimal backbone: it achieves $100\%$ success rate at the lowest per-run cost. Model cost is not a reliable predictor of optimization performance—the correlation between cost and success across the seven evaluated models is negative, and several high-cost models achieve only $\approx 75\%$ success. Task-specific hyperparameter optimization does not require the largest or most expensive LLMs.

\paragraph{Sample Efficiency.}
GA-Agent achieves strong sample efficiency advantages over both baselines. Relative to Regular~GA, it reduces average NFE by $33\%$ ($2{,}135$ vs.\ $3{,}210$) while improving the average success score by $18.7$ percentage points. Relative to Cascade-GA, GA-Agent+ (LLM~Init) reduces average NFE by $37\%$ ($1{,}147$ vs.\ $1{,}813$) at a comparable success rate ($95.0\%$ vs.\ $96.3\%$); on individual benchmarks the gains reach ${\approx}33\times$ (AUV), ${\approx}12\times$ (2DoF~Helicopter), and ${\approx}6\times$ (Ball~\&~Beam). This efficiency advantage reflects a fundamental asymmetry between evolutionary and semantic search: Cascade-GA must evaluate a full population of hyperparameter configurations at each outer generation, whereas the LLM performs directed reasoning over accumulated outcomes and proposes targeted adjustments, typically converging within one to three attempts. LLM-guided meta-optimization effectively replaces a blind outer-loop evolutionary search with goal-directed, outcome-conditioned reasoning.

\paragraph{Limitations and Future Directions.}
The current framework is evaluated exclusively on single-input, single-output (SISO) PID synthesis in closed-loop simulation with nominal plant models. Real-world deployment introduces challenges not yet addressed: process disturbances, measurement noise, unmodeled high-frequency dynamics, and actuator nonlinearities (saturation, hysteresis, dead zones) that differ from the hard-clipping saturation used in simulation. Robustness of the learned LLM policies to these practical perturbations remains an open question.

On the algorithmic side, the LLM currently observes only summary statistics from completed GA runs (best fitness, achieved metric values, population statistics). Richer feedback channels could accelerate adaptation: per-generation fitness convergence curves would reveal whether a run is stagnating early (suggesting premature population collapse), while elite solution statistics (e.g., the range of controller gains among the top-$k\%$ individuals) would help the LLM identify whether the GA is exploring a narrow region or searching broadly. Such enriched feedback is straightforward to compute and transmit but requires extended analysis to demonstrate tangible benefits.

The buffer size and prompt structure are currently selected via offline sensitivity analysis. Automatic online meta-configuration---where the agent learns to adapt these meta-hyperparameters during the optimization---is a natural extension that could eliminate manual tuning.

Extending the framework to multi-loop cascade control (e.g., inner velocity loop with outer position control) would address industrial practice, where controllers are typically nested. Similarly, enabling joint adaptation of controller structure (e.g., presence/absence of integral action) and parameters would broaden applicability. Multi-objective scenarios where Pareto-frontier weights themselves become adaptive, guided by user preferences, represent another promising direction.

Finally, the theoretical foundations of LLM-guided meta-optimization remain underdeveloped. Formal guarantees on convergence (e.g., that the LLM's proposals monotonically improve fitness), sample-complexity bounds relating the number of GA runs to convergence guarantees, and analysis of how the semantic reasoning capacity of the LLM translates to optimization-theoretic properties would strengthen the principled understanding of this approach.

\bibliographystyle{IEEEtran}
\bibliography{Refs}

@book{gulli.2025,
  author    = {Antonio Gulli},
  title     = {Agentic Design Patterns: A Hands-On Guide to Building Intelligent Systems},
  publisher = {Springer},
  year      = {2025}
}

@article{Wang.2412,
  title={Reinforcement Learning Enhanced LLMs: A Survey},
  author={Wang, Shuhe and Zhang, Shengyu and Zhang, Jie and Hu, Runyi and Li, Xiaoya and Zhang, Tianwei and Li, Jiwei and Wu, Fei and Wang, Guoyin and Hovy, Eduard},
  journal={arXiv preprint arXiv:2412.10400},
  year={2024}
}

@article{Cao.2404,
  title={Survey on large language model-enhanced reinforcement learning: Concept, taxonomy, and methods},
  author={Cao, Yuji and Zhao, Huan and Cheng, Yuheng and Shu, Ting and Chen, Yue and Liu, Guolong and Liang, Gaoqi and Zhao, Junhua and Yan, Jinyue and Li, Yun},
  journal={IEEE Transactions on Neural Networks and Learning Systems},
  year={2024},
  publisher={IEEE}
}

@article{Cai.2505,
  title={Large language model-enhanced reinforcement learning for low-altitude economy networking},
  author={Cai, Lingyi and Zhang, Ruichen and Zhao, Changyuan and Zhang, Yu and Kang, Jiawen and Niyato, Dusit and Jiang, Tao and Shen, Xuemin},
  journal={arXiv preprint arXiv:2505.21045},
  year={2025}
}

@article{Pternea.2402,
  title={The RL/LLM Taxonomy Tree: Reviewing Synergies Between Reinforcement Learning and Large Language Models},
  author={Pternea, Moschoula and Singh, Prerna and Chakraborty, Abir and Oruganti, Yagna and Milletari, Mirco and Bapat, Sayli and Jiang, Kebei},
  journal={Journal of Artificial Intelligence Research},
  volume={80},
  pages={1525--1573},
  year={2024}
}

@Inbook{Duriez.2017,
author="Duriez, Thomas
and Brunton, Steven L.
and Noack, Bernd R.",
title="Machine Learning Control (MLC)",
bookTitle="Machine Learning Control -- Taming Nonlinear Dynamics and Turbulence",
year="2017",
publisher="Springer International Publishing",
address="Cham",
pages="11--48",
isbn="978-3-319-40624-4",
doi="10.1007/978-3-319-40624-4_2",
url="https://doi.org/10.1007/978-3-319-40624-4_2"
}

@article{Wang.2508,
    title={When large language models meet evolutionary algorithms: Potential enhancements and challenges},
    author = {Chao Wang  and Jiaxuan Zhao  and Licheng Jiao  and Lingling Li  and Fang Liu  and Shuyuan Yang },
	journal = {Research},
	volume = {8},
	number = {},
	pages = {0646},
	year = {2025},
	doi = {10.34133/research.0646},
	URL = {https://spj.science.org/doi/abs/10.34133/research.0646},
	eprint = {https://spj.science.org/doi/pdf/10.34133/research.0646}
}

@article{Liu.2402,
  title={Large Language Model Agent for Hyper-Parameter Optimization},
  author={Liu, Siyi and Gao, Chen and Li, Yong},
  journal={arXiv preprint arXiv:2402.01881},
  year={2024}
}

@article{Xie.2309,
  title={Text2Reward: Reward Shaping with Language Models for Reinforcement Learning}, 
  author={Xie, Tianbao and Zhao, Siheng and Wu, Chen Henry and Liu, Yitao and Luo, Qian and Zhong, Victor and Yang, Yanchao and Yu, Tao},
  journal={arXiv preprint arXiv:2309.11489},
  year={2023},
  eprint={2309.11489},
  archivePrefix={arXiv},
  primaryClass={cs.LG},
  url={https://arxiv.org/abs/2309.11489}
}

@article{Ma.2310,
  title={Eureka: Human-Level Reward Design via Coding Large Language Models}, 
  eprint={2310.12931},
  archivePrefix={arXiv},
  primaryClass={cs.RO},
  url={https://arxiv.org/abs/2310.12931}, 
  author={Ma, Yecheng Jason and Liang, William and Wang, Guanzhi and Huang, De-An and Bastani, Osbert and Jayaraman, Dinesh and Zhu, Yuke and Fan, Linxi and Anandkumar, Anima},
  journal={arXiv preprint arXiv:2310.12931},
  year={2023}
}

@article{Eslami.2026,
  title={A Control-Theoretic Foundation for Agentic Systems},
  author={Eslami, Ali and Yu, Jiangbo},
  year={2026},
  eprint={2603.10779},
  archivePrefix={arXiv},
  primaryClass={eess.SY},
  url={https://arxiv.org/abs/2603.10779},
  journal={arXiv preprint arXiv:2603.10779},
}

@inproceedings{Raval.2025,
  title={Circuit-AI: A Self-Hosted AI-Agent Language Model Framework for Control Loop Implementation and Simulation},
  author={Raval, Vishwam and Zeid, Mohamed and Enjeti, Prasad},
  year={2025},
  doi={10.1109/INTELEC63987.2025.11214739},
  booktitle = {2025 IEEE International Communications Energy Conference (INTELEC)},
}

@article{Taheri.2025,
  title={BarrierBench: Evaluating Large Language Models for Safety Verification in Dynamical Systems},
  author={Ali Taheri and Alireza Taban and Sadegh Soudjani and Ashutosh Trivedi},
  year={2025},
  eprint={2511.09363},
  archivePrefix={arXiv},
  primaryClass={cs.AI},
  url={https://arxiv.org/abs/2511.09363},
  journal={arXiv preprint arXiv:2511.09363},
}

@article{Zhou.2025.ProPS,
  title={Prompted Policy Search: Reinforcement Learning through Linguistic and Numerical Reasoning in LLMs},
  author={Yifan Zhou and Sachin Grover and Mohamed El Mistiri and Kamalesh Kalirathnam and Pratyush Kerhalkar and Swaroop Mishra and Neelesh Kumar and Sanket Gaurav and Oya Aran and Heni Ben Amor},
  journal={Conference on Neural Information Processing Systems (NeurIPS)},
  year={2025}
}

@article{Aghaee.2026,
  title={RB-LLM Control: An intelligent control framework with rule-based large language model decision-making},
  author={Aghaee, Fateme and Shaker, Hamid Reza},
  journal={Aerospace Science and Technology},
  volume={168},
  year={2026},
  url={https://doi.org/10.1016/j.ast.2025.111259},
  doi={10.1016/j.ast.2025.111259}
}

@article{Aydin.2026,
  title = {MRAC-LLM Toolbox: An interactive model reference adaptive control enhanced with large language models},
  author = {Aydın, Merve Nilay and Okur, Halil Ibrahim and Gürsoy-Demir, Handan and Tohma, Kadir and Yeroğlu, Celaleddin},
  journal = {SoftwareX},
  volume = {34},
  year = {2026},
  url = {https://doi.org/10.1016/j.softx.2026.102556},
  doi = {10.1016/j.softx.2026.102556}
}

@article{Narimani.2025,
  title={AgenticControl: An Automated Control Design Framework Using Large Language Models},
  author={Narimani, Mohammad and Emami, Seyyed Ali},
  journal={arXiv preprint arXiv:2506.19160},
  year={2025},
  url={https://arxiv.org/abs/2506.19160}
}

@article{yao2022react,
  title={React: Synergizing reasoning and acting in language models},
  author={Yao, Shunyu and Zhao, Jeffrey and Yu, Dian and Du, Nan and Shafran, Izhak and Narasimhan, Karthik and Cao, Yuan},
  journal={arXiv preprint arXiv:2210.03629},
  year={2022}
}

@article{schick2023toolformer,
  title={Toolformer: Language models can teach themselves to use tools},
  author={Schick, Timo and Dwivedi-Yu, Jane and Dess{\`\i}, Roberto and Raileanu, Roberta and Lomeli, Maria and Hambro, Eric and Zettlemoyer, Luke and Cancedda, Nicola and Scialom, Thomas},
  journal={Advances in neural information processing systems},
  volume={36},
  pages={68539--68551},
  year={2023}
}

@article{wang2024survey,
  title={A survey on large language model based autonomous agents},
  author={Wang, Lei and Ma, Chen and Feng, Xueyang and Zhang, Zeyu and Yang, Hao and Zhang, Jingsen and Chen, Zhiyuan and Tang, Jiakai and Chen, Xu and Lin, Yankai and others},
  journal={Frontiers of Computer Science},
  volume={18},
  number={6},
  pages={186345},
  year={2024},
  publisher={Springer}
}

@book{stevens2015aircraft,
  title={Aircraft control and simulation: dynamics, controls design, and autonomous systems},
  author={Stevens, Brian L and Lewis, Frank L and Johnson, Eric N},
  year={2015},
  publisher={John Wiley \& Sons}
}

@article{esfandyari2013adaptive,
  title={Adaptive fuzzy tuning of PID controllers},
  author={Esfandyari, Morteza and Fanaei, Mohammad Ali and Zohreie, Hadi},
  journal={Neural Computing and Applications},
  volume={23},
  number={Suppl 1},
  pages={19--28},
  year={2013},
  publisher={Springer}
}

@article{ahmad2023modeling,
  title={Modeling and hybrid PSO-WOA-based intelligent PID and state-feedback control for ball and beam systems},
  author={Ahmad, Nur Syazreen},
  journal={Ieee Access},
  volume={11},
  pages={137866--137880},
  year={2023},
  publisher={IEEE}
}

@article{vahid2016modeling,
  title={Modeling and control of autonomous underwater vehicle (AUV) in heading and depth attitude via PPD controller with state feedback},
  author={Vahid, Soroush and Javanmard, Kaveh},
  journal={International Journal of Coastal and Offshore Engineering},
  volume={4},
  number={2016},
  pages={11--18},
  year={2016},
  publisher={International Journal of Coastal and Offshore Engineering}
}

@article{luo2017optimal,
  title={Optimal output regulation for model-free quanser helicopter with multistep Q-learning},
  author={Luo, Biao and Wu, Huai-Ning and Huang, Tingwen},
  journal={IEEE Transactions on Industrial Electronics},
  volume={65},
  number={6},
  pages={4953--4961},
  year={2017},
  publisher={IEEE}
}

@book{Ogata2010,
  author = {Ogata, Katsuhiko},
  title = {Modern Control Engineering},
  publisher = {Pearson},
  year = {2010},
  edition = {5th},
  address = {Upper Saddle River, NJ}
}

@misc{Tedrake2023,
  author = {Tedrake, Russ},
  title = {Underactuated Robotics: Algorithms for Walking, Running, Swimming, Flying, and Manipulation},
  year = {2023},
  howpublished = {Course Notes for MIT 6.832},
  url = {https://underactuated.csail.mit.edu},
  note = {Available at \url{https://underactuated.csail.mit.edu}}
}

@book{Spong2006,
  author = {Spong, Mark W. and Hutchinson, Seth and Vidyasagar, M.},
  title = {Robot Modeling and Control},
  publisher = {Wiley},
  year = {2006},
  edition = {1st}
}

@article{gad2024pygad,
  title={Pygad: An intuitive genetic algorithm python library},
  author={Gad, Ahmed Fawzy},
  journal={Multimedia tools and applications},
  volume={83},
  number={20},
  pages={58029--58042},
  year={2024},
  publisher={Springer}
}

@article{Nosrati.2026,
title = {When control meets large language models: From words to dynamics},
journal = {Engineering Applications of Artificial Intelligence},
volume = {178},
pages = {115119},
year = {2026},
issn = {0952-1976},
doi = {https://doi.org/10.1016/j.engappai.2026.115119},
url = {https://www.sciencedirect.com/science/article/pii/S0952197626014028},
author = {Komeil Nosrati and Aleksei Tepljakov and Juri Belikov and Eduard Petlenkov}
}

@article{Liu.2025,
title = {Large language model-based planning agent with generative memory strengthens performance in textualized world},
journal = {Engineering Applications of Artificial Intelligence},
volume = {148},
pages = {110319},
year = {2025},
issn = {0952-1976},
doi = {https://doi.org/10.1016/j.engappai.2025.110319},
url = {https://www.sciencedirect.com/science/article/pii/S0952197625003197},
author = {Junyang Liu and Wenning Hao and Kai Cheng and Dawei Jin}
}

@article{Udekwe.2024,
title = {Comparing actor-critic deep reinforcement learning controllers for enhanced performance on a ball-and-plate system},
journal = {Expert Systems with Applications},
volume = {245},
pages = {123055},
year = {2024},
issn = {0957-4174},
doi = {https://doi.org/10.1016/j.eswa.2023.123055},
url = {https://www.sciencedirect.com/science/article/pii/S0957417423035571},
author = {Daniel Udekwe and Ore-ofe Ajayi and Osichinaka Ubadike and Kumater Ter and Emmanuel Okafor}
}

\newpage
\appendix
\section{Case Study Descriptions}
\label{app:case_studies}

\paragraph{Transport Aircraft Longitudinal}
Medium-sized transport aircraft longitudinal dynamics with 5 states (true airspeed $V_T$, angle of attack $\alpha$, pitch angle $\theta$, pitch rate $Q$, altitude $H$) and 2 control inputs (throttle, elevator). For this MIMO system, PID control is applied to the elevator input (channel 1) to regulate pitch angle $\theta$ to a target of $0.0948$ rad, while throttle is held at trim. Performance targets: MSE $< 0.001$, settling time $< 7.0$ s, overshoot $< 10\%$, control effort $< 0.25$.

\paragraph{Autonomous Underwater Vehicle (AUV)}
Autonomous underwater vehicle pitch and depth dynamics with 3 states (pitch rate $q$, pitch angle $\theta$, depth $z$) controlled by stern plane deflection $f_s$. The control objective is depth regulation to $1.0$ m. Performance targets: MSE $< 0.1$, settling time $< 2.0$ s, overshoot $< 5\%$, control effort $< 0.1$.

\paragraph{Ball and Beam}
Ball and beam system where beam tilt angle $\alpha$ directly controls ball position. States are ball position $r$ and velocity $\dot{r}$. The control objective is to regulate ball position to $0.0$ m. Performance targets: MSE $< 0.05$, settling time $< 1.0$ s, overshoot $< 1\%$, control effort $< 0.1$.

\paragraph{CSTR Exothermic Reactor (CSTR)}
Continuous stirred tank reactor with exothermic reaction $A \rightarrow B$. States are concentration $x_1$, reactor temperature $x_2$, and cooling jacket temperature $x_3$. Control input is cooling jacket flow rate $q_c$. The objective is to regulate reactor temperature to $7.0$ K. Performance targets: MSE $< 0.01$, settling time $< 1.0$ s, overshoot $< 1\%$, control effort $< 0.1$.

\paragraph{DC Motor}
Armature-controlled DC motor with states angular velocity $\omega$ and armature current $i_a$. Control input is armature voltage $V_a$. The objective is to regulate angular velocity to $10.0$ rad/s. Performance targets: MSE $< 0.05$, settling time $< 2.0$ s, overshoot $< 5\%$, control effort $< 0.1$.

\paragraph{Inverted Pendulum}
Pendulum with torque input to balance at upright position. States are angle $\theta$ and angular velocity $\dot{\theta}$. The control objective is to stabilize the pendulum at $\theta = \pi$ rad. Performance targets: MSE $< 0.2$, settling time $< 1.0$ s, overshoot $< 5\%$, control effort $< 0.1$.

\paragraph{2-DoF Helicopter Yaw Control}
Quanser 2-DoF helicopter with pitch-yaw coupled dynamics. States are pitch, yaw, pitch rate, and yaw rate. Two control inputs are front and rear motor voltages. For this MIMO system, PID control is applied to the rear motor (input channel 2) to regulate yaw (output channel 2) to $0.1$ rad, while the front motor is held at trim. Performance targets: MSE $< 0.001$, settling time $< 1.0$ s, overshoot $< 1\%$, control effort $< 0.1$.

\paragraph{Two-Link Manipulator}
Two-link planar robotic manipulator with joint angles $\theta_1$, $\theta_2$ and angular velocities $\dot{\theta}_1$, $\dot{\theta}_2$. Two control inputs are joint torques $\tau_1$ and $\tau_2$. For this MIMO system, PID control is applied to $\tau_1$ (input channel 1) to regulate $\theta_1$ (output channel 1) to $0.0$ rad, while $\tau_2$ is held at trim. Performance targets: MSE $< 0.01$, settling time $< 1.0$ s, overshoot $< 1\%$, control effort $< 0.5$.










\section{Inside a Full Run --- What the Agent Actually Does}
\label{app:zoomed_run}

The aggregate tables in Section~\ref{subsec:results-b3} report what GA-Agent achieves across ten case studies; this appendix explains \textit{how}. We trace a single, complete four-attempt run in detail, surfacing the reasoning chain that drives each configuration decision, the search-space dynamics that follow from it, and the convergence behaviour that results. The selected run uses the \textbf{AUV} case study with DeepSeek-V4-Flash as the backbone LLM. 

\begin{figure*}[t]
    \centering
    \includegraphics[width=\linewidth]{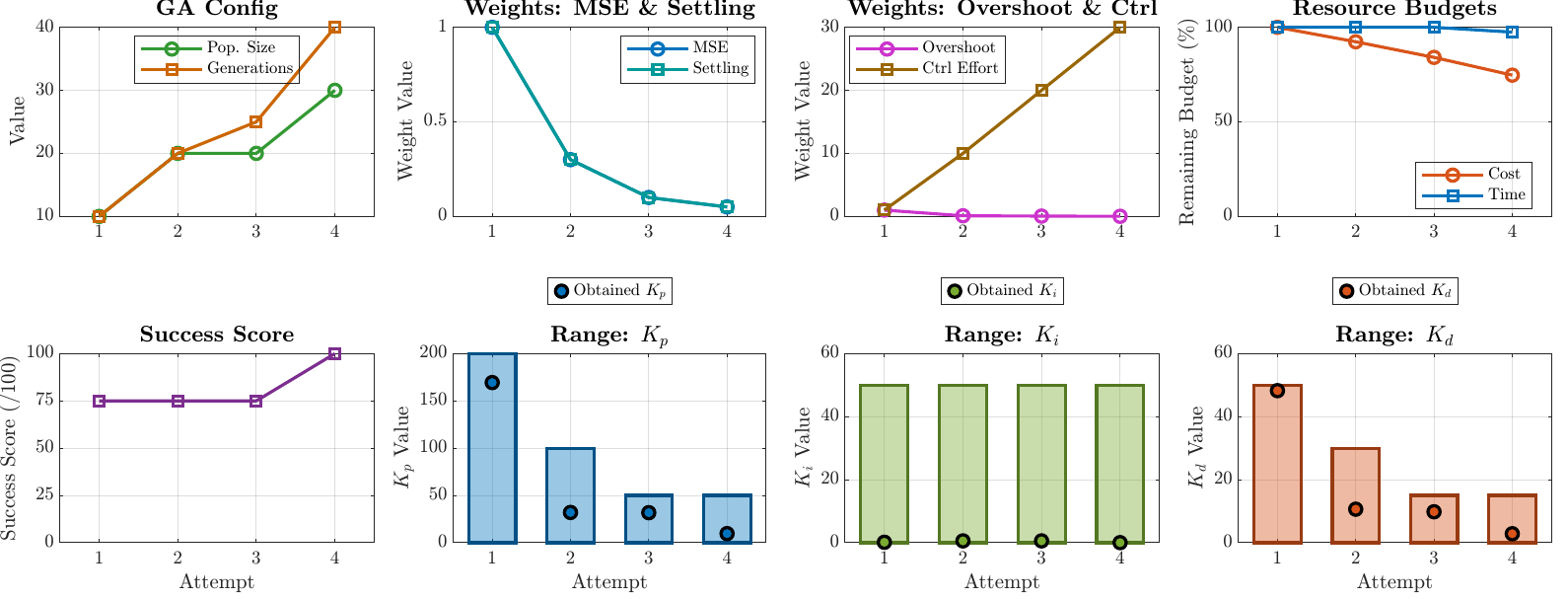}
    \caption{%
        \textbf{Per-attempt overview of the GA-Agent run (AUV, DeepSeek-V4-Flash).}
        Each column corresponds to one attempt.
        \textit{Top row, left to right:}
        (\textit{i})~GA hyperparameters: population size (green circles) and number of generations (orange squares) scale up from $(10,\,10)$ in attempt~1 to $(30,\,40)$ in attempt~4 as the agent gains confidence in the identified region.
        (\textit{ii})~Fitness weights for MSE and settling time, both collapsing from~$\approx$1 to~$\approx$0.05 once those objectives are satisfied.
        (\textit{iii})~Fitness weights for overshoot (near zero throughout) and control effort (rising to~30 by attempt~4), reflecting progressive re-prioritization toward the sole unsatisfied objective.
        (\textit{iv})~Remaining resource budgets: cost budget declines from 100\% to~$\approx$75\%; time budget stays above~93\%, indicating that the four attempts are inexpensive relative to the total allocation.
        \textit{Bottom row, left to right:}
        (\textit{v})~Success score (75/100 for attempts 1--3; 100/100 at attempt~4).
        (\textit{vi--viii})~Search-range evolution for $K_p$, $K_i$, and $K_d$; black markers show the best gain found per attempt.
        $K_p$ narrows from $[0,\,200]$ to $[0,\,50]$; $K_i$ remains stable at $[0,\,50]$; $K_d$ contracts from $[0,\,50]$ to $[0,\,15]$.%
    }
    \label{fig:b_summary}
\end{figure*}

Figure~\ref{fig:b_summary} serves as a reference map for the entire run: the eight panels trace, column by column across the four attempts, how the GA configuration, fitness weights, resource budgets, success score, and PID search bounds are updated. The reader may wish to consult it alongside the per-attempt descriptions below.

\subsection*{Attempt 1 --- Manual Initialization}

The first attempt uses the fixed configuration (Manual initialization); no meta-level LLM call is made. The GA runs with a small budget ($N_\mathrm{pop}{=}10$, $N_\mathrm{gen}{=}10$) over wide bounds: $K_p \in [0,\,200]$, $K_i \in [0,\,50]$, $K_d \in [0,\,50]$. Fitness weights are broadly distributed across MSE, settling time, and overshoot, with control effort effectively unpenalized.

Starting from a baseline cost of~16, the GA converges to a best cost of~$\approx 6.78$ by the end of the attempt. Three of the four targets are satisfied comfortably: MSE~$= 0.038$, settling time~$= 1.6$\,s, overshoot~$= 0.46\%$. Control effort, however, stands at~$0.95$---nearly $10\times$ above the target of~$0.1$. Two structural warning signs accompany this result: $K_d$ converges to~$48.2$ against an upper bound of~$50$, and $K_p$ settles at~$169$ against a bound of~$200$, both near their ceilings. The agent records a success score of~75/100 and passes control to the LLM.

\subsection*{Attempt 2 --- Diagnosing a Boundary-Constrained Local Minimum}

The LLM's first meta-level call produces the following reasoning:

\begin{quote}
\itshape
Attempt~1 achieved excellent MSE, settling, and overshoot, but control effort (0.95) dominates the cost. $K_d$ hit the upper boundary (48.2/50) and $K_p$ is high (169), both driving aggressive derivative and proportional action that inflates control effort. The GA appears trapped in a local minimum where high gains satisfy tracking at the cost of effort. To escape, restrict $K_p \in [0,\,100]$ and $K_d \in [0,\,30]$, forcing exploration of lower-gain regions. $K_i$ remains wide; its proximity to the lower boundary is acceptable given the low steady-state error. Increase the control-effort weight to~10 and reduce the others (MSE~$=0.3$, settling~$=0.3$, overshoot~$=0.1$) to de-emphasize objectives already met. With abundant resources (99.3\% time, 100\% cost remaining), use a moderate GA~($N_\mathrm{pop}{=}20$, $N_\mathrm{gen}{=}20$) for thorough search.
\end{quote}

The diagnosis is two-pronged: \textit{structural} (the GA is against its bounds and cannot explore lower-gain territory) and \textit{incentive-based} (the fitness landscape does not yet penalize high effort). Both are corrected simultaneously. The best cost drops from~$6.78$ to~$0.61$ by the end of attempt~2, and control effort falls to~$0.291$---a threefold improvement, yet still~$2.9\times$ the target. Success score: 75/100.

\subsection*{Attempt 3 --- Progressive Tightening of the Search Space}

The LLM observes that the trend is correct but the remaining space still accommodates solutions with excessive effort:

\begin{quote}
\itshape
Baseline cost improved significantly ($6.78 \to 0.61$) but control effort at~0.291 is still $2.9\times$ the target (0.1). MSE, settling, and overshoot are all well below their targets. The GA is exploring a reasonable region ($K_p{\approx}32$, $K_i{\approx}0.58$, $K_d{\approx}10.6$) but control effort remains high. To force the GA toward lower-effort solutions, tighten $K_p \in [0,\,50]$ and $K_d \in [0,\,15]$, restricting the search to lower-gain solutions. Increase the control-effort weight to~20 and reduce the others (MSE~$=0.1$, settling~$=0.1$, overshoot~$=0.05$). $K_i$ range kept wide since low $K_i$ is beneficial for control effort. With abundant resources (98.5\% cost, 96.8\% time), a moderate-sized GA ($N_\mathrm{pop}{=}20$, $N_\mathrm{gen}{=}25$) provides thorough exploration in the new parameter space.
\end{quote}

The bounds are tightened for the second consecutive time, cutting out the high-gain region that the GA might otherwise drift back toward. By the end of attempt~3, the best cost crosses into negative territory at~$-0.28$---meaning the weighted objective now exceeds the initial baseline---while control effort drops to~$0.212$. The success score remains at~75/100, since~$0.212 > 0.1$.

\subsection*{Attempt 4 --- Scaling Up Within the Identified Region}

The LLM's third and final call takes a different character: instead of restructuring the search space, it consolidates:

\begin{quote}
\itshape
Attempt~3 achieved a baseline cost of~$-0.28$, exceeding all targets except control effort (0.212 vs.\ target 0.1). The restriction strategy has worked: $K_p$ and $K_d$ no longer hit boundaries ($K_p{=}31.8/50$, $K_d{=}9.8/15$) and are exploring freely. Control effort has fallen steadily ($0.95 \to 0.291 \to 0.212$), confirming that the lower-gain region is correct. To push control effort toward~0.1, increase its weight ($20 \to 30$) and reduce others (MSE~$=0.05$, settling~$=0.05$, overshoot~$=0.02$) since those are already well below targets. Parameter ranges remain unchanged---no boundary hits indicate adequate space. With 96.8\% cost and 93.5\% time remaining, scale up the GA ($N_\mathrm{pop}{:}\,20 \to 30$, $N_\mathrm{gen}{:}\,25 \to 40$) for better convergence in the identified promising region.
\end{quote}

The absence of boundary violations is the decisive signal: the LLM treats it as confirmation that the search topology is correct and shifts its effort from exploration to exploitation, applying a larger GA budget to the same region. Control effort drops to~$0.092$, crossing below the target of~$0.1$ for the first time, and the best cost reaches~$-1.8$. The success score reaches~100/100.

\subsection*{Convergence Analysis and Comparison with Regular GA}

\begin{figure}[t]
    \centering
    \includegraphics[width=0.5\linewidth]{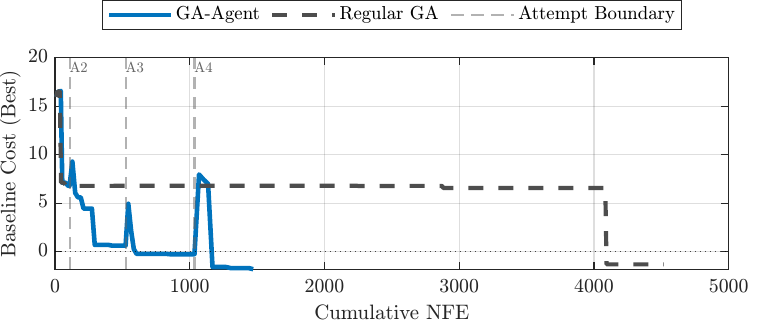}
    \caption{%
        \textbf{Cost convergence over cumulative NFE.}
        GA-Agent (solid blue) versus Regular GA (dashed black).
        Vertical markers \textbf{A2}, \textbf{A3}, \textbf{A4} mark the start of each new GA-Agent attempt; the horizontal dotted line near $y{=}0$ indicates the initial baseline cost.
        The sharp drops following A2 and A3 reflect the search-space restructuring triggered by LLM diagnosis after attempts 1 and~2, respectively.
        GA-Agent final cost: $-1.8$ (111\% below baseline of~16). Regular GA final cost: $-1.3$ (108\% below baseline).%
    }
    \label{fig:b_cost}
\end{figure}

\begin{figure}[t]
    \centering
    \includegraphics[width=0.8\linewidth]{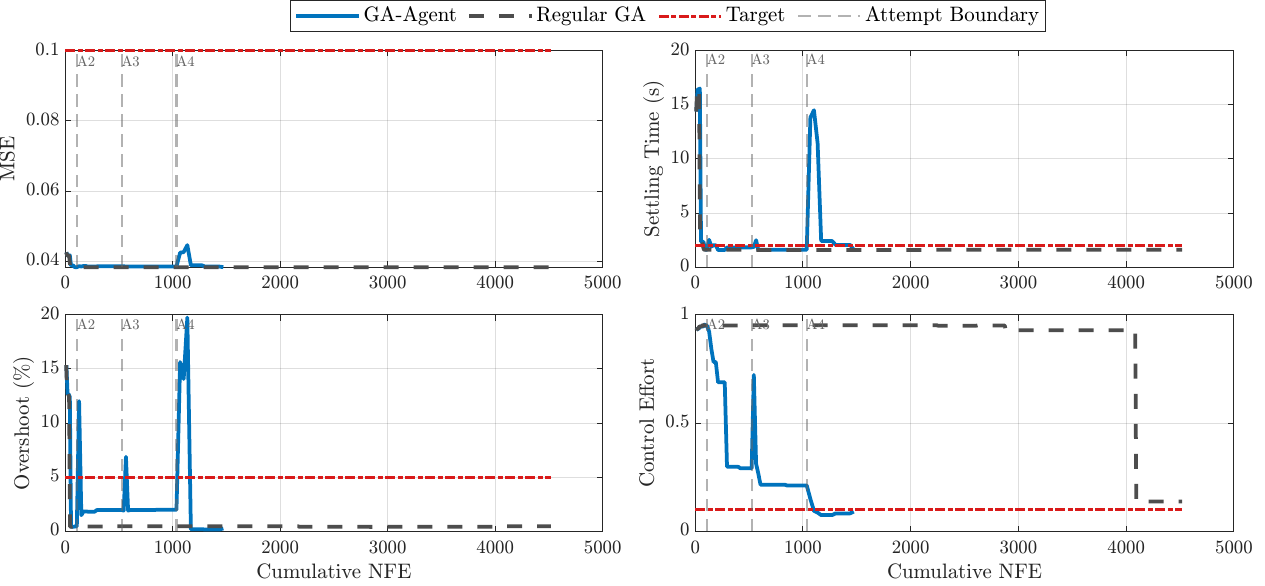}
    \caption{%
        \textbf{Individual performance metric convergence over cumulative NFE:}
        MSE (top-left), settling time (top-right), overshoot (bottom-left), control effort (bottom-right).
        Red dotted lines indicate performance targets.
        GA-Agent and Regular GA achieve near-identical MSE ($0.038$, $-10\%$) and settling time ($1.6$\,s, $-89\%$).
        Differences emerge in overshoot ($0.28\%$ vs.\ $0.49\%$) and, most markedly, in control effort ($0.092$, $-90\%$ vs.\ $0.14$, $-85\%$).
        The Regular GA never applies a fitness signal strong enough to escape the high-gain attractor; GA-Agent does so explicitly in attempt~2.%
    }
    \label{fig:b_metrics}
\end{figure}

\begin{figure}[t]
    \centering
    \includegraphics[width=\linewidth]{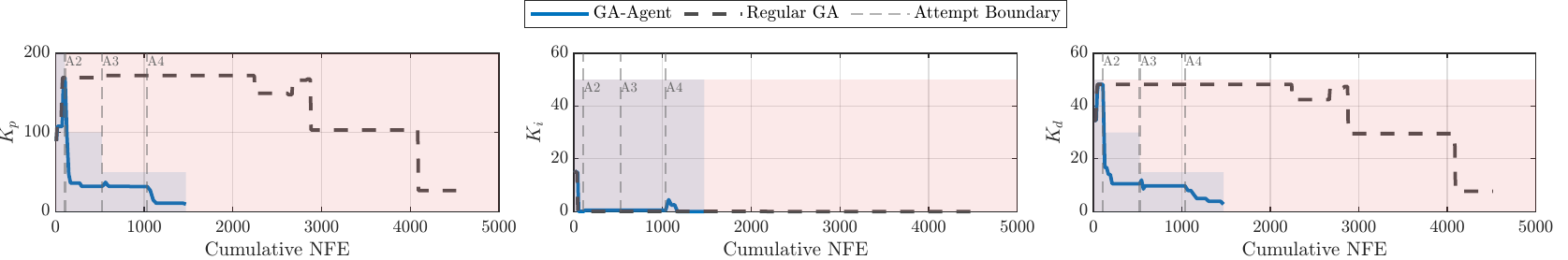}
    \caption{%
        \textbf{PID gain convergence over cumulative NFE:}
        $K_p$ (left), $K_i$ (middle), $K_d$ (right).
        Shaded regions denote the active search bounds in each attempt; vertical lines mark attempt boundaries A2--A4.
        GA-Agent converges to substantially lower gains ($K_p{=}9.6$, $K_i{=}0.020$, $K_d{=}2.8$) than Regular GA ($K_p{=}27$, $K_i{=}0.082$, $K_d{=}7.8$).
        The progressive contraction of the shaded bands across A2 and A3 reflects the LLM's deliberate narrowing of the search space; the absence of boundary violations in attempt~4 confirms that the GA explores freely within the identified low-gain region.%
    }
    \label{fig:b_gains}
\end{figure}

Figures~\ref{fig:b_cost}--\ref{fig:b_gains} complete the picture by placing GA-Agent's trajectory alongside that of a Regular GA operating with a single fixed configuration for the same total NFE budget.

\textbf{Overall cost (Figure~\ref{fig:b_cost}).}
GA-Agent achieves a final cost of~$-1.8$ (111\% improvement from the initial value of~16); Regular GA reaches~$-1.3$ (108\% improvement). The raw gap is modest, but the convergence profile is not: GA-Agent's descent is driven by discrete, sharp drops at each attempt boundary~(A2, A3) rather than by a slow continuous decline. The Regular GA, unable to redirect its search between runs, accumulates NFE without equivalent structural progress.

\textbf{Individual metrics (Figure~\ref{fig:b_metrics}).}
MSE and settling time are nearly identical across both methods (0.038 and~1.6\,s, respectively), reflecting that these objectives are accessible to any well-initialised GA. The difference is concentrated in overshoot (GA-Agent: 0.28\% vs.\ 0.49\%) and, decisively, in control effort (0.092 vs.\ 0.14). The Regular GA is never exposed to a fitness signal that strongly penalises control effort, so it settles in the same high-gain attractor that attempt~1 of GA-Agent revealed and then deliberately escaped.

\textbf{PID gain trajectories (Figure~\ref{fig:b_gains}).}
GA-Agent converges to significantly lower gains ($K_p{=}9.6$, $K_i{=}0.020$, $K_d{=}2.8$) than Regular GA ($K_p{=}27$, $K_i{=}0.082$, $K_d{=}7.8$). The shaded exploration bands shrink across attempts A2 and A3 as the LLM contracts the search bounds; by attempt~4, the GA operates in a narrow, well-targeted region with no boundary pressure. The Regular GA, by contrast, explores the full original bounds throughout, never concentrating its search where low-effort solutions reside.

Together, these four figures illustrate a qualitative point that aggregate metrics alone cannot convey: the value of GA-Agent lies not in any single GA run being superior, but in the agent's ability to observe failure modes, reason about their structural causes, and translate that reasoning into targeted configuration changes that redirect subsequent runs.

\section{Sensitivity Analysis}
\label{appx:d}

This appendix presents three complementary studies examining the
robustness of GA-Agent to axes of variation that are left fixed in the
main experiments: (\ref{app:sensitivity:models}) the choice of LLM backbone, (\ref{app:buffer_size}) the memory
buffer size, and (\ref{app:ablation}) the prompt formulation (ablation study).
Together, they characterize the practical operating envelope of the
framework and identify the configuration choices that most strongly
affect performance.

All experiments in this appendix use the system-blind prompt variant
(GA-Agent) with the Original prompt formulation unless stated
otherwise, and are conducted on two representative case studies:
\textbf{AUV} and \textbf{Aircraft Pitch}.
These two plants were selected because they exhibit contrasting
dynamics and difficulty profiles, providing a meaningful stress-test
for each axis of variation.
Each experimental condition is repeated over multiple independent runs;
error bars in all figures denote one standard deviation.

\subsection{LLM Models Comparison}
\label{app:sensitivity:models}

\begin{figure}[htbp]
    \centering
    \includegraphics[width=\linewidth]{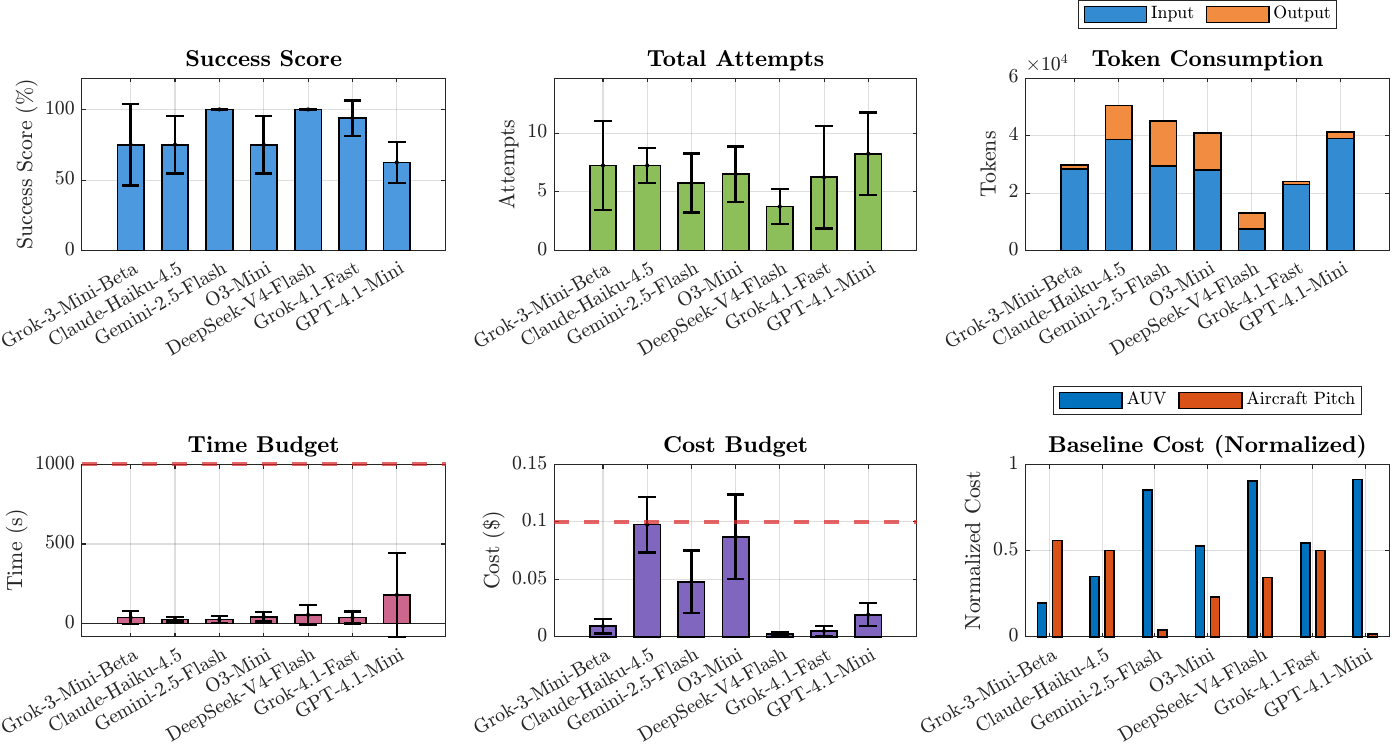}
    \caption{Comparative analysis of seven frontier LLMs acting as
    GA-Agent meta-optimizers across two case studies (AUV and Aircraft
    Pitch).
    \textit{Top row, left to right:} success score (\%), total attempts,
    and token consumption (input/output stacked).
    \textit{Bottom row, left to right:} time budget usage (s), monetary
    cost (\$), and min--max normalized baseline cost per case study.
    Red dashed lines indicate the predefined budget thresholds
    (\$0.10 and 1000\,s).
    Error bars represent one standard deviation across independent runs.}
    \label{fig:d1_llm}
\end{figure}

\paragraph{Setup.}
We evaluate seven frontier LLMs as drop-in replacements for the
meta-optimizer agent: \textbf{Gemini-2.5-Flash},
\textbf{DeepSeek-V4-Flash}, \textbf{Grok-4.1-Fast},
\textbf{Grok-3-Mini-Beta}, \textbf{Claude-Haiku-4.5},
\textbf{O3-Mini}, and \textbf{GPT-4.1-Mini}.
All other framework components (GA configuration space, prompt
formulation, buffer size) are held constant.
Performance is aggregated across the two case studies (AUV and
Aircraft Pitch) and reported on six metrics: success score,
total attempts, token consumption, time budget usage, monetary cost,
and per-case-study normalized baseline cost.
Figure~\ref{fig:d1_llm} summarises the results.

\paragraph{Success Score.}
The models span a 37.5 percentage-point range in mean success score.
Gemini-2.5-Flash and DeepSeek-V4-Flash both achieve a perfect
$100.0 \pm 0.0\%$, while GPT-4.1-Mini records the lowest mean at
$62.5 \pm 14.4\%$.
The remaining five models cluster between $75\%$ and $93.8\%$:
Grok-4.1-Fast ($93.8\%$), Grok-3-Mini-Beta and Claude-Haiku-4.5 and
O3-Mini (all $75.0\%$).
The spread suggests that, while every tested model can function as a
meta-optimizer, the ability to diagnose performance gaps and issue
targeted search-space adjustments varies substantially across model
families.

\paragraph{Convergence Speed.}
DeepSeek-V4-Flash is the most attempt-efficient model, requiring only
$3.8 \pm 1.5$ attempts on average, while GPT-4.1-Mini requires the
most ($8.2 \pm 3.5$ attempts).
Gemini-2.5-Flash combines perfect success with moderate convergence
speed ($5.8$ mean attempts), indicating that it reliably identifies
effective configurations without requiring extended iterative
refinement.
The high variability of GPT-4.1-Mini (std $= 3.5$) suggests that it
occasionally converges quickly but more often requires many correction
cycles before all performance targets are jointly satisfied.

\paragraph{Token Consumption and Cost.}
Token usage varies considerably across models and reflects both the
verbosity of the model's reasoning output and the number of attempts
required.
Claude-Haiku-4.5 consumes the most tokens on average ($50{,}433$
total: $38{,}632$ input + $11{,}800$ output), while DeepSeek-V4-Flash
is by far the most parsimonious ($13{,}013$ total: $7{,}530$ input +
$5{,}482$ output).
The input/output split is notably model-dependent: Grok-3-Mini-Beta and
Grok-4.1-Fast and GPT-4.1-Mini are predominantly input-heavy
(${\approx}94$--$95\%$ input tokens), whereas Gemini-2.5-Flash
generates a comparatively large fraction of output tokens ($35\%$),
consistent with more elaborate chain-of-thought reasoning.

The monetary cost range spans nearly two orders of magnitude:
DeepSeek-V4-Flash is the cheapest at \$$0.0026 \pm 0.0014$ per run,
while Claude-Haiku-4.5 is the most expensive at
\$$0.0976 \pm 0.0242$---a $37.7\times$ ratio.
Crucially, Claude-Haiku-4.5 consumes $98\%$ of the \$0.10 budget
despite not achieving the highest success score, making it a
poor cost--performance choice.
Viewed through the lens of cost-effectiveness (success score per
dollar), DeepSeek-V4-Flash leads at $38{,}619\%/\$$ followed by
Grok-4.1-Fast ($18{,}208\%/\$$) and Gemini-2.5-Flash
($2{,}089\%/\$$), while Claude-Haiku-4.5 ranks last among
competitive models at $768\%/\$$.

\paragraph{Wall-Clock Time and Time Budget Utilization.}
End-to-end optimization time (from WarmUp-Agent initialization through final controller convergence) varies across LLM backbones. Gemini-2.5-Flash and Claude-Haiku-4.5 are the fastest, with mean completion times of $25.7$ and $26.8$ seconds respectively across all experiments, utilizing only $2.6\%$ and $2.7\%$ of the 1000-second per-experiment budget. GPT-4.1-Mini requires substantially more time at $180.2 \pm 94.1$ seconds (mean $\pm$ std), utilizing $18\%$ of the budget and showing high variance across runs. All tested models complete well within the computational time constraint, indicating that wall-clock time is not a limiting factor in this evaluation and that the dominant cost is GA function evaluations, not LLM inference or overhead.

\paragraph{Per-Case-Study Performance.}
The normalized baseline cost (min--max normalized per model to enable
cross-model comparison) reveals that model rankings are not universal
across case studies.
On AUV, Grok-3-Mini-Beta achieves the lowest normalized cost ($0.20$),
while GPT-4.1-Mini and DeepSeek-V4-Flash record the highest
($0.91$ and $0.90$ respectively).
On Aircraft Pitch the ranking inverts: GPT-4.1-Mini achieves the
lowest normalized cost ($0.02$), while Gemini-2.5-Flash is best
after it ($0.04$).
This case-study dependence underscores that no single model dominates
across all plant dynamics, and that the choice of backbone may benefit
from problem-specific considerations when maximum performance is
required on a specific plant.

\paragraph{Overall Assessment.}
Taken together, \textbf{DeepSeek-V4-Flash} offers the best overall cost--performance trade-off: it achieves $100\%$ success at the fewest attempts and the lowest cost per run, making it the recommended choice.

\textbf{Gemini-2.5-Flash} is the strongest option when success rate and \emph{execution speed} are prioritized over monetary cost, completing experiments in $\approx 26$ seconds on average. This model is suitable for latency-sensitive applications where per-token cost is less critical than wall-clock time.

Surprisingly, \textbf{GPT-4.1-Mini}, despite being marketed as a cost-efficient model, is outperformed on both success rate and per-run cost by DeepSeek-V4-Flash and other alternatives in this evaluation. Its higher monetary cost per token and longer execution time ($\approx 180$ seconds) make it suboptimal across both dimensions.

These findings highlight that GA-Agent is robust to backbone choice across the upper half of the model ranking (100\% or near-$100\%$ success on DeepSeek, Gemini, Grok, O3-Mini, Claude variants), with meaningful performance degradation only for the weakest tested model. This robustness suggests the meta-optimization task does not require top-tier reasoning capacity, creating practical opportunity for cost reduction without sacrificing controller quality.

\subsection{Effect of Buffer Size}
\label{app:buffer_size}

\paragraph{Setup.}
The memory buffer controls how many prior GA-run summaries the LLM receives when proposing the next hyperparameter configuration. A buffer of size $B$ provides the $B$ most recent attempts (in chronological order), giving the LLM a rolling context window over the optimization trajectory and allowing it to observe performance trends without the overhead of processing the entire run history.

To characterize how in-context reasoning capacity affects buffer-size sensitivity, we selected two representative LLM backbones from the same vendor (xAI) differing primarily in model scale: \textbf{Grok-4.1-Fast} (heavy-weight, higher reasoning capacity) and \textbf{Grok-3-Mini-Beta} (lightweight, optimized for efficiency). This pairing provides a controlled comparison isolating the effect of model capacity on buffer sensitivity, while vendor-internal consistency minimizes confounding factors due to API behavior, output formats, or prompt engineering biases.

\begin{figure}[htbp]
    \centering
    \includegraphics[width=\linewidth]{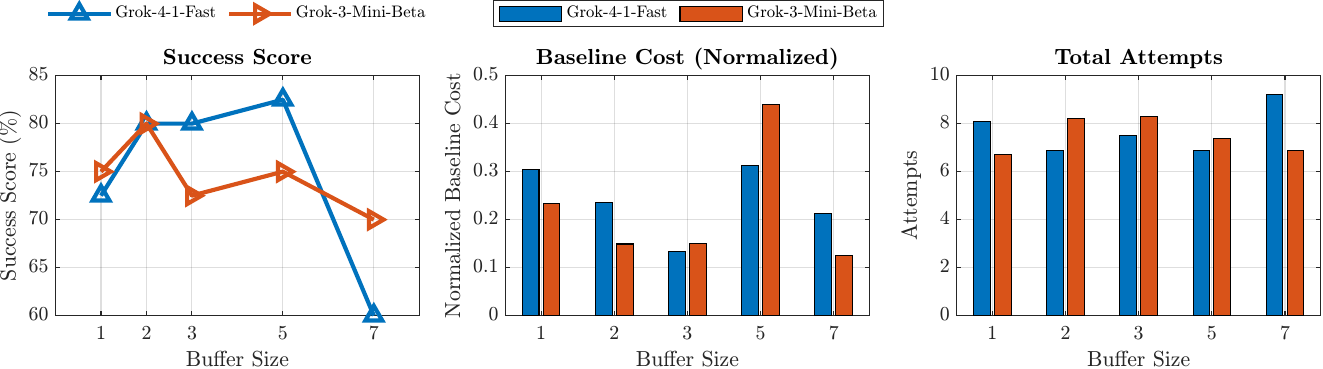}
    \caption{Effect of memory buffer size ($B \in \{1,2,3,5,7\}$) on
    GA-Agent performance for Grok-4.1-Fast (blue) and Grok-3-Mini-Beta
    (orange).
    \textit{Left:} success score (\%).
    \textit{Middle:} min--max normalized baseline cost.
    \textit{Right:} total number of GA attempts.
    Error bars represent one standard deviation across independent runs.}
    \label{fig:d2_buffer}
\end{figure}

We evaluate five buffer sizes $B \in \{1, 2, 3, 5, 7\}$ for each DeepSeek-V4-Flash backbone over the two selected case studies with five independent runs per (case study, $B$) combination. Results are reported on three metrics: success score (percentage of targets jointly satisfied), normalized baseline cost $\mathcal{L}$ (lower is better), and cumulative function evaluations (NFE). Figure~\ref{fig:d2_buffer} summarizes per-case and aggregated findings.

\paragraph{Success Score.}
For both tested models, success score does not improve monotonically with buffer size; instead, performance peaks at intermediate values and exhibits a mild declining trend at larger buffer sizes.

\emph{Grok-4.1-Fast} (heavy-weight) peaks at $B=5$ with $82.5 \pm 18.2\%$ success and remains robust across the range $B \in \{3, 5\}$ (above 80\%). Performance declines more sharply beyond $B=5$, falling to $60.0 \pm 21.1\%$ at $B=7$—a 22.5 percentage-point drop. The general tolerance of this model to buffer size in the range $B \in \{3, 5\}$ reflects stronger in-context learning capacity.

\emph{Grok-3-Mini-Beta} (lightweight) peaks much earlier at $B=2$, achieving $80.0 \pm 21.8\%$ success, and shows steady decline thereafter, dropping to $70.0 \pm 25.8\%$ at $B=7$ compared to $75.0 \pm 33.3\%$ at $B=1$. The early peak and subsequent degradation suggests that lightweight models benefit from a constrained context window.

The divergent optimal buffer sizes between the two models provide mechanistic insight: larger buffers can introduce information overload for smaller models, where noisy or less-relevant older runs dilute the signal from recent performance trends. Heavier models have greater capacity to filter noisy historical data, but even they show degradation at $B=7$. Based on these results, we recommend $B=2$ for lightweight models and $B=3$ for heavy-weight models as practical defaults, balancing performance and computational cost. The main experiments use DeepSeek-V4-Flash (heavy-weight), hence the default buffer size is $B=3$.

\paragraph{Normalized Baseline Cost.}
Larger buffers are associated with lower (better) normalized baseline
cost for both models: Grok-4.1-Fast decreases from $0.304 \pm 0.369$
at $B=1$ to $0.213 \pm 0.292$ at $B=7$, and Grok-3-Mini-Beta from
$0.233 \pm 0.344$ to $0.126 \pm 0.118$.
This pattern at first appears to contradict the success score trend.
However, baseline cost is a continuous aggregate metric that can
decrease even when individual performance targets are not jointly
satisfied (Section~\ref{subsec:results-c3}).
The dissociation between the two metrics here therefore reflects the
same nuance observed in the Cascade-GA comparison: a lower baseline
cost with a larger buffer does not imply that more targets are jointly
satisfied, but rather that the controller is, on average, closer to
each target in isolation.

\paragraph{Total Attempts.}
The number of attempts is broadly stable across buffer sizes for both
models, ranging from approximately $6.7$ to $9.2$ and showing no
strong monotonic trend.
Grok-4.1-Fast exhibits a slight increase at $B=7$ ($9.2 \pm 1.8$
vs.\ $8.1 \pm 2.2$ at $B=1$), consistent with the hypothesis that
a large context window can make each individual proposal less decisive,
requiring more iterations to converge.

\paragraph{Practical Recommendation.}
Buffer size selection depends on LLM capacity and computational constraints. For cost-sensitive deployment with lightweight models (Grok-3-Mini-Beta, GPT-4.1-Mini), $B=2$ is the clear optimal choice. For stronger models capable of leveraging richer context (Gemini-2.5-Flash, Grok-4.1-Fast, DeepSeek-V4-Flash), we recommend $B=3$ as the default, balancing performance gains from slightly expanded history against token cost.

\subsection{Ablation Study}
\label{app:ablation}

\begin{figure}[htbp]
    \centering
    \includegraphics[width=\linewidth]{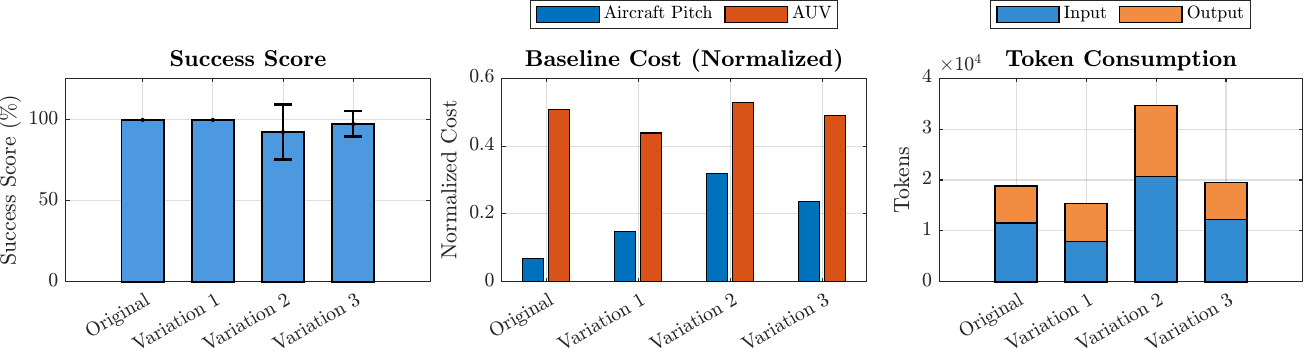}
    \caption{Ablation study over four prompt variants (Original,
    Variation~1--3) evaluated on two case studies (AUV and Aircraft
    Pitch).
    \textit{Left:} success score (\%) with error bars (std).
    \textit{Middle:} min--max normalized baseline cost, broken down by
    case study (Aircraft Pitch in blue, AUV in orange).
    \textit{Right:} token consumption (input in blue, output in orange),
    averaged across runs.
    Note that lower token consumption reflects earlier convergence of
    the agent rather than a shorter prompt: all variants encode
    identical information and differ only in presentational format.}
    \label{fig:d3_ablation}
\end{figure}

\paragraph{Setup.}
We evaluate four prompt variants described in Appendix~\ref{app:prompts}:
\textbf{Original} (Markdown-formatted, used in all main experiments),
\textbf{Variation~1} (Prose), \textbf{Variation~2} (XML), and
\textbf{Variation~3} (Enumerated).
All four variants encode identical semantic content; only the
presentational encoding differs (see Table~\ref{tab:prompt_variants}
for a structural overview).
Experiments use DeepSeek-V4-Flash as the backbone and are conducted on the AUV and Aircraft Pitch
case studies.
Figure~\ref{fig:d3_ablation} summarises the results.

\paragraph{Success Score.}
All four variants achieve high success rates, with a modest 7.5
percentage-point spread between the best and worst performers.
The Original and Variation~1 (Prose) both attain a perfect
$100.0 \pm 0.0\%$, demonstrating that the core semantic content of the
prompt is sufficient for reliable meta-optimization regardless of
whether it is presented with Markdown decoration or as continuous prose.
Variation~3 (Enumerated) reaches $97.5 \pm 7.9\%$, a marginal
degradation, while Variation~2 (XML) records the lowest success score
at $92.5 \pm 16.9\%$---also the most variable across runs.
The elevated variance of Variation~2 suggests that XML-structured
prompts occasionally lead the agent to over-interpret the tag
hierarchy, producing configurations that satisfy some targets while
neglecting others.

\paragraph{Baseline Cost.}
The Original prompt achieves the lowest mean normalized baseline cost
($0.288$), confirming it as the most consistent overall formulation.
The per-case-study breakdown reveals a noteworthy interaction: on
Aircraft Pitch, the ranking follows the same order as the success score.
On AUV, however, \textbf{Variation~1 slightly outperforms the Original}
(normalized cost $0.438$ vs.\ $0.504$, respectively), suggesting that
a prose formulation may produce marginally better decisions on this particular plant.
Variation~2 has the highest normalized cost on both case studies
($0.320$ on Aircraft Pitch; $0.530$ on AUV), consistent with its
lowest success score.

\paragraph{Token Consumption.}
Token usage provides an indirect measure of convergence speed: an
agent that satisfies all targets in fewer attempts will accumulate
fewer tokens over the same run.
Variation~1 (Prose) is the most economical at $15{,}410$ total tokens
(input $7{,}916$ + output $7{,}494$), followed by the Original
($18{,}766$ total).
Variation~2 consumes by far the most tokens ($34{,}629$ total;
$2.2\times$ more than Variation~1), consistent with its lower success
rate requiring more corrective iterations.
The input-to-output ratio is also informative: Variation~1 generates
nearly as many output tokens as input tokens ($1.1$:$1$ ratio),
indicating that the prose formulation elicits longer chain-of-thought
responses, while Grok variants and GPT-based models in the LLM
comparison (\ref{app:sensitivity:models}) tend toward input-heavy profiles.

\paragraph{Overall Assessment.}
The ablation confirms that GA-Agent is largely robust to prompt
format: a 7.5 percentage-point success spread across four
substantially different surface formats indicates that the framework's
performance is driven primarily by its semantic content rather than by Markdown conventions or tag syntax.
The Original Markdown formulation is recommended as the default for
its combination of perfect success, lowest baseline cost, and moderate
token efficiency.
Variation~1 (Prose) is a viable alternative for settings where
Markdown rendering is unavailable or undesirable, with no success
penalty and slightly better performance on the AUV plant.
Variation~2 (XML) is the least recommended formulation due to its
higher variance and elevated token cost.

\subsubsection{Effect of System Context: GA-Agent vs.\ GA-Agent+}
\label{subsubsec:context-ablation}

GA-Agent and GA-Agent+ share identical optimization configurations; the only structural difference is whether the LLM prompt includes a structured plant description (system name, control objective, and a brief description of the system dynamics). Table~\ref{tab:gaagent_vs_gaagentplus_both} compares final baseline cost~$\mathcal{L}$ across all three variants and all eight case studies. 

Three contrasting patterns stand out across the benchmarks.

\textit{Aircraft~Pitch} is the clearest case in favor of Manual~Init: it achieves the lowest
cost ($-2.02 \pm 0.15$, $100\%$ success at $896$~NFE) while LLM~Init fails with a large
positive cost ($2.70 \pm 5.75$, $75\%$ success at $4{,}264$~NFE).
GA-Agent (system-blind) lies between the two ($-1.86 \pm 0.28$, $100\%$ at $2{,}366$~NFE).
The failure of LLM~Init here is consistent with Aircraft~Pitch requiring high PID gains due to
low control authority: without prior performance data, the LLM proposes moderate gain ranges
that prove insufficient for this plant, resulting in high-variance, seed-dependent behavior.

\textit{DC~Motor} shows the opposite pattern: LLM~Init achieves the best cost ($-3.38 \pm 0.49$)
and $100\%$ success at only $1{,}426$~NFE, while GA-Agent+ (Manual~Init) records a large
positive cost ($4.64 \pm 17.72$) and only $80.0 \pm 32.6\%$ success at $11{,}660$~NFE.
GA-Agent falls in between ($-3.33 \pm 0.53$, $100\%$ at $6{,}911$~NFE).
For a system with typical, smaller gain values, the LLM can propose practically reasonable
gain ranges from the very first attempt; it is Manual~Init that introduces seed-dependent
failure by starting with broad gain bounds that include large-gain regions unsuitable for
DC~Motor.

\textit{2DoF~Helicopter~$\psi$} reveals a cost of LLM initialization even on a trivially easy
plant: GA-Agent and GA-Agent+ (Manual~Init) both converge to $100\%$ success at only $46$~NFE,
because the fixed preset already satisfies all performance targets in the very first GA run.
LLM~Init, by contrast, uses $3{,}953$~NFE to reach the same $100\%$ success, because the LLM,
lacking feedback history, proposes complex configurations (e.g., larger populations) that
require many unnecessary evaluations for a system that a minimal configuration would solve
immediately.

The success score comparison confirms that GA-Agent (system-blind) achieves $100\%$ on all eight benchmarks, making it the most robust variant overall. 
GA-Agent+ (LLM~Init) achieves the highest NFE efficiency on three benchmarks: AUV ($88$~NFE),
Ball~\&~Beam ($422$~NFE), and DC~Motor ($1{,}426$~NFE)---but fails to reach $100\%$ on
Aircraft~Pitch ($75\%$) and CSTR ($85\%$), and is highly inefficient on 2DoF~Helicopter.
GA-Agent+ (Manual~Init) occupies a middle ground: reliable enough to reach $100\%$ on six of
eight benchmarks, with the notable exception of DC~Motor ($80\%$).

The fundamental asymmetry is therefore as follows: LLM~Init works well when the LLM's prior
knowledge about the plant yields gain ranges consistent with the true optimal region; it
degrades when the plant requires atypical gains (either unusually high or unusually small)
that the LLM cannot anticipate without observational data.
Manual~Init is more robust precisely because the fixed preset uses broad initial bounds that
span a wide region of gain space, deferring gain-range specialization to the LLM adaptation
loop rather than front-loading it.
A deeper discussion of these trade-offs and their practical implications is provided in
Section~\ref{sec:discussion}.

\begin{figure}[htbp]
    \centering
    \includegraphics[width=0.95\linewidth]{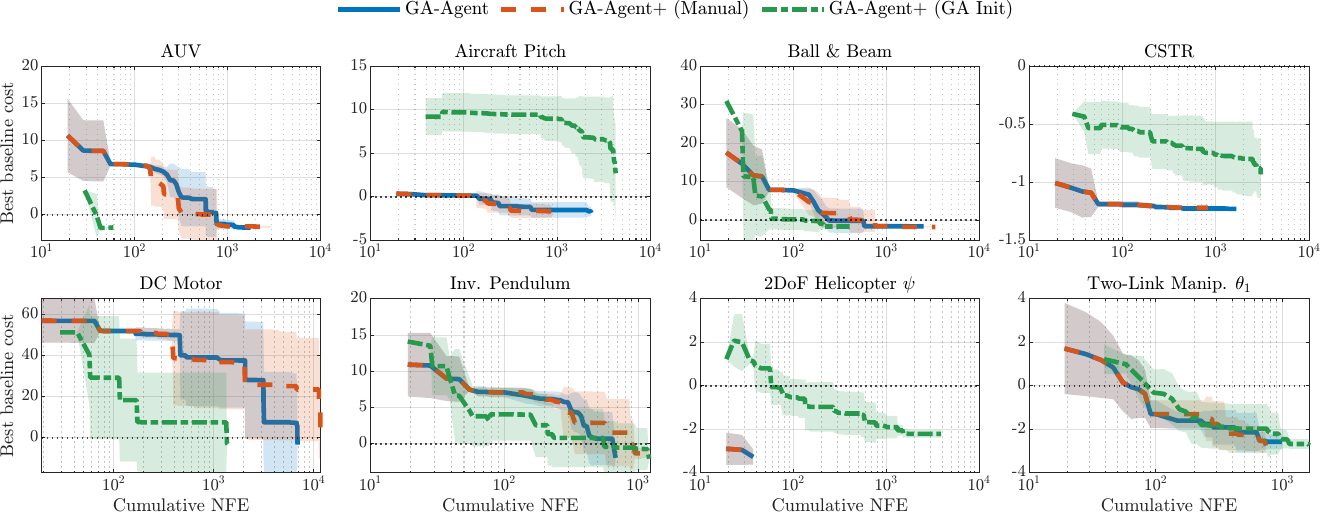}
    \caption{Per-attempt baseline cost trajectories for GA-Agent (system-blind),
    GA-Agent+ (Manual~Init), and GA-Agent+ (LLM~Init) across all eight case studies.
    Trajectories are reproduced from Figures~\ref{fig:batch-baseline}
    and~\ref{fig:cascade-multicase}.
    DC~Motor illustrates the failure mode of Manual~Init under system context (high variance,
    positive costs) and the success of LLM~Init (rapid convergence to negative cost).
    Aircraft~Pitch shows the complementary pattern: LLM~Init fails while Manual~Init stabilizes.
    2DoF~Helicopter illustrates that LLM~Init can be unnecessarily expensive even on easy
    systems when no performance history is available.}
    \label{fig:context-ablation}
\end{figure}

Figure~\ref{fig:context-ablation} overlays the per-attempt baseline cost trajectories from
all three variants across all eight case studies, providing a direct visual comparison.

\begin{table*}[htbp]
    \centering
    \caption{GA-Agent (system-blind, Manual~Init only) vs.\ GA-Agent+ (Manual~Init) vs.\
    GA-Agent+ (LLM~Init) -- lower baseline cost $\mathcal{L}$ is better.
    \colorbox{lightgreen}{\textbf{Bold green}} indicates the best result per row; ties are
    left unhighlighted.}
    \label{tab:gaagent_vs_gaagentplus_both}
    \small
    \resizebox{\textwidth}{!}{%
    \begin{tabular}{lccc}
        \toprule
        \textbf{Case Study}
          & \textbf{GA-Agent}
          & \textbf{GA-Agent+ (Manual~Init)}
          & \textbf{GA-Agent+ (LLM~Init)} \\
        & \textbf{Cost $\pm$ Std (NFE)}
          & \textbf{Cost $\pm$ Std (NFE)}
          & \textbf{Cost $\pm$ Std (NFE)} \\
        \midrule
        AUV
          & \cellcolor{lightgreen}\textbf{$-1.72 \pm 0.12$~(1230)}
          & $-1.62 \pm 0.13$~(1405)
          & $-1.70 \pm 0.13$~(48) \\
        Aircraft Pitch
          & $-1.86 \pm 0.28$~(749)
          & \cellcolor{lightgreen}\textbf{$-2.02 \pm 0.15$~(401)}
          & $2.70 \pm 5.75$~(3053) \\
        Ball \& Beam
          & $-1.63 \pm 0.17$~(1201)
          & \cellcolor{lightgreen}\textbf{$-1.79 \pm 0.13$~(1437)}
          & $-1.67 \pm 0.19$~(256) \\
        CSTR
          & \cellcolor{lightgreen}\textbf{$-1.23 \pm 0.01$~(811)}
          & $-1.22 \pm 0.01$~(395)
          & $-0.94 \pm 0.19$~(2201) \\
        DC Motor
          & $-3.33 \pm 0.53$~(3156)
          & $4.64 \pm 17.72$~(7413)
          & \cellcolor{lightgreen}\textbf{$-3.38 \pm 0.49$~(365)} \\
        Inv.\ Pendulum
          & $-1.95 \pm 0.18$~(523)
          & $-1.67 \pm 0.41$~(661)
          & \cellcolor{lightgreen}\textbf{$-2.06 \pm 0.16$~(443)} \\
        2DoF Helicopter $\psi$
          & \cellcolor{lightgreen}\textbf{$-3.27 \pm 0.35$~(26)}
          & $-3.27 \pm 0.35$~(26)
          & $-2.22 \pm 0.17$~(1810) \\
        Two-Link Manip.\ $\theta_1$
          & $-2.59 \pm 0.33$~(550)
          & \cellcolor{lightgreen}\textbf{$-2.75 \pm 0.24$~(495)}
          & $-2.74 \pm 0.12$~(818) \\
        \bottomrule
    \end{tabular}
    }
\end{table*}

\section{Reproducibility}
\label{app:repro}


\subsection{Regular GA}
\label{app:repro:ga:regular}

This section provides the complete numerical specification of the Regular~GA baseline used in Section~\ref{subsec:results-b3}.

\paragraph{Operator Parameters}
\label{app:repro:ga:operators}

All values in Table~\ref{tab:ga_regular_params} are fixed across all eight case studies and all random seeds; no parameter is adapted between runs.

\begin{table}[h]
  \centering
  \caption{Regular~GA operator parameters, fixed across all case studies
    and seeds. The inner GA of GA-Agent uses the same operators; only the
    hyperparameters $\theta = (N_{\mathrm{pop}},\,N_{\mathrm{gen}},\,
    w_i,\,[\underline{p},\overline{p}])$ are adapted by the LLM between
    attempts.}
  \label{tab:ga_regular_params}
  \small
  \setlength{\tabcolsep}{5pt}
  \resizebox{\textwidth}{!}{%
  \begin{tabular}{lll}
    \toprule
    \textbf{Parameter} & \textbf{Value} & \textbf{Notes} \\
    \midrule
    \multicolumn{3}{l}{\textit{Population and generation budget}} \\
    \quad $N_{\mathrm{pop}}$
      & $10$
      & \\
    \quad $N_{\mathrm{gen}}$
      & $500$
      & Subject to early termination \\
    \midrule
    \multicolumn{3}{l}{\textit{Selection, crossover, and mutation}} \\
    \quad Selection (\texttt{parent\_selection\_type})
      & Steady-state (\texttt{"sss"})
      & Truncation-based; top-$k$ by fitness \\
    \quad Mating parents (\texttt{num\_parents\_mating})
      & $\max(2,\lfloor N_{\mathrm{pop}}/5\rfloor)=2$
      & \\
    \quad Elites (\texttt{keep\_parents})
      & $k=2$
      & Carried forward without re-evaluation \\
    \quad Crossover (\texttt{crossover\_type})
      & Single-point (\texttt{"single\_point"})
      & \\
    \quad Mutation
      & Uniform random, 1 gene
      & \texttt{mutation\_type="random"},
        \texttt{mutation\_num\_genes=1} \\
    \midrule
    \multicolumn{3}{l}{\textit{Fitness weights}} \\
    \quad $w_i$, all $i \in \mathcal{M}$
      & $1.0$
      & Equal; see fitness~\eqref{eq:fitness} \\
    \quad Divergence penalty
      & $-10{,}000$
      & Returned on ODE failure \\
    \midrule
    \multicolumn{3}{l}{\textit{Gain search bounds}} \\
    \quad $K_p$
      & $[0.0,\;200.0]$
      & \multirow{3}{*}{Identical to GA-Agent manual init. bounds} \\
    \quad $K_i$
      & $[0.0,\;50.0]$  & \\
    \quad $K_d$
      & $[0.0,\;50.0]$  & \\
    \midrule
    \multicolumn{3}{l}{\textit{Budget}} \\
    \quad Wall-clock limit $T_{\max}$
      & $1{,}000\;\text{s}$
      & Checked after every generation \\
    \bottomrule
  \end{tabular}
  }
\end{table}

\paragraph{Termination Conditions}
\label{app:repro:termination}

The GA halts as soon as the \emph{first} of the following conditions is
met, checked inside \texttt{on\_generation} after every generation:

\begin{enumerate}
  \item \textbf{Generation budget:} all $N_{\mathrm{gen}} = 500$
        generations completed (normal \texttt{PyGAD} exit).
  \item \textbf{All targets met:} success score $100$; a
        \texttt{StopIteration} is raised immediately.
  \item \textbf{Wall-clock budget:} $t_{\mathrm{elapsed}} \geq
        T_{\max} = 1{,}000\;\text{s}$, measured from a shared
        experiment timer that spans all GA-Agent attempts; a
        \texttt{StopIteration} is raised.
\end{enumerate}

\noindent
Cases where NFE falls below the maximum (${\approx}4{,}518$) are
terminated by condition~2 or~3.
In Table~\ref{tab:llm_ga_vs_regular_ga_score}, CSTR terminates at
${\approx}294$ generations ($\mathrm{NFE} = 2{,}656$) and Aircraft
Pitch at ${\approx}418$ generations ($\mathrm{NFE} = 3{,}772$) under
Regular~GA, consistent with condition~3 (wall-clock) or early target
satisfaction in a subset of seeds.



\paragraph{Manual Initialization Configuration}
\label{app:repro:warmstart}

When comparing GA-Agent against the Regular~GA
(Section~\ref{subsec:results-b3}), GA-Agent's first attempt uses a
fixed initial configuration, giving both methods an identical
starting point before the LLM takes over.
Table~\ref{tab:warmstart_vs_regular} contrasts the two configurations.

\begin{table}[h]
  \centering
  \caption{Manual~Init. (GA-Agent, attempt~1) vs.\ Regular~GA.
    From attempt~2 onward the LLM is free to modify all parameters.}
  \label{tab:warmstart_vs_regular}
  \small
  \setlength{\tabcolsep}{8pt}
  \renewcommand{\arraystretch}{1.1}
  \begin{tabular}{lcc}
    \toprule
    \textbf{Parameter}
      & \textbf{Manual~Init. (attempt~1)}
      & \textbf{Regular GA} \\
    \midrule
    $N_{\mathrm{pop}}$   & $10$            & $10$ \\
    $N_{\mathrm{gen}}$   & $10$            & $500$ \\
    $w_i$, all $i$       & $1.0$ (equal)   & $1.0$ (equal) \\
    $K_p$ bounds         & $[0.0,\;200.0]$ & $[0.0,\;200.0]$ \\
    $K_i$ bounds         & $[0.0,\;50.0]$  & $[0.0,\;50.0]$ \\
    $K_d$ bounds         & $[0.0,\;50.0]$  & $[0.0,\;50.0]$ \\
    \bottomrule
  \end{tabular}
\end{table}

\noindent
The Manual~Init. budget ($\mathrm{NFE} = 100$, approximately $2.2\%$
of the Regular~GA maximum) is intentionally lightweight, providing
a quick anchor so that both methods share an identical search space
at the outset.


\subsection{Cascade GA}
\label{app:repro:ga:cascade}

The Cascade GA implements a bilevel optimization structure (Section~\ref{sec:bilevel},
paragraph~(B)): an \emph{outer} (high-level) GA searches the hyperparameter
space~$\Theta$, while an \emph{inner} (low-level) GA optimizes PID gains for each
candidate configuration.
The outer chromosome simultaneously encodes fitness weights
$\{w_i\}_{i\in\mathcal{M}}$ and the upper bounds of the PID gain search ranges;
the inner GA then searches within the bounds specified by that chromosome.
The fitness function~$J$ and baseline cost~$\mathcal{L}$ are defined by
Eqs.~\eqref{eq:fitness}--\eqref{eq:baseline}.

\paragraph{Outer (high-level) GA.}
The outer GA evolves seven-gene chromosomes (Table~\ref{tab:cascade_hl_genes}).
Genes 1--4 encode the fitness weights $\{w_i\}$; genes 5--7 encode the
\emph{upper bounds} $(\bar{K}_p,\bar{K}_i,\bar{K}_d)$ of the PID search ranges,
so the inner GA searches over
$[0,\bar{K}_p]\times[0,\bar{K}_i]\times[0,\bar{K}_d]$.
Fixed operator parameters are given in Table~\ref{tab:cascade_hl_params}.

\begin{table}[h]
  \centering
  \caption{Outer (high-level) GA operator parameters.
           PyGAD~\cite{gad2024pygad} keyword names in parentheses.}
  \label{tab:cascade_hl_params}
  \resizebox{\textwidth}{!}{%
  \begin{tabular}{lll}
    \toprule
    \textbf{Parameter} & \textbf{Value} & \textbf{Notes} \\
    \midrule
    Population size $N_{\mathrm{pop}}^{\mathrm{out}}$ & 5
        & \texttt{sol\_per\_pop} \\
    Generations $N_{\mathrm{gen}}^{\mathrm{out}}$      & 10
        & \texttt{num\_generations}; subject to early stop \\
    Chromosome length                                   & 7
        & 4 weights $+$ 3 PID range upper bounds \\
    \midrule
    Selection    & Steady-state (SSS)
        & \texttt{parent\_selection\_type="sss"} \\
    Parents mated $k$  & $\max\!\left(2,\lfloor N_{\mathrm{pop}}^{\mathrm{out}}/5\rfloor\right)=2$
        & \texttt{num\_parents\_mating} \\
    Elitism      & 1 individual
        & \texttt{keep\_parents=1} \\
    \midrule
    Crossover    & Single-point
        & \texttt{crossover\_type="single\_point"} \\
    Mutation     & Uniform random, $25\%$ of genes ($\approx$2 genes)
        & \texttt{mutation\_type="random"}, \texttt{mutation\_percent\_genes=25} \\
    Mutation probability & $0.3$
        & \texttt{mutation\_probability=0.3} \\
    Duplicate genes      & Allowed
        & \texttt{allow\_duplicate\_genes=True} \\
    \midrule
    \multicolumn{3}{l}{\textit{Early termination (first condition reached)}} \\
    $\quad$ All targets met & score $= 100\%$ in any inner run
        & Propagated from inner GA; outer loop halts via \texttt{return "stop"} \\
    $\quad$ Generations exhausted & $g = N_{\mathrm{gen}}^{\mathrm{out}}$
        & Normal termination \\
    \bottomrule
  \end{tabular}
  }
\end{table}

\begin{table}[h]
  \centering
  \caption{Outer GA chromosome: gene index, encoded variable, and search range
           used in all experiments.
           Weight ranges and PID range limits are \emph{uniform} across all
           ten case studies.}
  \label{tab:cascade_hl_genes}
  \setlength{\tabcolsep}{5pt}
  \begin{tabular}{clll}
    \toprule
    \textbf{Gene} & \textbf{Variable} & \textbf{Search range} & \textbf{Role} \\
    \midrule
    1 & $w_{\mathrm{MSE}}$           & $[0,\;10]$ & Fitness weight for MSE \\
    2 & $w_{t_s}$                    & $[0,\;10]$ & Fitness weight for settling time \\
    3 & $w_{M_p}$                    & $[0,\;10]$ & Fitness weight for overshoot \\
    4 & $w_{u_{\mathrm{eff}}}$       & $[0,\;10]$ & Fitness weight for control effort \\
    \midrule
    5 & $\bar{K}_p$  & $[20,\;200]$
        & Upper bound of $K_p$ search range \\
    6 & $\bar{K}_i$  & $[0.5,\;50]$
        & Upper bound of $K_i$ search range \\
    7 & $\bar{K}_d$  & $[0.5,\;50]$
        & Upper bound of $K_d$ search range \\
    \bottomrule
  \end{tabular}
\end{table}

\paragraph{Inner (low-level) GA.}
Each unique outer chromosome triggers a dedicated inner GA run
with the weights and PID bounds encoded in that chromosome.
The inner GA's PID gain search range is always
$[0,\bar{K}_p]\times[0,\bar{K}_i]\times[0,\bar{K}_d]$, where the lower bound is
fixed at zero by construction.
Table~\ref{tab:cascade_ll_params} lists the inner GA's fixed configuration.

\begin{table}[h]
  \centering
  \caption{Inner (low-level) GA configuration within the Cascade GA.
           The fitness and baseline cost formulas are identical to those of
           the Regular GA (Appendix~\ref{app:repro:ga:regular}); the
           differences are the population size, generation budget, and
           mutation operator.}
  \label{tab:cascade_ll_params}
  \resizebox{\textwidth}{!}{%
  \begin{tabular}{lll}
    \toprule
    \textbf{Parameter} & \textbf{Value} & \textbf{Notes} \\
    \midrule
    Population size $N_{\mathrm{pop}}$ & 10
        & \texttt{sol\_per\_pop}; fixed, not tuned by outer GA \\
    Generations $N_{\mathrm{gen}}$     & 10
        & \texttt{num\_generations}; subject to early stop \\
    Chromosome                         & 3 genes ($K_p$, $K_i$, $K_d$)
        & Same gene space as Regular GA \\
    \midrule
    Selection    & Steady-state (SSS)
        & \texttt{parent\_selection\_type="sss"} \\
    Parents mated $k$  & $\max\!\left(2,\lfloor N_{\mathrm{pop}}/5\rfloor\right)=2$
        & \texttt{num\_parents\_mating} \\
    Elitism      & 2 individuals
        & \texttt{keep\_parents=2} \\
    Crossover    & Single-point & \texttt{crossover\_type="single\_point"} \\
    Mutation     & Uniform random, \textbf{1 gene} per individual
        & \texttt{mutation\_type="random"}, \texttt{mutation\_num\_genes=1} \\
    \midrule
    PID search range & $[0,\bar{K}_p]\times[0,\bar{K}_i]\times[0,\bar{K}_d]$
        & Set by outer chromosome genes 5--7 \\
    Fitness weights $\{w_i\}$ & From outer chromosome genes 1--4
        & Updated per outer evaluation \\
    Fixed targets $\{\mathrm{target}_i\}$ & Case-study specific
        & Used for baseline cost and success score only \\
    \midrule
    \multicolumn{3}{l}{\textit{Early termination (first condition reached)}} \\
    $\quad$ All targets met & score $= 100\%$
        & Halts inner run; signals outer GA to stop via \texttt{return "stop"} \\
    $\quad$ Time budget    & $t \geq T_{\max}$
        & Checked at end of each generation \\
    \bottomrule
  \end{tabular}
  }
\end{table}

\noindent\textbf{Remark (mutation operator difference).}
The inner GA uses a fixed count of $1$ mutated gene per individual
(\texttt{mutation\_num\_genes=1}), while the outer GA uses percentage-based mutation
($25\%$ of 7 genes $\approx 2$ genes) with an additional application probability of
$0.3$.  This asymmetry reflects the different chromosome lengths and search
sensitivities: the 7-gene outer chromosome benefits from exploratory multi-gene
perturbations, while the 3-gene inner chromosome is better served by targeted
single-gene mutations to avoid excessive disruption of well-performing PID solutions.

\paragraph{Evaluation caching.}
To avoid redundant inner GA runs, each call to the outer fitness function is
memoized.
The cache key is a 7-tuple formed by rounding the encoded gene values to six
decimal places:
\begin{equation*}
  \mathrm{key}
    = \bigl(
        \lfloor w_{\mathrm{MSE}}\rceil_6,\;
        \lfloor w_{t_s}\rceil_6,\;
        \lfloor w_{M_p}\rceil_6,\;
        \lfloor w_{u_{\mathrm{eff}}}\rceil_6,\;
        \lfloor\bar{K}_p\rceil_6,\;
        \lfloor\bar{K}_i\rceil_6,\;
        \lfloor\bar{K}_d\rceil_6
      \bigr),
\end{equation*}
where $\lfloor\cdot\rceil_6$ denotes rounding to 6 decimal places.
If the outer GA presents a chromosome whose key has been evaluated in a prior
generation, the cached result is returned immediately without re-running the inner GA.
Cache hits are logged per outer generation. 

\paragraph{Per-chromosome random seeding.}
To ensure that each unique outer chromosome explores a distinct region of the PID
search landscape, a deterministic but chromosome-specific seed is derived for every
inner GA run:
\begin{equation*}
  \mathrm{seed}(\mathrm{key})
    = \mathrm{MD5}\!\bigl(\mathrm{str}(\mathrm{key})\bigr)_{\text{hex}[0:8]}
      \!\!\mod 2^{31}.
\end{equation*}
This guarantees that (i)~results are fully reproducible given the same outer
chromosome and (ii)~different chromosomes do not share the same inner-GA
initialization, preventing the cache from masking genuine performance differences
between configurations.


\paragraph{Outer fitness function.}
The outer GA maximizes the negated baseline cost of the inner GA's best solution:
\begin{equation*}
  F_{\mathrm{out}}(\theta)
    = -\,\mathcal{L}\!\bigl(p^{*}(\theta)\bigr)
    = -\sum_{i\in\mathcal{M}}
      \frac{\mathrm{achieved}_i\!\bigl(p^{*}(\theta)\bigr) - \mathrm{target}_i}
           {\mathrm{target}_i},
\end{equation*}
where $p^{*}(\theta)$ is the best PID solution returned by the inner GA for
configuration~$\theta$, and $\mathrm{target}_i$ are the fixed case-study targets
(Appendix~\ref{app:case_studies}).
If the inner GA fails to produce a valid simulation result, a large penalty
$F_{\mathrm{out}}=-10^{6}$ is assigned.

\subsection{ODE Solver and Simulation}
\label{app:ode}

All closed-loop simulations use a custom \textbf{fixed-step forward Euler}
integrator implemented directly in NumPy, without any adaptive step-size
control or error estimation. For each candidate PID gain vector~$p$, the
state is advanced as
\begin{equation}
    x_{k+1} = x_k + \Delta t \cdot f\!\bigl(t_k,\, x_k,\, u_k\bigr),
    \qquad k = 0, 1, \ldots, N-1,
    \label{eq:euler}
\end{equation}
where $f(\cdot)$ is the plant dynamics function~\eqref{eq:dynamics},
$\Delta t$ is the fixed time step, and $N = T/\Delta t$ is the number
of simulation steps. The control input $u_k$ is computed once per step
and held constant over the interval $[t_k, t_{k+1})$ (zero-order hold).

\paragraph{PID discretization.}
The integral and derivative terms are discretized as follows:
\begin{align}
    \text{(integral)} \quad
        I_{k+1} &= I_k + e_k\,\Delta t, \\
    \text{(derivative)} \quad
        D_k &= \frac{e_k - e_{k-1}}{\Delta t},
\end{align}
where $e_k = r - y_k$ is the tracking error at step~$k$.
No derivative filtering or anti-windup is applied; the control signal
is saturated by hard clipping to the actuator limits
$[u_{\min},\, u_{\max}]$ before being passed to the dynamics.

\paragraph{Instability detection.}
At each step, the state vector is checked for numerical blow-up.
A simulation is declared failed if any state component becomes
non-finite (\texttt{NaN} / \texttt{Inf}) or exceeds the threshold
$\lVert x_k \rVert_\infty > 10^6$.
A failed simulation returns an infinite cost and is discarded by the
GA without contributing to population statistics.

\paragraph{Step sizes and simulation horizons.}
All simulation parameters are fixed per case study and are listed in Table~\ref{tab:sim_params}. The choice of $\Delta t$ was tested to yield stable Euler integration for the nonlinear dynamics of each plant at the trim conditions.

\begin{table}[h]
    \centering
    \caption{Simulation parameters per case study.}
    \label{tab:sim_params}
    \begin{tabular}{lcccc}
        \toprule
        Case Study & $\Delta t$ (s) & $T$ (s) & $u_{\min}$ & $u_{\max}$ \\
        \midrule
        Aircraft Pitch        & 0.01  & 20.0 & $-25.0$ & $25.0$ \\
        AUV                   & 0.01  & 20.0 & $-0.5$  & $0.5$ \\
        Ball \& Beam          & 0.01  & 20.0 & $-0.5$  & $0.5$ \\
        CSTR                  & 0.001 & 10.0 & $0.0$  & $3.0$ \\
        DC Motor              & 0.01  & 20.0 & $-24.0$ & $24.0$ \\
        Inverted Pendulum     & 0.01  & 20.0 & $-1.0$  & $1.0$ \\
        Quanser Helicopter    & 0.01  & 10.0 & $-24.0$ & $24.0$ \\
        Two-Link Manipulator  & 0.01  & 20.0 & $-1.0$  & $1.0$ \\
        \bottomrule
    \end{tabular}
\end{table}

\paragraph{Remark on stiff systems.}
Forward Euler is only conditionally stable: for a system with dominant eigenvalue~$\lambda$, stability requires $|\lambda|\,\Delta t \leq 2$. For the CSTR case study, which involves fast reaction kinetics, the step size $\Delta t = 0.001$\,s was selected conservatively (an order of magnitude smaller compared to other plants) to ensure Euler stability. No implicit or stiff solver is used in any experiment; all results are therefore subject to the truncation error of the first-order Euler scheme, $\mathcal{O}(\Delta t)$. This is a deliberate design choice to keep each fitness evaluation computationally cheap and deterministic, which is essential for the large number of evaluations performed by the GA.

\subsection{Implementation Details}
\label{app:impl}

This appendix documents the complete software stack, hardware configuration, LLM setup, and experimental execution environment for all experiments reported in the main paper.

\subsubsection{Software Dependencies}
\label{app:impl:software}

All code is written in Python and requires the following packages:

\begin{table}[h] 
    \centering
    \caption{Software versions used in all experiments.}
    \label{tab:software_versions}
    \begin{tabular}{lc}
        \toprule
        Package & Version \\
        \midrule
        Python & 3.11.10 \\
        LangGraph & 0.3.27 \\
        OpenAI Python SDK & 1.84.0 \\
        NumPy & 2.2.4 \\
        PyGAD & 3.5.0 \\
        \bottomrule
    \end{tabular}
\end{table}

\paragraph{Role of each package:}
\begin{itemize}
    \item \textbf{LangGraph} orchestrates the agentic workflow, including state management, routing between agents (WarmUp-Agent, Planning-Agent, Review-Agent), memory buffer updates, and termination logic.
    \item \textbf{OpenAI SDK} provides HTTP client for LLM API calls to the \texttt{/v1/chat/completions} endpoint; this is used for all models tested (both primary and sensitivity analysis).
    \item \textbf{NumPy} implements the forward Euler integrator and all numerical operations for state simulation, control law computation, and fitness function evaluation.
    \item \textbf{PyGAD} implements the genetic algorithm operators (selection, crossover, mutation, elitism) and population management for both the Regular GA and Cascade GA baselines.
\end{itemize}

\subsubsection{Hardware Configuration}
\label{app:impl:hardware}

\begin{table}[h] 
    \centering
    \caption{Hardware specifications.}
    \label{tab:hardware}
    \begin{tabular}{lc}
        \toprule
        Component & Specification \\
        \midrule
        CPU Model & Intel Core i7-14700HX (2.10 GHz base, up to 5.5 GHz) \\
        Physical Cores & 20 (8P + 12E) \\
        Logical Cores & 28 (with hyperthreading) \\
        Total RAM & 32.0 GB (31.7 GB usable) \\
        \bottomrule
    \end{tabular}
\end{table}

\paragraph{Computational environment notes:}
\begin{itemize}
    \item No GPU-accelerated computation was performed. All experiments are strictly CPU-bound.
    \item GA evaluations do \emph{not} use multiprocessing; all candidates within a generation are evaluated sequentially on a single CPU core.
    \item The main bottleneck in GA-Agent workflows is the LLM inference latency and API round-trip time, not the fitness evaluations themselves.
\end{itemize}

\subsubsection{LLM Configuration}
\label{app:impl:llm}

\paragraph{Primary backbone.}
All main experiments use \textbf{DeepSeek-V4-Flash} as the LLM backbone. This model was chosen for its cost-effectiveness, 
fast inference, and strong performance on the control domain reasoning tasks required by GA-Agent.

\paragraph{Decoding parameters.}
\begin{table}[h] 
    \centering
    \caption{Fixed LLM decoding parameters (DeepSeek-V4-Flash and all alternatives).}
    \label{tab:llm_params}
    \begin{tabular}{lc}
        \toprule
        Parameter & Value \\
        \midrule
        Temperature & 0.0 (deterministic / greedy) \\
        Top-$p$ & Default (not modified) \\
        Frequency penalty & Default (not modified) \\
        Presence penalty & Default (not modified) \\
        Max output tokens & Not explicitly capped \\
        \bottomrule
    \end{tabular}
\end{table}

Deterministic decoding (temperature = 0) was chosen to ensure reproducibility across repeated runs with the same prompt and memory state. All other hyperparameters are left at SDK defaults to minimize confounding factors and simplify tuning.

\paragraph{Output format enforcement.}
GA-Agent outputs are required to be valid JSON objects (containing fields such as \texttt{param\_ranges}, \texttt{weights}, \texttt{ga\_population\_size}, \texttt{ga\_generations}, \texttt{reasoning}, etc.). Compliance is enforced via \emph{prompt instruction only}; no constrained-decoding API feature (e.g., OpenAI's JSON mode or similar) is used. This choice allows the framework to remain backend-agnostic and port to any LLM provider.

\paragraph{Alternative models (sensitivity analysis).}
The following models were evaluated in Appendix~\ref{app:sensitivity:models} to assess robustness to model choice:
\begin{itemize}
    \item Gemini-2.5-Flash (Google)
    \item Grok-4.1-Fast (xAI)
    \item Grok-3-Mini-Beta (xAI)
    \item Claude-Haiku-4.5 (Anthropic)
    \item O3-Mini (OpenAI)
    \item GPT-4.1-Mini (OpenAI)
\end{itemize}

All models used identical decoding parameters (temperature = 0, defaults for all other settings) to isolate the effect of model backbone from decoding strategy.

\subsubsection{Token Consumption and Cost Breakdown}
\label{app:impl:cost}

\paragraph{Per-run token and cost profile (DeepSeek-V4-Flash).}
Across 40 complete optimization runs (5 trials $\times$ 8 case studies), GA-Agent with system-blind configuration and manual initialization exhibits the following aggregate metrics:
\begin{itemize}
    \item \textbf{Total tokens per run}: 4730.4 $\pm$ 4969.6 input tokens, 4287.1 $\pm$ 3927.6 output tokens (mean $\pm$ std).
    \item \textbf{LLM calls per run}: 1.9 $\pm$ 1.5 calls.
    \item \textbf{Tokens per call}: approximately 2557 input, 2317 output (averages).
    \item \textbf{Cost per run}: \$$0.0019 \pm 0.0020$ at DeepSeek-V4-Flash pricing at the time of experimentation.
\end{itemize}
These figures include the system prompt, structured problem specification, memory buffer of prior GA-run summaries, and task-specific reasoning instructions. Token consumption varies substantially across case studies due to variability in convergence speed: simple systems requiring fewer iterations incur lower token costs, while challenging plants requiring multiple refinement attempts.

\paragraph{Variants and sensitivity.}
GA-Agent+ (system-aware, manual initialization) exhibits higher token consumption: 6946.1 $\pm$ 10945.8 input, 5426.1 $\pm$ 7746.3 output per run, with 2.4 $\pm$ 2.8 LLM calls. GA-Agent+ (LLM initialization) requires the most: 7963.8 $\pm$ 13230.6 input, 6169.9 $\pm$ 8313.5 output per run, with 3.3 $\pm$ 3.5 LLM calls, corresponding to approximately \$$0.0027 \pm 0.0035$ per run on the same pricing model. The increase reflects the additional context needed for system-aware reasoning and the greater number of meta-optimization iterations required.

\paragraph{Cost-efficiency perspective.}
All tested configurations remain well within the \$$0.05$ per-experiment budget constraint, with typical runs costing 3.8--5.4\% of the budget. This budget utilization is dominated by GA function evaluations (which carry no API cost), with LLM inference representing a small marginal cost. The efficiency of LLM-based meta-optimization relative to Cascade-GA stems from the reduced number of GA attempts required (typically 1--3) rather than token-level efficiencies.

\paragraph{Aggregate costs.}
\begin{table}[h] 
    \centering
    \caption{Total API costs by experimental subset.}
    \label{tab:api_costs}
    \begin{tabular}{lrrr}
        \toprule
        Experiment & Runs & Total Cost & Per-Run Avg \\
        \midrule
        Main Comparison (GA-Agent vs. Regular GA) & 40 & \$0.0745 & \$0.0019 \\
        Extended Comparison (GA-Agent+ vs. Cascade-GA) & 80 & \$0.2134 & \$0.0027 \\
        Sensitivity: Prompt Ablation & 36 & \$0.1594 & \$0.0044 \\
        Sensitivity: Memory Buffer Size & 100 & \$0.7625 & \$0.0076 \\
        Sensitivity: LLM Backbone & 38 & \$1.0754 & \$0.0384 \\
        \midrule
        \textbf{Total Across All Experiments} & 294 & \$2.2852 & \$0.0110 \\
        \bottomrule
    \end{tabular}
\end{table}

The higher per-run cost in the LLM backbone sensitivity analysis reflects the use of more expensive models; the memory buffer experiment uses more runs but maintains the same model, hence similar per-run cost.

\subsubsection{Reproducibility and Seeding}
\label{app:impl:seeds}

\paragraph{Random seed management.}
All experiments use explicit random seeds. For each case study, 5 random seeds are used in the main comparison (GA-Agent vs. Regular GA) and the extended comparison (GA-Agent+ vs. Cascade-GA). Seeds are applied to:
\begin{itemize}
    \item NumPy's global random state (for Euler integration, candidate initialization, selection randomness).
    \item PyGAD's internal random generator (for crossover, mutation, and selection).
    \item Any LLM sampling (though temperature = 0, so sampling is deterministic).
\end{itemize}






\section{Full Prompts Used}
\label{app:prompts}

Each LLM call comprises two parts: a \textbf{system prompt} fixed across
all attempts, encoding the agent's role and decision rules, and a
\textbf{user prompt} assembled at runtime from the goal, memory, and
budget modules (Section~\ref{sec:prompt}).

The four prompt variants evaluated in this paper are as follows:
\begin{itemize}
    \item \textbf{Original} (default; used in all main experiments and as
          the reference in the ablation study,
          Appendix~\ref{app:ablation}):
          full heuristic guidelines delivered in rich Markdown --- bold
          headers, bullet lists, and pipe tables --- covering boundary
          diagnosis, local-minimum escape, and resource-aware scheduling.
    \item \textbf{Variation~1 (Prose):} the identical information
          conveyed as continuous prose paragraphs, with all Markdown
          decoration removed; hierarchy and emphasis are expressed
          exclusively through sentence structure.
    \item \textbf{Variation~2 (XML):} the identical information wrapped
          in semantic XML elements (\texttt{<parameter>},
          \texttt{<rule>}, \texttt{<resource\_strategy>}, etc.); the
          tag vocabulary encodes the logical structure that Markdown
          symbols encode in the Original.
    \item \textbf{Variation~3 (Enumerated):} the identical information
          organised as a strict decimal-numbered hierarchy
          (\texttt{1.} / \texttt{1.1.} / \texttt{1.1.1.}); every fact
          is addressable by a coordinate that the LLM can cite in its
          \texttt{"reasoning"} field.
\end{itemize}

\paragraph{Information parity.}
All four variants encode the same semantic content: the two cost
formulae, all four parameter descriptions with their boundary-hit rules
and the ceiling-chasing exception, resource-strategy phases, the JSON
output schema, and all decision guidelines.  What differs is solely the
\emph{presentational encoding}, isolating the effect of surface
formatting on LLM reasoning from the effect of information content.

\paragraph{Context variants.}
Independently of the formatting variant, two \emph{context} variants
exist for the user prompt:
\begin{itemize}
    \item \textbf{GA-Agent}: system name and physical description are
          \emph{withheld}; the agent receives only numerical targets and
          prior results.
    \item \textbf{GA-Agent+}: system name and physical description are
          \emph{included}, giving the agent domain context to inform its
          reasoning.
\end{itemize}
The system prompt is shared by both context variants; only the user
prompt header differs.  All formatting variants support both context
variants through identical placeholder substitution.

\paragraph{Structural overview.}
Table~\ref{tab:prompt_variants} summarises the key structural
differences across the four prompt variants.

\begin{table}[h]
\centering
\caption{Structural comparison of the four prompt variants.
  ``Format primitive'' denotes the primary visual/syntactic device used
  to signal hierarchy and emphasis.  All variants share the same
  placeholders, output schema, and semantic content.}
\label{tab:prompt_variants}
\small
\begin{tabular}{lllp{3.6cm}}
\toprule
\textbf{Variant} & \textbf{Format primitive} &
\textbf{Hierarchy signal} &
\textbf{Metric / param tables replaced by} \\
\midrule
Original
  & Markdown
  & \texttt{\#\#} headers, \texttt{**bold**}
  & Pipe tables with \texttt{{\char`\⚠}} boundary flags \\[2pt]
Variation 1 (Prose)
  & Plain text
  & Sentence structure
  & Inline enumeration within paragraphs \\[2pt]
Variation 2 (XML)
  & XML elements
  & Tag nesting + attributes
  & Attribute-carrying \texttt{<metric>} / \texttt{<param>} elements \\[2pt]
Variation 3 (Enum.)
  & Plain text
  & Decimal numbering
  & Decimal-numbered sub-items \\
\bottomrule
\end{tabular}
\end{table}

The complete prompts are reproduced in Sections~\ref{app:prompts:F1}
through~\ref{app:prompts:F4} below.

\subsection{Original Variant}
\label{app:prompts:F1}

\subsubsection{System Prompt}
\label{app:prompts:system}

The system prompt is identical for every attempt, every case study, and
both context variants.  It is transmitted as the \texttt{system} role
in the OpenAI-compatible Messages API.

\begin{lstlisting}[style=promptstyle,
                   caption={System prompt (Original formulation).},
                   label={lst:sysprompt}]
You are a Genetic Algorithm expert designing GA hyperparameters and
parameter search ranges for PID controller optimization.

Your task is to specify GA configuration parameters and weights that
will help the optimizer find good PID controller parameters efficiently.

**GA Optimization Context:**
The GA minimizes the following fitness cost function:

  GA Fitness Cost = sum_i  w_i * (achieved_i)^2
  for i in {MSE, settling time, overshoot, control effort}.

You choose the weights w_i to emphasize different objectives.
- Higher weight -> GA will prioritize minimizing that objective
- Lower weight  -> GA will deprioritize that objective
- Weights are relative: doubling all weights has no effect, only
  ratios matter.

For cross-run comparisons, you will be provided with **baseline cost**:

  Baseline Cost = sum_i  (achieved_i - target_i) / target_i

Fixed targets are predefined and NOT adjustable by you. Baseline cost
measures how far the achieved metrics are from the targets -- lower is
better, negative means target exceeded.

You can only adjust: weights, GA configuration (population/generations),
and param_ranges (PID gain ranges).

**GA Parameters to Configure:**
1. population_size (typically 10-100)
   - Larger -> better exploration but slower
   - Smaller -> faster but may miss good solutions
2. generations (typically 20-200)
   - More -> better convergence but slower; too few -> premature convergence
3. param_ranges: search space for PID parameters [min, max]
   - Parameters hitting boundaries -> ranges may be too tight (expand) or
     GA may be in a local minimum (restrict). Direction matters critically:
     * Hits upper boundary AND metric improving     -> expand upper bound
     * Hits upper boundary AND metric stagnant/
       worsening                                    -> restrict upper bound
   - Restriction Convergence Failure (override rule): if the upper bound
     has been restricted 2+ consecutive times and the parameter still hits
     the new ceiling -> switch strategy: EXPAND by 50-100%.
4. weights: relative importance of each objective.
   - Weight increase alone may be insufficient if the GA is trapped in a
     local minimum -- consider restricting the range of the responsible
     gain in addition to or instead of increasing its weight.

**RESOURCE-AWARE OPTIMIZATION:**
Two budgets constrain the optimization:
1. Cost Budget (Money): per LLM call, independent of GA config size.
2. Time Budget (Computation): proportional to population x generations.

Strategic Resource Planning:
- cost < 30% & time > 70%: minimize LLM calls; can use larger GA configs
- time < 30% & cost > 70%: small GA configs; can afford more LLM calls
- both < 30%:              conservative strategy
- Early (abundant resources): pop=5-10, gen=10-20 for weight exploration
- Refinement phase:           scale up GA config if time budget allows
- Final 1-2 attempts:         use remaining time for one larger run

Return ONLY valid JSON in this exact format:
{
  "weights": {
    "mse": <float>,
    "settling_time": <float>,
    "overshoot": <float>,
    "control_effort": <float>
  },
  "ga_population_size": <int>,
  "ga_generations": <int>,
  "param_ranges": {
    "PID": {
      "Kp": [<float>, <float>],
      "Ki": [<float>, <float>],
      "Kd": [<float>, <float>]
    }
  },
  "reasoning": "<brief explanation of configuration choices>"
}

Guidelines:
- Cost trajectory converges early -> reduce generations
- Cost oscillates/stagnates      -> increase population or generations
- Boundary hit within 5% of min/max: diagnose why before acting
  * Metric improved       -> expand range 50-100%
  * Metric stagnant/worse -> restrict upper bound 30-50%
  * Restriction applied 2+ consecutive times, parameter still hits new
    ceiling -> switch to expansion (ceiling-chasing detected)
- Parameters cluster away from boundaries -> tighten ranges 20-30%
- Metric fails target -> increase that metric's weight
- Metric failed 2+ attempts despite weight increases AND range changes
  -> local-minimum escape: restrict the responsible gain's range
- Explain resource-aware reasoning in the "reasoning" field
- simulation_params (dt, max_time) are FIXED; do not modify them
\end{lstlisting}

\subsubsection{User Prompt Templates}
\label{app:prompts:user}

Two user prompt templates exist: one for the \emph{initial} attempt
($k = 1$, no prior history) and one for all \emph{subsequent} attempts
($k > 1$).  Each template has two context variants: \textbf{GA-Agent}
(without system description) and \textbf{GA-Agent+} (with system
description).  Placeholders in \texttt{\{braces\}} are filled at
runtime.

\paragraph{GA-Agent+ --- initial attempt.}

\begin{lstlisting}[style=promptstyle,
                   caption={User prompt: GA-Agent+ (with context), initial attempt.},
                   label={lst:user_plus_init}]
## System Information

**System Name:** {system_name}
**Control Objective:** {control_objective}
**System Description:** {system_description}

**Fixed Targets (for Baseline Cost - NOT adjustable by you):**
- Target MSE: {fixed_target_mse}
- Target Settling Time: {fixed_target_settling_time}s
- Target Overshoot: {fixed_target_overshoot}%
- Target Control Effort: {fixed_target_control_effort}
- Simulation Parameters (FIXED): dt={dt}s, max_time={max_time}s

**THIS IS ATTEMPT {current_attempt} OF {max_attempts}**
{feedback_section}

Design initial GA configuration, weights, and parameter ranges in JSON
format (consider resource constraints):
\end{lstlisting}

\paragraph{GA-Agent+ --- subsequent attempts.}

\begin{lstlisting}[style=promptstyle,
                   caption={User prompt: GA-Agent+ (with context), subsequent attempts.},
                   label={lst:user_plus}]
## System Information

**System Name:** {system_name}
**Control Objective:** {control_objective}
**System Description:** {system_description}

**Fixed Targets (for Baseline Cost - NOT adjustable by you):**
- Target MSE: {fixed_target_mse}
- Target Settling Time: {fixed_target_settling_time}s
- Target Overshoot: {fixed_target_overshoot}%
- Target Control Effort: {fixed_target_control_effort}
- Simulation Parameters (FIXED): dt={dt}s, max_time={max_time}s

**Current Weights (adjustable by you for GA Fitness):**
- MSE Weight: {weight_mse}
- Settling Weight: {weight_settling}
- Overshoot Weight: {weight_overshoot}
- Control Effort Weight: {weight_control}

**Current GA Configuration (adjustable):**
- Population Size: {current_pop_size}
- Generations: {current_generations}
- Kp Range: {current_kp_range}
- Ki Range: {current_ki_range}
- Kd Range: {current_kd_range}

**THIS IS ATTEMPT {current_attempt} OF {max_attempts}**

{feedback_section}

Provide your GA configuration and weights in JSON format (consider
resource constraints):
\end{lstlisting}

\paragraph{GA-Agent (no context).}
The GA-Agent templates are identical to
Listings~\ref{lst:user_plus_init} and~\ref{lst:user_plus}, with the
\texttt{System Information} header replaced by the following generic
preamble; all remaining fields are unchanged.

\begin{lstlisting}[style=promptstyle,
                   caption={Replacement header for GA-Agent (no context) templates.},
                   label={lst:user_blind}]
## Problem Context
You are tasked with tuning a PID controller for a dynamic system.
The goal is to optimize performance across multiple objectives
including tracking accuracy, transient response, and control effort
efficiency.

## Optimization Targets
**Fixed Targets (for Baseline Cost - NOT adjustable by you):**
... (remainder identical to GA-Agent+ template)
\end{lstlisting}

\subsubsection{Feedback Section Structure}
\label{app:prompts:feedback}

The \texttt{\{feedback\_section\}} placeholder is populated at runtime.
For the initial attempt ($k=1$) it contains only the budget status
block.  For attempt $k > 1$, it contains the last
$\min(k-1,\, B)$ per-attempt summaries (where $B$ is the memory buffer
size, Appendix~\ref{app:buffer_size}) followed by a cross-attempt
trend block.  Each per-attempt summary includes the following fields, in
order:

\begin{enumerate}
    \item \textbf{Scalar summary:} baseline cost~$\mathcal{L}$
          \eqref{eq:baseline}, GA fitness cost~$J$ \eqref{eq:fitness},
          wall-clock run time, number of function evaluations, LLM call
          cost (USD and token counts), and percentage of each resource
          budget consumed.

    \item \textbf{Configuration used:} fixed targets, simulation
          parameters, weights~$\{w_i\}$, and GA settings
          ($N_{\mathrm{pop}}$, $N_{\mathrm{gen}}$, per-gain ranges) that
          produced these results.

    \item \textbf{Achieved metrics:} point estimates for MSE, settling
          time, overshoot, and control effort, plus rise time and
          steady-state error.

    \item \textbf{LLM reasoning trace:} the natural-language
          \texttt{"reasoning"} field returned by the LLM at this attempt,
          reproduced verbatim.  The reasoning trace is included in the
          next prompt's feedback block for context, but is not used as an
          instruction.

    \item \textbf{Parameter exploration statistics:} a table with one
          row per PID gain reporting the best value found, the allowed
          range, and population-level mean~$\pm$\,std, minimum, and
          maximum.  Gains within $5\%$ of a range boundary are flagged
          (\texttt{{\char`\⚠}}) with an inline diagnostic note.

    \item \textbf{Objective metric statistics:} a table with one row per
          metric reporting the best achieved value, the fixed target, and
          population-level statistics.
\end{enumerate}

Following all per-attempt blocks, the feedback section appends a
\textbf{cross-attempt trend analysis} containing: the best and
most-recent baseline cost with a directional trend label
(\texttt{IMPROVING\,/\,STABLE\,/\,DEGRADING}); running averages of GA
run time and LLM call cost; a resource status block with remaining
budget percentages, last-call consumption, and estimated remaining
calls; a resource-aware strategy recommendation; and
boundary-analysis instructions reiterating the local-minimum-escape
rules, closing with the directive for the current attempt.

In the Original variant, the parameter exploration statistics
(item~5) and the objective metric statistics (item~6) are rendered as
Markdown pipe tables, with boundary flags delivered as inline
\texttt{{\char`\⚠}} symbols followed by a natural-language diagnostic
note.  The cross-attempt trend label is typeset in bold.  Variations~1--3
replace these tables with format-native representations as detailed in
Sections~\ref{app:prompts:F2}--\ref{app:prompts:F4}.

\subsection{Variation 1 (Prose)}
\label{app:prompts:F2}

\subsubsection{Overview}

Variation~1 removes all Markdown decorators --- no \texttt{**bold**},
no \texttt{\#\#} headers, no bullet lists, no pipe tables --- and
re-expresses the entire system prompt as continuous prose paragraphs.
Hierarchy and emphasis are conveyed exclusively through sentence
structure (e.g., subordinate clauses for conditions, explicit
transitional phrases for priority ordering).  The hypothesis under test
is whether Markdown formatting provides a meaningful inductive signal to
the LLM, or whether semantic content alone is sufficient.

The user prompt templates retain the same placeholders as the Original
(Listings~\ref{lst:user_plus_init}--\ref{lst:user_blind}) but remove
Markdown headers and bold markers; targets, weights, and GA
configuration are presented as plain declarative sentences.  The
feedback section preserves all six informational fields of
Section~\ref{app:prompts:feedback} but replaces pipe tables and
bulleted boundary flags with inline enumeration within paragraphs; the
trend label is embedded in a sentence rather than set in bold.

\subsubsection{System Prompt (Variation 1 -- Prose)}

The following listing shows the complete system prompt.  Structurally,
it follows the same logical order as Listing~\ref{lst:sysprompt}:
role statement, cost definitions, adjustable parameters (including
boundary rules and the ceiling-chasing exception), resource budgets,
output schema, and decision guidelines.

\begin{lstlisting}[style=promptstyle,
                   caption={System prompt (Variation~1: Prose).},
                   label={lst:sysprompt_prose}]
You are a Genetic Algorithm expert designing GA hyperparameters and
parameter search ranges for PID controller optimization. Your task is
to specify GA configuration parameters and weights that will help the
optimizer find good PID controller parameters efficiently.

The GA minimizes a fitness cost defined as the weighted sum of squared
achieved values across four objectives: MSE, settling time, overshoot,
and control effort. Specifically, GA Fitness Cost equals the sum over
each objective i of w_i multiplied by the square of achieved_i. You
choose the weights w_i to steer the optimizer. A higher weight causes
the GA to prioritize minimizing that objective, while a lower weight
causes it to deprioritize it. Weights are relative, so doubling all of
them simultaneously has no effect; only their ratios matter.

For cross-run comparisons you will receive a baseline cost computed as
the sum over all objectives i of (achieved_i minus target_i) divided by
target_i. Fixed targets are predefined constants that you cannot change.
The baseline cost measures how far the achieved metrics are from the
targets: lower is better, and a negative value means the target was
exceeded. You are only permitted to adjust weights, GA configuration
parameters (population size and generations), and param_ranges (the
PID gain search bounds).

You have four categories of adjustable parameters. The first is
population_size, the number of candidate solutions evaluated in each
generation, which typically ranges from 10 to 100. A larger population
provides better exploration of the search space but makes each
generation slower, while a smaller population runs faster but may miss
good solutions. The second is generations, the number of evolutionary
cycles the GA performs, typically between 20 and 200. More generations
improve convergence quality but increase runtime, and too few can lead
to premature convergence. The third is param_ranges, which defines the
search bounds for each PID gain as a two-element [minimum, maximum]
pair. When parameters consistently reach a boundary their range may be
too tight and should be expanded; when parameters cluster well away
from boundaries the ranges may be too wide and can be tightened. The
direction in which a boundary is hit matters critically. When a
parameter hits its upper boundary and the associated performance metric
is improving, the optimum likely lies beyond the current boundary and
you should expand the upper bound by 50 to 100 percent. When a
parameter hits its upper boundary but a metric is stagnant or
worsening, the GA is likely trapped in a local minimum; in that case
you should restrict the upper bound by 30 to 50 percent to force the
optimizer to explore lower-gain regions. There is an important
exception: if you have restricted the upper bound two or more
consecutive times and the parameter still reaches the new ceiling each
time, the restriction strategy has been falsified and you should switch
to expanding the upper bound by 50 to 100 percent. The fourth
adjustable element is the weights vector. When a metric has persistently
failed its target across multiple attempts despite weight increases, the
problem is likely structural: the parameter ranges may be forcing the
GA into a region where the metric cannot improve, and you should
restrict the upper bound of the gain most physically responsible for
that metric in addition to or instead of further weight increases.

You are subject to two limited budgets. The cost budget covers each LLM
call and is independent of GA configuration size. The time budget covers
wall-clock time consumed by GA runs, where larger configurations consume
more time. When cost remaining is below 30 percent but time remaining is
above 70 percent, minimize LLM calls and use larger GA configurations.
When time remaining is below 30 percent but cost remaining is above 70
percent, use small GA configurations but afford more LLM iterations.
When both are below 30 percent, adopt a conservative strategy. During
early exploration use small configurations (population 5 to 10,
generations 10 to 20) to test weight configurations quickly. During
refinement scale up the GA configuration if the time budget permits.
For the final one or two attempts, commit remaining time to a larger GA
run if the weights are already well tuned.

Return only valid JSON in exactly this format:
{
  "weights": {
    "mse": <float>,
    "settling_time": <float>,
    "overshoot": <float>,
    "control_effort": <float>
  },
  "ga_population_size": <int>,
  "ga_generations": <int>,
  "param_ranges": {
    "PID": {
      "Kp": [<float>, <float>],
      "Ki": [<float>, <float>],
      "Kd": [<float>, <float>]
    }
  },
  "reasoning": "<brief explanation of configuration choices>"
}

Reduce generations if the cost trajectory shows early convergence, and
increase population or generations if the cost oscillates or stagnates.
When a parameter is within 5 percent of a boundary, diagnose the cause
before acting: expand the range if the associated metric is improving,
or restrict the upper bound if the metric is stagnant or worsening,
unless the ceiling-chasing exception described above applies. Tighten
ranges by 20 to 30 percent if parameters cluster away from boundaries.
Increase a metric's weight if it is not meeting its fixed target, but
if a metric has failed its target across two or more attempts despite
weight increases, treat this as a structural problem and adjust the
corresponding parameter range. The simulation parameters dt and
max_time are fixed and cannot be changed. Always explain your
resource-aware reasoning in the reasoning field.
\end{lstlisting}

\subsubsection{User Prompt Templates}

The user prompt templates for Variation~1 are structurally identical
to Listings~\ref{lst:user_plus_init}--\ref{lst:user_blind} (same
placeholders, same logical sections, same GA-Agent / GA-Agent+ context
split) but with all Markdown markers removed.  Targets, weights, and GA
configuration fields are presented as plain declarative sentences (e.g.,
``Target MSE is \texttt{\{fixed\_target\_mse\}}.'') rather than bullet
lists.  The closing instruction replaces the Markdown bold directive
with a plain imperative sentence.

\subsection{Variation 2 (XML)}
\label{app:prompts:F3}

\subsubsection{Overview}

Variation~2 wraps every logical block in a semantic XML element.
The tag vocabulary directly encodes the ontology of the prompt:
\texttt{<role>}, \texttt{<parameter name="...">}, \texttt{<rule
id="...">}, \texttt{<budget name="...">}, \texttt{<phase name="...">},
\texttt{<output\_format>}, and \texttt{<guidelines>} at the system
level; \texttt{<fixed\_targets>}, \texttt{<current\_weights>}, and
\texttt{<current\_ga\_config>} at the user level.  Quantitative values
and diagnostic flags are delivered as XML attributes
(e.g., \texttt{boundary\_issue="true"}, \texttt{pct\_remaining="62.5"})
rather than as inline text.  The hypothesis under test is whether
explicit structural labelling via tag names helps the LLM apply the
correct rule to the correct context, compared with Markdown headers or
prose cues.

The user prompt templates use self-closing or short-content XML
elements for each field (e.g., \texttt{<target metric="mse">0.01
</target>}, \texttt{<param\_range gain="Kp">[0.5, 50.0]</param\_range>})
and replace the closing instruction with a \texttt{<task>} element.
The feedback section preserves all six informational fields of
Section~\ref{app:prompts:feedback}: scalar summaries are encoded as
XML attributes on a \texttt{<summary>} element; the pipe tables in
items~5--6 are replaced by \texttt{<param>} and \texttt{<metric>}
elements with \texttt{pop\_mean}, \texttt{pop\_std}, \texttt{pop\_min},
\texttt{pop\_max}, and \texttt{boundary\_issue} attributes; and the
trend label is the content of a \texttt{<baseline\_trend
direction="...">} element.

\subsubsection{System Prompt (Variation 2 -- XML)}

\begin{lstlisting}[style=promptstyle,
                   caption={System prompt (Variation~2: XML).},
                   label={lst:sysprompt_xml}]
<role>
You are a Genetic Algorithm expert designing GA hyperparameters and
parameter search ranges for PID controller optimization. Your task is
to specify GA configuration parameters and weights that will help the
optimizer find good PID controller parameters efficiently.
</role>

<cost_functions>
  <ga_fitness_cost>
    GA Fitness Cost = sum over i of (w_i * achieved_i^2),
    where i in {MSE, settling_time, overshoot, control_effort}.
    You choose the weights w_i. Higher weight: GA prioritizes that
    objective. Lower weight: GA deprioritizes it. Weights are relative:
    doubling all weights has no effect; only their ratios matter.
  </ga_fitness_cost>
  <baseline_cost>
    Baseline Cost = sum over i of ((achieved_i - target_i) / target_i).
    Fixed targets are NOT adjustable by you. Lower is better;
    negative means target exceeded.
  </baseline_cost>
  <adjustable_by_you>weights, ga_population_size, ga_generations,
    param_ranges</adjustable_by_you>
</cost_functions>

<adjustable_parameters>

  <parameter name="population_size">
    <description>Number of candidates per generation. Typical: 10-100.
    </description>
    <tradeoff>Larger: better exploration, slower. Smaller: faster, may
    miss good solutions.</tradeoff>
  </parameter>

  <parameter name="generations">
    <description>Number of evolutionary cycles. Typical: 20-200.
    </description>
    <tradeoff>More: better convergence, higher runtime. Too few:
    premature convergence.</tradeoff>
  </parameter>

  <parameter name="param_ranges">
    <description>Search bounds for each PID gain as [min, max].
    </description>
    <rule id="upper_hit_improving">
      Parameter hits upper boundary AND metric is improving:
      expand upper bound by 50-100%.
    </rule>
    <rule id="upper_hit_stagnant">
      Parameter hits upper boundary AND metric is stagnant/worsening:
      restrict upper bound by 30-50% to escape local minimum.
    </rule>
    <rule id="ceiling_chasing">
      If upper bound restricted 2+ consecutive times and parameter still
      hits new ceiling: switch strategy, expand by 50-100%.
    </rule>
    <rule id="loose_range">
      Parameters cluster away from boundaries: tighten by 20-30%.
    </rule>
  </parameter>

  <parameter name="weights">
    <rule id="weight_increase">Metric not meeting target: increase its
    weight.</rule>
    <rule id="weight_insufficient">Metric fails target across 2+ attempts
    despite weight increases: restrict the range of the responsible gain
    in addition to or instead of increasing its weight.</rule>
  </parameter>

</adjustable_parameters>

<resource_management>
  <budget name="cost">Per LLM call; independent of GA config size.
  </budget>
  <budget name="time">Proportional to population x generations.</budget>
  <strategy id="low_cost_high_time">cost &lt; 30% and time &gt; 70%:
    minimize LLM calls; use larger GA configs.</strategy>
  <strategy id="low_time_high_cost">time &lt; 30% and cost &gt; 70%:
    small GA configs; afford more LLM iterations.</strategy>
  <strategy id="both_low">both &lt; 30%: conservative strategy.</strategy>
  <phase name="early">pop=5-10, gen=10-20 for weight exploration.
  </phase>
  <phase name="refinement">scale up GA config if time budget allows.
  </phase>
  <phase name="final">use remaining time for one larger convergence run.
  </phase>
</resource_management>

<output_format>
Return ONLY valid JSON in this exact structure:
{
  "weights": {
    "mse": <float>,
    "settling_time": <float>,
    "overshoot": <float>,
    "control_effort": <float>
  },
  "ga_population_size": <int>,
  "ga_generations": <int>,
  "param_ranges": {
    "PID": {
      "Kp": [<float>, <float>],
      "Ki": [<float>, <float>],
      "Kd": [<float>, <float>]
    }
  },
  "reasoning": "<brief explanation of configuration choices>"
}
</output_format>

<guidelines>
  <guideline>Early convergence in cost trajectory: reduce generations.
  </guideline>
  <guideline>Oscillating/stagnating cost: increase population or
  generations.</guideline>
  <guideline>Parameter within 5% of boundary: apply param_ranges rules
  above.</guideline>
  <guideline>Metric exceeds target: increase that metric's weight.
  </guideline>
  <guideline>Metric fails 2+ attempts despite weight and range changes:
  local-minimum escape -- restrict the responsible gain's range.
  </guideline>
  <guideline>Explain resource-aware reasoning in the reasoning field.
  </guideline>
  <guideline>simulation_params (dt, max_time) are FIXED.</guideline>
</guidelines>
\end{lstlisting}

\subsubsection{User Prompt Templates}

The user prompt templates for Variation~2 wrap each field in a
corresponding XML element.  The GA-Agent+ initial template opens with
\texttt{<system\_info>} containing \texttt{<name>}, \texttt{<control\_
objective>}, and \texttt{<description>} children; fixed targets appear
inside \texttt{<fixed\_targets>} as \texttt{<target metric="...">}
elements with a \texttt{status="FIXED"} attribute on the simulation
parameters node.  The GA-Agent (no-context) template replaces
\texttt{<system\_info>} with a \texttt{<task>} preamble element.  The
subsequent-attempt template adds \texttt{<current\_weights>} and
\texttt{<current\_ga\_config>} blocks with attribute-tagged children.
The attempt counter is encoded as \texttt{<attempt number="\{current\_
attempt\}" of="\{max\_attempts\}"/>}.  The closing instruction is a
\texttt{<task>} element directing the agent to return valid JSON.
All placeholder strings (\texttt{\{...\}}) are identical to those in
the Original.

\subsection{Variation 3 (Enumerated)}
\label{app:prompts:F4}

\subsubsection{Overview}

Variation~3 organises the entire system prompt as a strict decimal-
numbered outline (e.g., \texttt{1.} / \texttt{1.1.} / \texttt{1.1.1.}).
Every rule, guideline, and tradeoff is addressable by a coordinate
that the LLM can cite verbatim in its \texttt{"reasoning"} field,
making it possible to trace which rule drove each decision.  No prose
paragraphs, Markdown markers, or XML tags appear outside of the JSON
schema block.  The hypothesis under test is whether explicit
addressability of rules improves the consistency and traceability of
the agent's decisions compared with Markdown or prose formulations.

The user prompt templates adopt the same decimal scheme: section
headers are numbered top-level items (\texttt{1. SYSTEM INFORMATION},
\texttt{2. FIXED TARGETS}, \texttt{3. CURRENT WEIGHTS}, etc.) with
individual fields as sub-items.  The closing instruction is the final
numbered section (\texttt{N. TASK}).  The feedback section preserves
all six informational fields of Section~\ref{app:prompts:feedback}:
scalar summaries appear as numbered sub-items; boundary flags are
delivered as a \texttt{[B]} marker in the parameter label (e.g.,
\texttt{Kp [B]}) with a diagnostic note as the following sub-item; the
pipe tables of items~5--6 become decimal-numbered sub-items with one
row per gain or metric; and the trend label is a numbered item reading
\texttt{Baseline Cost Trend: IMPROVING / WORSENING / STAGNANT}.

\subsubsection{System Prompt (Variation 3 -- Enumerated)}

\begin{lstlisting}[style=promptstyle,
                   caption={System prompt (Variation~3: Enumerated).},
                   label={lst:sysprompt_enum}]
1. ROLE AND TASK
   1.1. You are a Genetic Algorithm expert designing GA hyperparameters
        and parameter search ranges for PID controller optimization.
   1.2. Your task is to specify GA configuration parameters and weights
        that will help the optimizer find good PID controller parameters
        efficiently.

2. COST FUNCTIONS
   2.1. GA Fitness Cost (what the optimizer minimizes):
        GA Fitness Cost = sum over i of (w_i * achieved_i^2),
        where i in {MSE, settling_time, overshoot, control_effort}.
   2.2. Weight interpretation:
        2.2.1. Higher weight: GA prioritizes minimizing that objective.
        2.2.2. Lower weight: GA deprioritizes that objective.
        2.2.3. Weights are relative: doubling all weights has no effect.
   2.3. Baseline Cost (cross-run comparison only):
        Baseline Cost = sum over i of ((achieved_i - target_i) / target_i).
   2.4. Baseline cost interpretation:
        2.4.1. Fixed targets are NOT adjustable by you.
        2.4.2. Lower is better; negative means target exceeded.
   2.5. You can only adjust: weights, ga_population_size,
        ga_generations, param_ranges.

3. ADJUSTABLE PARAMETERS
   3.1. population_size
        3.1.1. Number of candidates per generation. Typical: 10-100.
        3.1.2. Larger: better exploration, slower runtime.
        3.1.3. Smaller: faster, may miss good solutions.
   3.2. generations
        3.2.1. Number of evolutionary cycles. Typical: 20-200.
        3.2.2. More: better convergence, higher runtime.
        3.2.3. Too few: premature convergence.
   3.3. param_ranges
        3.3.1. Search bounds for each PID gain as [min, max].
        3.3.2. Tight range: parameter reaches boundary. Action: expand.
        3.3.3. Loose range: parameters cluster from boundaries.
               Action: tighten 20-30%.
        3.3.4. Upper boundary hit, metric IMPROVING:
               a. Optimum likely lies beyond current boundary.
               b. Action: expand upper bound 50-100%.
        3.3.5. Upper boundary hit, metric STAGNANT or WORSENING:
               a. GA is trapped in a local minimum.
               b. Action: restrict upper bound 30-50%.
        3.3.6. Ceiling-chasing (restriction override):
               a. Signal: upper bound restricted 2+ times in a row AND
                  parameter still hits new ceiling each time.
               b. Interpretation: restriction strategy falsified.
               c. Action: expand upper bound 50-100%.
   3.4. weights
        3.4.1. Metric not meeting target: increase that metric's weight.
        3.4.2. Metric fails target across 2+ attempts despite weight
               increases: problem is structural (local minimum).
               Action: restrict the range of the responsible gain in
               addition to or instead of increasing its weight.

4. RESOURCE MANAGEMENT
   4.1. Cost budget: per LLM call; independent of GA config size.
   4.2. Time budget: proportional to population x generations.
   4.3. Strategy selection:
        4.3.1. cost < 30% and time > 70%: minimize LLM calls;
               use larger GA configs.
        4.3.2. time < 30% and cost > 70%: small GA configs;
               can afford more LLM iterations.
        4.3.3. both < 30%: conservative strategy, small configs.
   4.4. Phase-based approach:
        4.4.1. Early (abundant): pop=5-10, gen=10-20; explore weights.
        4.4.2. Refinement: scale up GA config if time permits.
        4.4.3. Final (last 1-2 attempts): commit time to larger run.

5. OUTPUT FORMAT
   5.1. Return ONLY valid JSON in exactly this structure:
        {
          "weights": {
            "mse": <float>,
            "settling_time": <float>,
            "overshoot": <float>,
            "control_effort": <float>
          },
          "ga_population_size": <int>,
          "ga_generations": <int>,
          "param_ranges": {
            "PID": {
              "Kp": [<float>, <float>],
              "Ki": [<float>, <float>],
              "Kd": [<float>, <float>]
            }
          },
          "reasoning": "<brief explanation>"
        }

6. DECISION GUIDELINES
   6.1. Cost converges early: reduce generations.
   6.2. Cost oscillates/stagnates: increase population or generations.
   6.3. Parameter within 5% of boundary: apply rules 3.3.4-3.3.6.
   6.4. Metric exceeds target: increase that metric's weight.
   6.5. All metrics meet targets but one falls far short: reduce others.
   6.6. Metric fails 2+ attempts despite weight and range changes:
        local-minimum escape -- restrict the responsible gain's range
        (rules 3.3.5 and 3.4.2).
   6.7. Explain resource-aware reasoning in the reasoning field,
        citing applicable rule numbers.
   6.8. simulation_params (dt, max_time) are FIXED; do not modify.
\end{lstlisting}

\subsubsection{User Prompt Templates}

The user prompt templates for Variation~3 adopt the same decimal scheme as the system prompt.

The GA-Agent+ initial template opens with \texttt{1.SYSTEM INFORMATION} (sub-items: system name, control objective, system description), followed by \texttt{2.FIXED TARGETS}, \texttt{3.ATTEMPT STATUS}, \texttt{4.FEEDBACK}, and \texttt{5.TASK}.

The subsequent-attempt template inserts \texttt{3.CURRENT WEIGHTS} and \texttt{4.CURRENT GA CONFIGURATION} before \texttt{5.ATTEMPT STATUS}. The GA-Agent (no-context) template replaces section~1 with \texttt{1.PROBLEM CONTEXT}.  All placeholder strings (\texttt{\{...\}}) are identical to those in the Original.

The closing task section (e.g., \texttt{7.TASK}) instructs the agent to return valid JSON, to modify both the GA configuration and the weights, and to account for resource constraints, with a sub-item numbering that allows the agent to reference the instruction coordinate in its reasoning trace.

\end{document}